\documentclass[%
 dvipdfmx,
 reprint,
superscriptaddress,
showpacs,preprintnumbers,
amsmath,amssymb,
aps, 
prd,
10pt,
floatfix
]{revtex4-1}

\usepackage[dvips]{graphicx}
\usepackage{dcolumn}
\usepackage{bm}
\usepackage[caption=false,subrefformat=parens,labelformat=parens]{subfig}
\usepackage{color}

\def\be{\begin{equation}}
\def\ee{\end{equation}}

\def\bes{\begin{equation*}}
\def\ees{\end{equation*}}

\def\({\left(}
\def\){\right)}

\def\[{\left[}
\def\]{\right]}

\def\lm{\lambda}

\begin{document}


\title{Diluted equilibrium sterile neutrino dark matter}

\author{Amol V. Patwardhan}
\email{apatward@ucsd.edu}
\affiliation{Department of Physics, University of California, San Diego, La Jolla, CA 92093-0319, USA}
\author{George M. Fuller}
\email{gfuller@ucsd.edu}
\affiliation{Department of Physics, University of California, San Diego, La Jolla, CA 92093-0319, USA}
\author{Chad T. Kishimoto}
\email{ckishimo@physics.ucsd.edu}
\affiliation{Department of Physics, University of California, San Diego, La Jolla, CA 92093-0319, USA}
\affiliation{Department of Physics, University of San Diego, San Diego, California 92110, USA}
\author{Alexander Kusenko}
\email{kusenko@ucla.edu}
\affiliation{Department of Physics and Astronomy, University of California, Los Angeles, California 90095}
\affiliation{Kavli Institute for the Physics and Mathematics of the Universe (WPI), University of Tokyo, Kashiwa, Chiba 277-8568, Japan}





\date{\today}

\begin{abstract}
We present a model where sterile neutrinos with rest masses in the range $\sim$ keV to $\sim$ MeV can be the dark matter and be consistent with all laboratory, cosmological, large-scale structure, as well as x-ray constraints. These sterile neutrinos are assumed to freeze out of thermal and chemical equilibrium with matter and radiation in the very early Universe, prior to an epoch of prodigious entropy generation (\lq\lq dilution\rq\rq) from out-of-equilibrium decay of heavy particles. In this work, we consider heavy, entropy-producing particles in the $\sim$ TeV to $\sim$ EeV rest-mass range, possibly associated with new physics at high-energy scales. The process of dilution can give the sterile neutrinos the appropriate relic densities, but it also alters their energy spectra so that they could act like cold dark matter, despite relatively low rest masses as compared to conventional dark matter candidates. Moreover, since the model does not rely on active-sterile mixing for producing the relic density, the mixing angles can be small enough to evade current x-ray or lifetime constraints. Nevertheless, we discuss how future x-ray observations, future lepton number constraints, and future observations and sophisticated simulations of large-scale structure could, in conjunction, provide evidence for this model and/or constrain and probe its parameters.
\end{abstract}




\maketitle


\section{Introduction} \label{sec:intro}

In this paper we show how sterile neutrinos with rest masses in the $\sim$ keV to $\sim$ MeV range could evade all existing cosmological and laboratory bounds, comprise all or a significant component of the dark matter, and behave as cold dark matter (CDM). The idea of an electroweak singlet (sterile neutrino) as dark matter is not new~\cite{Dodelson:1994rt,Shi:1999lq,Abazajian:2001lr,Dolgov:2002ve, Abazajian:2002bh,Asaka:2005fx,Asaka:2005fj,Abazajian:2006qf,Asaka:2006qy, Kusenko:2006rh,Shaposhnikov:2006fj,Boyanovsky:2007lr, Boyanovsky:2007fk,Asaka:2007fk,Shaposhnikov:2007qy,Gorbunov:2007uq,Kishimoto:2008pd, Laine:2008kx, Petraki:2008yq,Petraki:2008vn, Kusenko:2009lr,Kusenko:2010uq,Bezrukov:2010rm,Nemevsek:2012zl,Canetti:2013qy,Bezrukov:2013qv, Canetti:2013lr,Merle:2013ty, Abazajian:2014zl, Tsuyuki:2014uq,Merle:2014qy,Bezrukov:2014nr,Roland:2014lr,Abada:2014fk,Lello:2015lr, Humbert:2015jk, Humbert:2015xy, Adulpravitchai:2015qv}. Some of these models run afoul of x-ray observations or large-scale structure considerations or both, as discussed in Refs.~\cite{Abazajian:2001lr,Abazajian:2001fk,Hansen:2002lr,Abazajian:2009pd,Merle:2014fk} and also in Sec.~\ref{sec:obs} below. However, many of them still remain viable. 

Most of these models posit no sterile neutrinos at extremely high temperatures in the early Universe and engineer a build-up of sterile neutrino density via active neutrino scattering-induced decoherence, lepton number-driven medium enhancement of that process, or particle decay. In a different class of models~\cite{Shaposhnikov:2006fj,Kusenko:2006rh,Petraki:2008vn,Kusenko:2010uq,Merle:2014qy}, a population of thermally decoupled sterile neutrinos forms in the early Universe with a density below the equilibrium density, and the subsequent expansion of the Universe reduces both the density and the momenta of sterile neutrinos, making them acceptable dark matter candidates. It is also possible that scattering-induced decoherence produces a population of sterile neutrinos with a number density that is initially too high, but is subsequently reduced to an acceptable level by entropy generation (\lq\lq dilution\rq\rq) effected by out-of-equilibrium decay of heavy sterile neutrinos different from the dark matter candidate sterile neutrino~\cite{Asaka:2006qy}. 

Finally, one can start with the dark matter candidate sterile neutrinos in thermal and chemical equilibrium at some high temperature~\cite{Bezrukov:2010rm,Nemevsek:2012zl,Bezrukov:2013qv,Tsuyuki:2014uq}. These sterile neutrinos eventually decouple, and, after decoupling, have their number density reduced to the required dark matter density by dilution due to out-of-equlibrium heavy particle decay. Our paper follows this approach.

In this paper we consider heavy, unstable dilution generators, henceforth referred to as \lq\lq dilutons,\rq\rq\ with rest masses in the $\sim$ TeV to $\sim$ EeV range. There is nothing sacred about this diluton mass range. It could be higher and it could be somewhat lower, though cosmological considerations discussed below may limit how low. In any case, our model requires large dilution and therefore, the diluton lifetime against decay has to be long enough for the dilutons to survive until the Universe has cooled to temperatures well below their rest mass.

A diluton particle with those properties could be, for example, another sterile neutrino, with an extremely small vacuum mixing with active neutrino flavors. The small vacuum mixing would enable evasion of laboratory neutrino mass and accelerator bounds, and cause a relatively long lifetime~\cite{Pal:1982qy,Fuller:2011lr}. This sterile neutrino diluton perhaps could be one of the heavy right-handed neutrino species invoked in the seesaw mechanism for explaining neutrino mass phenomenology. 

Another possibility is that the diluton is a supersymmetric particle that decays into standard-model particles via R-parity-violating couplings. The relatively long lifetime required for this particle could be effected by having R-parity be a nearly respected, but ultimately broken symmetry~\cite{Mohapatra:2015lr}. However, this would also mean that dark matter would not be a lightest supersymmetric particle (LSP). A supersymmetric diluton which is an LSP would require that the scale of supersymmetry be very high.

As mentioned above, the idea of diluted-equilibrium sterile neutrino dark matter (DESNDM) has been discussed before for particular kinds of dilutions. For example, gauge extensions of the standard model containing right-handed neutrinos, i.e., left-right symmetric models, can furnish diluton candidates decaying around the weak decoupling epoch (temperature $T \sim 1$ MeV)~\cite{Bezrukov:2010rm}, or the QCD scale ($T\sim 100$ MeV)~\cite{Nemevsek:2012zl}. These models can produce $\sim$ keV rest-mass-scale sterile neutrino warm dark matter. Variants of this model posit more massive dilutons, decaying above the electroweak scale ($T\sim 100$ GeV), and can produce heavier, colder dark matter~\cite{Bezrukov:2013qv,Tsuyuki:2014uq}.

In this work we shall remain agnostic about the identity and rest mass of the diluton, with the only assumptions being that it also has an equilibrium distribution prior to its decoupling and subsequent decay, and that all of its final decay products thermalize. This allows us to make sterile neutrinos in the rest-mass range $\sim$ keV to $\sim$ MeV be CDM.

Entropy generation in the early Universe can have constrainable consequences. In particular, it has been shown that entropy injection at or near the weak decoupling scale ($T \sim$ MeV) can be constrained by big bang nucleosynthesis (BBN) and radiation energy density ($N_\text{eff})$ considerations~\cite{Scherrer:1985qv,Dolgov:2000fr,Fuller:2011lr,Menestrina:2012fj,Grohs:2015lr}. However, there would be no effect on BBN and $N_\text{eff}$ if particle decay-generated entropy injection occurs sufficiently prior to weak decoupling, so long as all of the diluton decay products thermalize in the plasma of the early Universe. Evading other potential cosmological bounds may argue for an even higher temperature scale for a significant dilution event. For example, though we do not know where the baryon number is made, it could be produced at or above the electroweak scale. Our dilution event may have to occur above the temperature epoch associated with baryogenesis; otherwise, a higher pre-dilution baryon number would have to be produced, placing potentially unattainable demands on baryon generation mechanisms. Nevertheless, we also consider cases where the dilution event occurs at temperatures much lower than the electroweak scale, and these would require an appropriate accompanying baryogenesis scheme.




In Sec.~\ref{sec:tl}, we briefly discuss various mechanisms for sterile neutrino production and thermalization and describe how dilution can be incorporated into the thermal history of the Universe. This is followed by an assessment of the effects of dilution in Sec.~\ref{sec:consq}. Section \ref{sec:obs} gives an overview of the various possible means to observationally or experimentally probe our model and constrain its parameters. We conclude in Sec.~\ref{sec:concl}.

\section{Sterile neutrinos and the history of the early Universe} \label{sec:tl}

\begin{figure}[htb]
	\includegraphics[width=	0.45\textwidth]{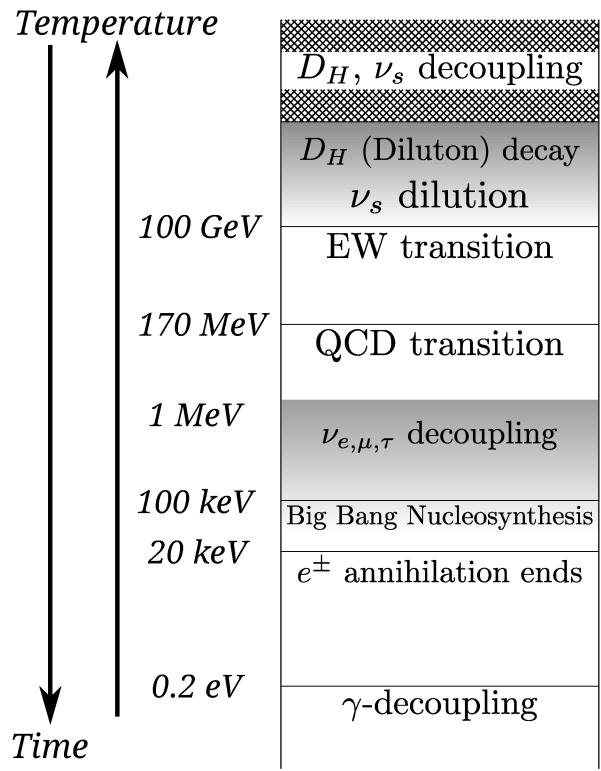}
	\caption{Cartoon illustrating how the DESNDM model fits together with the thermal history of the Universe. So long as the diluton $D_H$ decays exclusively into standard model particles, and the process goes to completion before weak decoupling, the decay products can completely thermalize. In such a scenario, all standard model particle spectra remain thermal, and neither big bang nucleosynthesis nor $N_\text{eff}$ are affected in any way. However, since the sterile neutrinos $\nu_s$ are already decoupled, their number densities are diluted relative to the particles still in equilibrium. The active neutrinos eventually decouple, and subsequently, the $e^\pm$ pairs annihilate away as the Universe cools, further diluting the active and sterile neutrino number densities relative to the photons. Scenarios with dilution built in prior to the electroweak scale, such as the one shown here, are likely better suited to meeting baryogenesis requirements.}
	\label{fig:dilution}
\end{figure}

Figure~\ref{fig:dilution} illustrates the thermal history of the early Universe and highlights key epochs in the history of active and sterile neutrino species. This figure gives a particular example of when dilution events might occur in viable scenarios of diluted equilibrium sterile neutrino dark matter (DESNDM). The dilution event is depicted in this figure to occur before the electroweak phase transition, which might be more favorable from the point of view of some baryogenesis models. However, we also consider longer-lived dilutons in this paper.

\subsection{Sterile neutrino production in the early Universe} \label{sec:prod}

Sterile neutrinos can interact subweakly with ordinary matter via their mixing with active neutrinos \cite{Pontecorvo:1968fk}. For example, in a simplified model where the sterile neutrino mixes with only one of the active species, the interaction rate of a sterile neutrino with the background plasma in the early Universe can be estimated as
\be \label{eq:scatt}
\Gamma_{\nu_s} \sim G_F^2\,T^5\,\sin^2{2\theta_m},
\ee
where $G_F$ and $T$ are the Fermi coupling constant and the plasma temperature, respectively, and $\theta_m$ is the effective active-sterile in-medium mixing angle. At high temperatures, these in-medium mixing angles can be heavily suppressed due to active neutrino interactions with background matter. At a plasma temperature $T$, the effective in-medium mixing angle for a neutrino state with momentum $p$ satisfies
\be \label{eq:mixing}
\sin^2{2\theta_m} = \frac{\Delta^2(p) \sin^2{2\theta_v}}{\Delta^2(p) \sin^2{2\theta_v} + \[\Delta(p)\cos{2\theta_v} - V_D - V_T\]^2},
\ee
where $\theta_v$ is the active-sterile mixing angle in vacuum, and $\Delta(p) = \Delta m^2/2p$, with $\Delta m^2$ being the appropriate mass-squared splitting in vacuum. $V_D$ and $V_T$ are the finite-density and finite-temperature potentials felt by the active neutrino. In the absence of an appreciable net lepton number, the finite-density potential is negligible. The finite-temperature potential, for $T \lesssim M_W$ ($M_W$ is the $W$-boson mass), is given by $V_T = -G_\text{eff}^2\,p\,T^4$, where $G_\text{eff}$ can be taken to represent some overall neutrino interaction strength, summed over all the particle species in the background plasma. 

In the limit of negligible lepton number, i.e., $V_D \sim 0$, the thermal potential $V_T$ causes heavy suppression of the effective in-medium mixing angle at high temperatures. Numerically, the thermal potential $V_T$ and the vacuum oscillation term $\Delta(p)$ can be calculated to be
\begin{align}
V_T &\sim -10 \text{ eV} \(\frac{p}{\text{GeV}}\)\(\frac{T}{\text{GeV}}\)^4, \\
\Delta(p) &\sim 0.05 \text{ eV} \(\frac{m_s}{10\text{ keV}}\)^2 \(\frac{p}{\text{GeV}}\)^{-1}.
\end{align}

For a keV--MeV rest-mass-scale sterile neutrino, $\vert V_T\vert \sim \Delta(p)$ at $T \sim 0.1\text{--}1$ GeV. At higher temperatures, the thermal potential dominates, and the scattering rate goes like $\Gamma_{\nu_s} \propto T^{-7}$, whereas at lower temperatures, where the vacuum oscillation term dominates, $\Gamma_{\nu_s} \propto T^5$. The sterile neutrino scattering rate is therefore maximal in these intermediate temperature regimes, where the vacuum and thermal terms are comparable in magnitude. For equilibration, this rate has to be greater than the expansion rate, which assuming radiation-dominated conditions is given by $H = (8\pi^3/90)^{1/2}\, g^{1/2}\, T^2/m_\text{pl}$. Here $g$ is the statistical weight in relativistic particles and the Planck mass is $m_\text{pl}$. With these considerations it can be shown that sterile neutrinos with vacuum mixing angles smaller than $\sin^2{2\theta_v} \sim 10^{-6} \, (10\text{ keV}/m_s)$ can never be in thermal equilibrium with the plasma, so long as their interactions arise solely through these mixings. Therefore, in such scenarios, their relic density cannot be set via a freeze-out process, unlike ordinary active neutrinos.

Such mixing with active neutrinos can, however, lead to sterile neutrinos being produced athermally in the early Universe via scattering-induced decoherence, as was first pointed out by Dodelson and Widrow \cite{Dodelson:1994rt}. Active neutrino scattering-induced decoherence into sterile states is driven by the active neutrino scattering rate $\Gamma_{\nu_\alpha} \sim G_F^2\,T^5$. However, this scattering also gives rise to the active neutrino matter potentials described above, as well as quantum damping, both of which serve to inhibit active-sterile neutrino conversion. The sterile neutrino production rate via this mechanism can also be shown to peak at temperatures of $\sim 0.1\text{--}1$ GeV. 

Combined constraints from x-ray and Lyman-$\alpha$ observations rule out a Dodelson-Widrow-type sterile neutrino being all of the dark matter (although it may comprise some fraction of the total $\Omega_\text{DM}$, depending on the mass and mixing angle). If the lepton number is sizable, however, the scattering-induced conversion rate can be resonantly enhanced \cite{Shi:1999lq,Abazajian:2001lr}. Sterile neutrinos which are resonantly produced in this way can have considerably lower vacuum mixing angles for a given relic density and can have \lq\lq colder\rq\rq\ energy spectra, as compared to the Dodelson-Widrow case, and can therefore evade some of these bounds. The Lyman-alpha bounds are also relaxed if dilution takes place after the production of sterile neutrinos through this mechanism~\cite{Asaka:2006qy}.

Alternatively, one could envision ways to bring the sterile neutrino into equilibrium early on by invoking some additional interactions, such as, e.g., left-right symmetric models, suitably broken at some energy scale, above which a right-handed $Z$-boson effects equilibration of this sterile neutrino with the plasma \cite{Bezrukov:2010rm,Nemevsek:2012zl,Bezrukov:2013qv,Tsuyuki:2014uq}. We adopt this approach in our paper, assuming that the dark matter candidate sterile neutrino attains an equilibrium distribution via some beyond-standard-model interactions, before freezing out. Additionally, for simplicity, we also assume the diluton to be thermally populated, either through similar or through different interactions as the dark matter candidate sterile neutrino.


There are a number of models that would produce an equilibrium population of sterile neutrinos through processes other than oscillations. In the context of grand unified theories (GUT) with the popular $SO(10)$ gauge group, the right-handed neutrino transforms as a component of a 16-dimensional representation that also includes all the standard model fermions. At the GUT scale, the right-handed neutrinos can be produced in equilibrium by the exchange of the GUT scale bosons. These interactions freeze out at lower temperatures, and the sterile neutrinos go out of equilibrium, but their distribution functions remain the same and scale with temperature. Depending on the mode of $SO(10)$ breaking, there may or may not be any entropy production between the GUT scale and the electroweak scale. 

As an example, if the breaking occurs along the route $SO(10)\rightarrow SU(5)\times U(1)$ with a subsequent breaking of $SU(5)$ down to the standard model at a scale close to the $SO(10)$ breaking scale, the population of sterile neutrinos can exist with near-thermal distribution functions. In the case of an alternative symmetry-breaking route, leading to a left-right symmetric $SU(2)_L \times SU(2)_R \times$... model, the right-handed $SU(2)_R$ gauge bosons can keep the sterile neutrinos in equilibrium down to the scale a few orders of magnitude below the $SU(2)_R$ breaking scale, which has to be at the TeV-scale or higher~\cite{Langacker:1980js}. If the right-handed neutrinos are doublets of $SU(2)_R$, the Majorana mass cannot be greater than the $SU(2)_R$ breaking scale; hence, we \lq\lq naturally\rq\rq\ get a Majorana mass much smaller than the Planck scale. Furthermore, one might expect to find the right-handed $Z$ and $W$ at the Large Hadron Collider (LHC)~\cite{Langacker:2008yv}.

More generally, any high-energy theory that includes a gauge $U(1)_{B-L}$ either as a subgroup of the GUT group (e.g., $SO(10)$), or as a stand-alone feature (e.g., split seesaw model~\cite{Kusenko:2010uq}) generates an approximately thermal population of sterile neutrinos through the exchange of the $U(1)_{B-L}$ gauge boson. This is because our sterile neutrinos carry a lepton number, and hence a nonzero $B-L$. In the case of a split seesaw model of Ref.~\cite{Kusenko:2010uq}, a first-order phase transition was employed to dilute the density of sterile neutrinos, but such dilution can be small or none if the $U(1)_{B-L}$ breaking transition is not strongly first order.

In all of these scenarios with additional interactions at high temperatures, equilibration of a massive sterile neutrino species will lead to their \lq\lq overproduction\rq\rq\ in the early Universe, if the sterile mass is greater than $m_s \sim 100$ eV. In that case, entropy generation after the sterile neutrinos have decoupled can help dilute their relic densities to a level consistent with astrophysical data.

\subsection{Thermal decoupling}

The DESNDM model assumes that at very early times, the keV--MeV mass sterile neutrino, as well as the diluton are in thermal and chemical equilibrium with matter and radiation. A particle species thermally decouples from this plasma when its scattering rate falls below the expansion rate of the Universe (or equivalently, its mean free path becomes longer than the Hubble length). This could happen, for example, if the interactions responsible for keeping these particles in equilibrium in the very early Universe weaken considerably, or cease to operate below a certain energy scale, e.g., following the breaking of some symmetry as in the above examples.

\subsection{Decay-induced dilution}

In our model, we assume that the heavy diluton decays into standard model particles (such as photons, pions and a variety of charged and neutral leptons) after both the diluton and the dark matter sterile neutrino have decoupled, thus pumping vast amounts of entropy into the plasma. Let us define $S \equiv s \cdot a^3$ as the comoving entropy of the plasma, where $s = (2 \pi^2/45) \,g_s\, T^3$ is the entropy density. Here, $g_s$, $a$ and $T$ are the effective entropic degrees of freedom, the scale factor, and the plasma temperature, respectively. The symmetries of a Friedmann-Lema\^itre-Robertson-Walker (FLRW) spacetime imply that the comoving entropy is conserved as long as there are not any timelike heat flows. However, out-of-equilibrium particle decays can source such heat flows. The rate of change of comoving entropy resulting from diluton decay is given by
\be \label{eq:dsdt}
\frac{dS}{dt} = \frac{n_H \cdot a^3}{\tau_H} \cdot \frac{m_H}{T} \cdot f_T,
\ee
where $m_H$, $\tau_H$, and $n_H$ are the mass, lifetime, and number density, respectively, of the diluton, and $f_T$ is the fraction of the total mass energy of decay products which thermalizes in the plasma. Here, we are assuming $T \ll m_H$, so that the decaying particle's energy can be approximated as its rest mass. Assuming that the dilutons were initially in thermal equilibrium, their number density relative to photons is given by
\be \label{eq:nhnb}
\frac{n_H}{n_\gamma} = \frac{3}{4} \(\frac{T_H}{T}\)^3 e^{-t/\tau_H},
\ee
where we have assumed a relativistic Fermi-Dirac shaped energy/momentum distribution function for the diluton with a temperature parameter $T_H$, and with $g = 2$ spin degrees of freedom. This assumption about the shape of the diluton distribution function is tantamount to an assumption that the diluton particles are relativistic when they decouple. Putting together Eqs. (\ref{eq:dsdt}) and (\ref{eq:nhnb}), we can write
\be \label{eq:entrt}
\frac{1}{S}\frac{dS}{dt} = \frac{135 \, \zeta(3)}{4\pi^4\, g_s} \cdot f_T \cdot \frac{m_H}{T} \cdot \frac{1}{\tau_H} \(\frac{T_H}{T}\)^3 e^{-t/\tau_H},
\ee
where we have used $S/(a^3 \, n_\gamma) = \pi^4\,g_s/(45 \,\zeta(3))$. This allows us to numerically compute the entropy added to the plasma at each time-step, for a given diluton mass and lifetime. Assuming that the diluton decays exclusively into standard-model particles, and that the decay occurs well before active neutrino (i.e., weak) decoupling, all the decay products can fully thermalize (i.e., $f_T = 1$), thus preventing some of the entropy from \lq\lq leaking away\rq\rq\ into decoupled particles. Defining the \lq\lq dilution factor\rq\rq\ $F \equiv S_\text{final}/S_\text{initial}$, as the ratio of the comoving entropies of the plasma after and before the dilution event, we can write
\be \label{eq:entro}
	g_{s}\, a^3\, T^3 = g_{s,i}\, a_i^3 \,T_i^3 \,F.
\ee
Here, the subscript \lq\lq$i$\rq\rq\ (initial) appearing on the right-hand side of the equation is supposed to indicate the onset of dilution, whereas the quantities on the left-hand side (without this subscript) are to represent any postdilution epoch. The dillution factor can be $F \gtrsim 10$, so that most of the entropy of the Universe is generated in this decay-induced dilution scenario. The DM candidate steriles, which are assumed to decouple prior to diluton decay, do not benefit from any of this entropy injection, and the temperature parameter $T_{\nu_s}$ that characterizes their energy distribution simply redshifts inversely with the scale factor, i.e., $ a \, T_{\nu_s} = a_i\, T_{\nu_s,i}$. Assuming $T_{\nu_s,i} = T_i$ (i.e., steriles initially in equilibrium), we can write
\be
	\frac{T_{\nu_s}}{T} = \(\frac{g_s}{g_{s,i}}\cdot\frac{1}{F}\)^{1/3}.
\ee

The \lq\lq cooling\rq\rq\ of the sterile neutrino sea relative to the plasma can, therefore, be seen to be a combined effect of dilution from particle decay, as well as the disappearance of statistical degrees of freedom as the Universe cools. The latter effect is most prominent across the quark-hadron transition at $T \approx 170$ MeV, where $g_s$ drops sharply by a factor of $\sim 3$. Overall, $g_s$ decreases from $\sim 100$ at $T \gtrsim 100$ GeV, to $g_s \approx 10.75$ by the time the active neutrinos decouple at $T \sim$ MeV, at which point the only relativistic degress of freedom are the photons, electrons, positrons, neutrinos and anti-neutrinos. Subsequently, as the temperature drops significantly below the electron rest mass, the $e^\pm$ pairs annihilate away transferring most of their entropy to photons, which offsets the temperature of active and sterile neutrinos relative to the photons by a factor of $\approx (4/11)^{1/3}$. As a result, $g_s \approx 43/11$ at late times, taking into account the entropy in photons, neutrinos, and anti-neutrinos.

Since the number densities of sterile neutrinos and thermally coupled particles are $\sim T_{\nu_s}^3$ and $\sim T^3$, respectively, this process results in an effective decrease in the number of sterile neutrinos relative to the plasma, diluting their relic density down from an initial thermal value. And as long as the diluton decay happens sufficiently early and all decay products completely thermalize in the plasma, no imprint is left on $N_\text{eff}$ or BBN.

\begin{figure}[htb]
	\centering
	\subfloat{\label{fig:PeV}\includegraphics[width=0.45\textwidth]{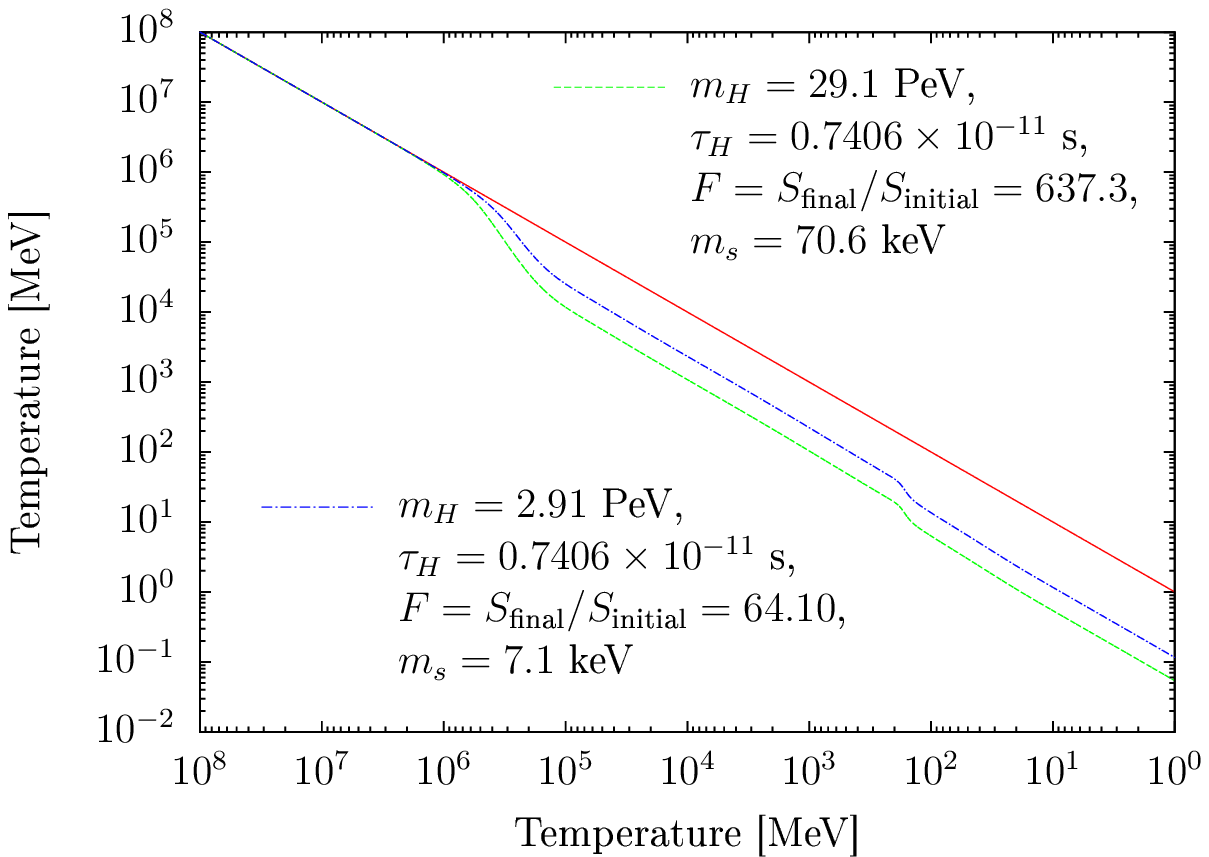}} \\ 
	\subfloat{\label{fig:TeV}\includegraphics[width=0.45\textwidth]{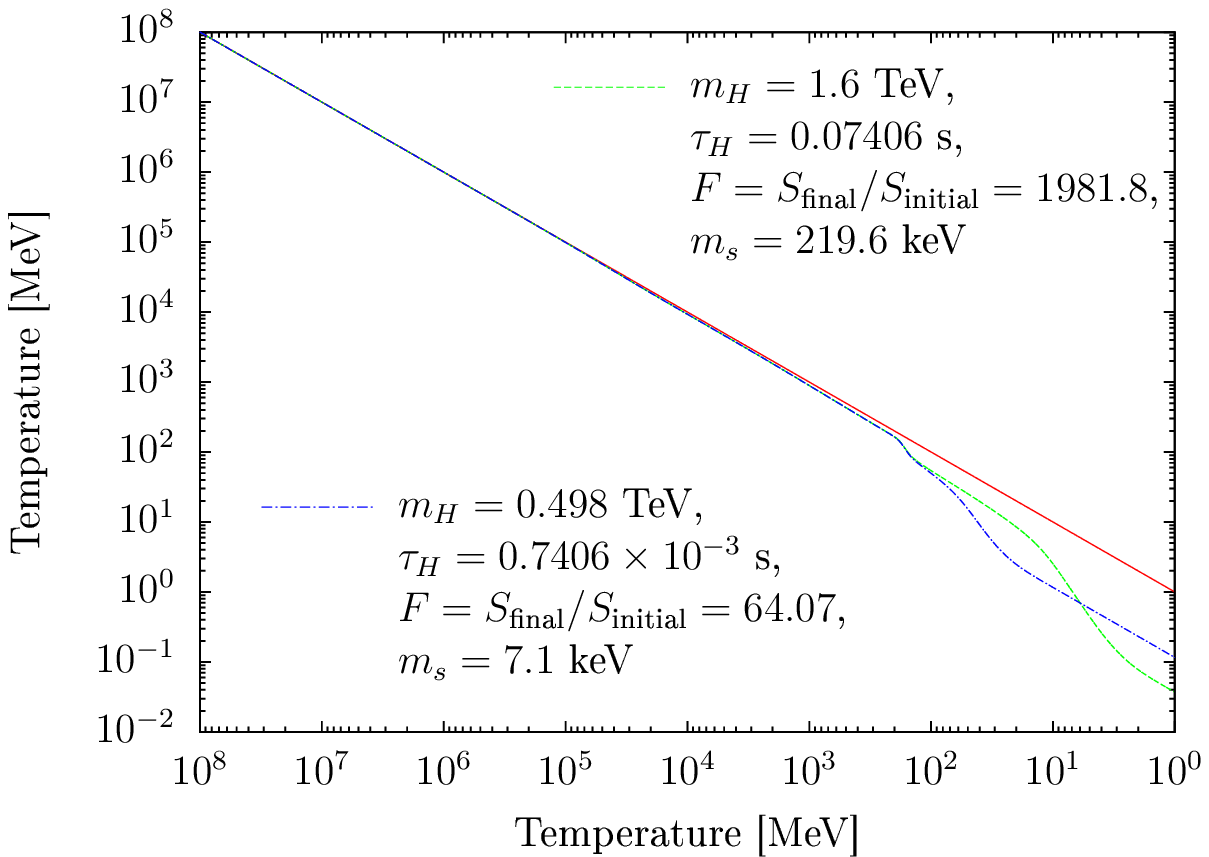}} 
	\caption{Active neutrino (solid, red) and sterile neutrino (dashed, green and dot-dashed, blue) cooling curves as a function of plasma temperature, illustrating how a combination of out-of-equilibrium particle decay and loss of statistical degrees of freedom can dilute decoupled particles relative to species still in thermal equilibrium. The curves begin to separate as the diluton starts decaying, pumping entropy into the plasma and diluting the the number density of the sterile neutrino sea. The figures are for the cases with diluton rest mass $m_H$ and lifetime $\tau_H$, and associated dilution factor $F$, and dark matter sterile neutrino rest mass $m_s$, as labeled. The sterile neutrino mass, for a given diluton mass and lifetime, is chosen to give closure fraction $\Omega_s = 0.258$ for Hubble parameter $h = 0.6781$ (in units of $100\,\text{km}\,\text{s}^{-1}\,\text{Mpc}^{-1}$).}
	\label{fig:dilcurves}
\end{figure}

Figure \ref{fig:dilcurves} illustrates the process of dilution by plotting the active and sterile neutrino temperature curves against the plasma temperature. Figure \ref{fig:PeV} (top) gives examples of dilution events occuring prior to the electroweak scale, whereas Fig.~\ref{fig:TeV} (bottom) considers postelectroweak dillution cases. The sterile neutrino mass $m_s$ is calculated in each case so as to get the appropriate relic density for them to be the dark matter, as explained in the following section. In particular, the case of $m_s \approx 7.1$ keV has attracted recent interest in light of certain x-ray observations (see Sec.~\ref{sec:xray}). However, the range of applicability of this model extends beyond just that particular case.

\section{Consequences of dilution} \label{sec:consq}

\subsection{Dark matter particle mass and relic density}

We have seen how dilution from out-of-equilibrium particle decay subsequent to sterile neutrino decoupling, but prior to active neutrino decoupling, will dilute the dark matter steriles relative to the active neutrinos. In turn, $e^\pm$-pair annihilation subsequent to active neutrino decoupling cools the active and sterile neutrinos relative to the photons by the usual factor of $\approx (4/11)^{1/3}$, so that the ratio of sterile neutrino temperature to photon temperature $T_\gamma$ at much later epochs is
\be \label{eq:Tnus/T}
\frac{T_{\nu_s}}{T_\gamma} = \[\frac{4}{11} \cdot \frac{g_{s,\text{wd}}}{g_{s,i}} \cdot \frac{1}{F} \]^{1/3},
\ee
where $g_{s,\text{wd}} \approx 10.75$ is the statistical weight for entropy at the active neutrino decoupling epoch.

Since the sterile neutrinos would be expected to be nonrelativistic at the present epoch, their energy density would simply be the product of their number density and rest mass, $\rho_{\nu_s} = n_{\nu_s}\,m_s$. However, since these particles would decouple while they were still relativisitc, their energy distribution would retain its relativistic Fermi-Dirac shape, with a temperature parameter $T_{\nu_s}$ that redshifts inversely with the scale factor. We can, therefore, write $\rho_{\nu_s} = {\left[ (3\, \zeta(3) T_{\nu_s}^3)/(2 \pi^2) \right]} \cdot m_s$, where $\zeta(3) \approx 1.20206$ is the zeta function of argument $3$, and where we add in both right- and left-handed sterile states. This implies that the sterile neutrino rest-mass contribution to closure is $\Omega_s = \rho_{\nu_s}/\rho_\text{crit}$, where the critical density is $\rho_\text{crit} = 3\,H_0^2\,m_\text{pl}^2/ 8 \pi$, and where $H_0$ and $m_\text{pl}$ are the Hubble parameter at the current epoch, and the Planck mass, respectively.

Using Eq.~(\ref{eq:Tnus/T}), and given the observed cosmic microwave background (CMB) temperature, $T_{\gamma0} = 2.725\,\text{K}$, and a Hubble parameter $h \equiv H_0/\(100\,\text{km}\text{ s}^{-1}\,\text{Mpc}^{-1}\)$, the sterile neutrino rest mass which would account for a closure parameter $\Omega_s$ at the current epoch is
\be \label{eq:ms}
\begin{split}
	m_s &= \frac{11\pi}{16\zeta(3)} \cdot \frac{m_\text{pl}^2\,H_0^2}{T_{\gamma0}^3} \cdot \frac{g_{s,i}}{g_{s,\text{wd}}} \cdot F \cdot \Omega_s \\
		&\approx 2.26\text{ keV} \(\frac{g_{s,i}/g_{s,\text{wd}}}{10}\) \(\frac{F}{20}\) \(\frac{\Omega_sh^2}{0.12}\).
\end{split}
\ee

\begin{figure}[htb]
	\centering
	\subfloat{\label{fig:dilctr912}\includegraphics[width=0.9\linewidth]{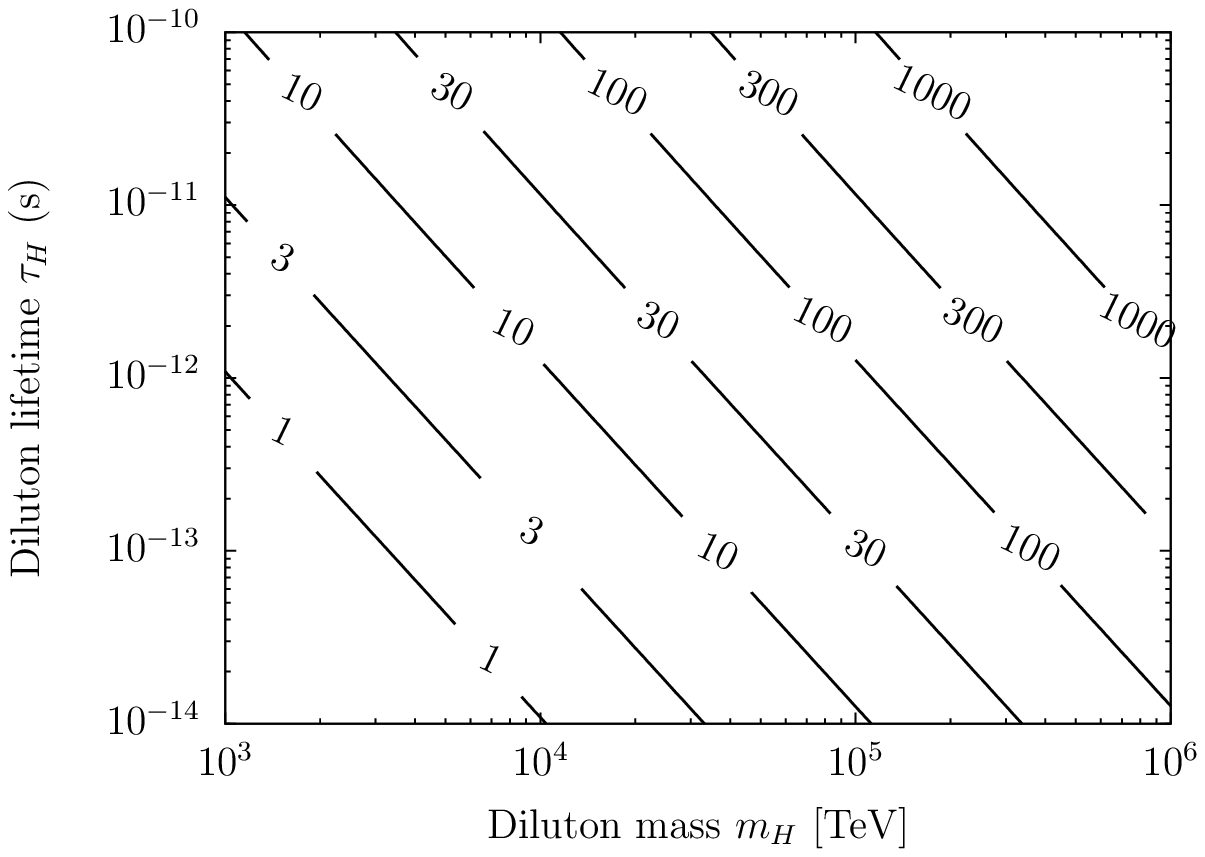}} \\
	\vskip 20pt
	\subfloat{\label{fig:dilctr69}\includegraphics[width=0.9\linewidth]{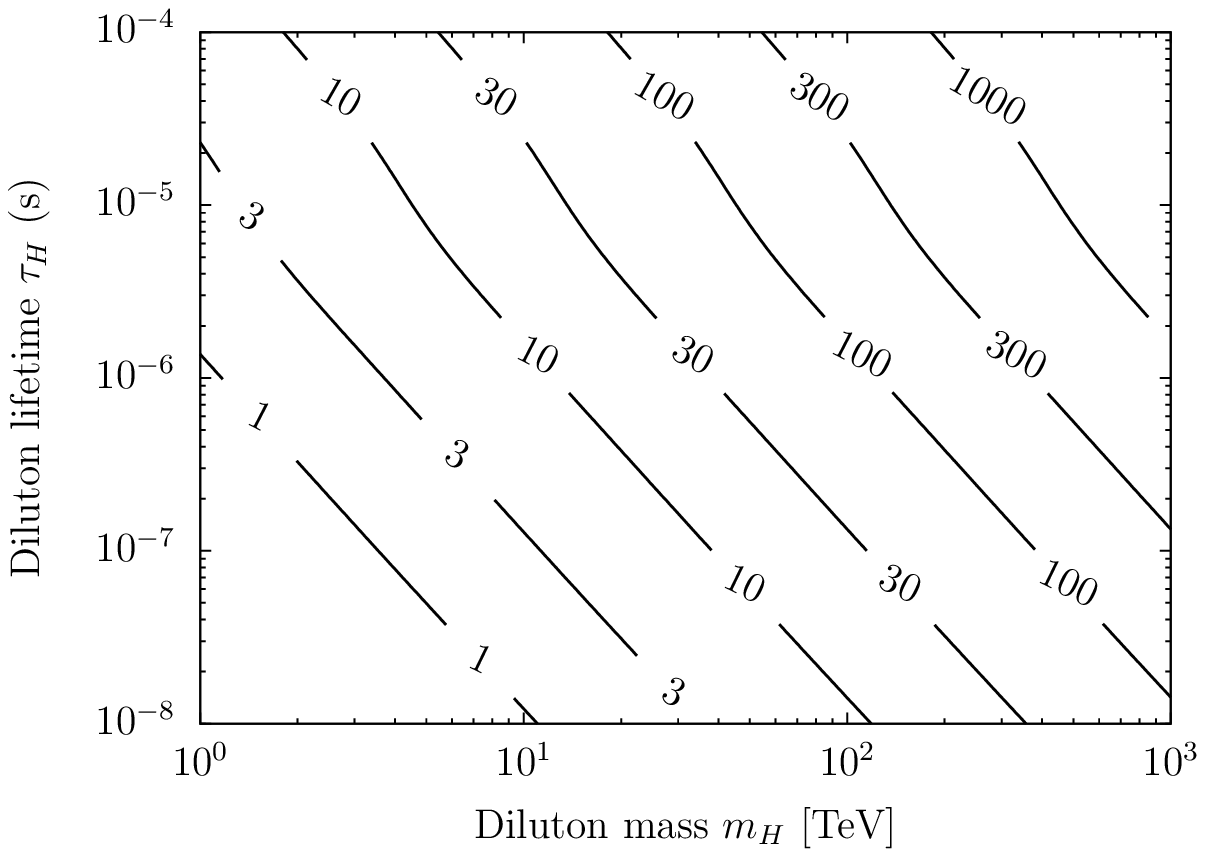}}
	\caption{Figures showing contours of the sterile neutrino rest mass $m_s$ (in keV) which would give a current relic abundance of $\Omega_\text{s} = 0.258$ for Hubble parameter $H_0 = 67.81$ km/s Mpc$^{-1}$ in the DESNDM model, plotted against a parameter space spanned by diluton rest mass $m_H$ and decay lifetime $\tau_H$.}
	\label{fig:dilctr}
\end{figure}

Figure \ref{fig:dilctr} shows contours of sterile neutrino rest mass (in keV) that would give the measured dark matter relic abundance, for different diluton rest masses and lifetimes. Figure \ref{fig:dilctr912} (top) explores diluton rest masses in the PeV--EeV range, whereas Fig.~\ref{fig:dilctr69} (bottom) has dilutons with TeV--PeV rest masses, but longer lifetimes.

In summary, the diluton rest mass and lifetime together determine the dilution factor (i.e., the ratio of final to initial comoving entropy), which in turn picks out an appropriate sterile neutrino rest mass for them to be the dark matter. 

\subsection{Dark matter collisionless damping scale} \label{sec:dmcd}

Dark matter particles can be classified as \lq\lq hot,\rq\rq\ \lq\lq warm,\rq\rq\ or \lq\lq cold,\rq\rq depending on their collisionless damping scale. For a sterile neutrino distribution characterized by a temperature parameter $T_{\nu_s}$, the collisionless damping (free-streaming) length scale, comoving to the current epoch, can be estimated as
\be
\begin{split}
\lm_\text{FS} &\approx 0.27\,\text{Mpc} \(\frac{\text{keV}}{m_s}\) \(\frac{T_{\nu_s}}{T}\) \(\frac{\langle p/T\rangle_{\nu_s}}{3.15}\) \\ &\times \[7 + \ln\(\frac{m_s}{\text{keV}} \cdot \frac{T}{T_{\nu_s}} \cdot \frac{3.15}{\langle p/T\rangle_{\nu_s}} \cdot \frac{0.14}{\Omega_mh^2}\)\],
\end{split}
\ee
where we have adopted the analysis from Ref.~\cite{Kolb:1990bs}, with slight modifications. Here, $\Omega_m$ is the fraction of energy density at the current epoch contributed by all nonrelativistic matter (dark matter $+$ baryons). In the DESNDM model, the average ratio of momentum to temperature for the decoupled sterile neutrino energy distribution is given by the thermal value, $ \langle p/T\rangle_{\nu_s} \approx 3.15$, since the spectrum retains its thermal, Fermi-Dirac shape postdilution, albeit with a significantly reduced temperature parameter.

Although some of the sterile neutrino rest masses discussed here may appear low enough to be in trouble with some of the more stringent bounds based on observed structure in the Lyman-$\alpha$ forest \cite{Seljak:2006fk}, this is not the case. The reason is that these diluted sterile neutrinos would have small collisionless damping lengths, since the temperature parameter that characterizes their energy distribution is lower than the photon temperature. Therefore, these sterile neutrinos effectively behave as if they were thermal particles with a higher mass, $m_s^\text{cd} \equiv m_s\,\(T/T_{\nu_s}\)$, so that
\be
\lm_\text{FS} \approx 0.27\,\text{Mpc} \(\frac{\text{keV}}{m_s^\text{cd}}\) \[7 + \ln\(\frac{m_s^\text{cd}}{\text{keV}} \cdot \frac{0.14}{\Omega_mh^2}\)\].
\ee

Here, the superscript \lq\lq cd,\rq\rq\ short for \lq\lq collisionless damping,\rq\rq\ is being used to emphasize that this thermally adjusted particle mass is the relevant parameter that determines the collisionless damping scale. Using Eqs. (\ref{eq:Tnus/T}) and (\ref{eq:ms}), this can be expressed as
\be
\begin{split}
m_s^\text{cd} &= m_s\, \(\frac{11}{4}\)^{1/3} \(\frac{g_{s,i}}{g_{s,\text{wd}}}\)^{1/3} \,F^{1/3} \\
&\approx 18.5\,\text{keV}\, \(\frac{g_{s,i}/g_{s,\text{wd}}}{10}\)^{\frac{4}{3}} \(\frac{F}{ 20}\)^{\frac{4}{3}} \(\frac{\Omega_sh^2}{0.12}\).
\label{mscd}
\end{split}
\ee

For all but the lightest sterile neutrinos in the rest-mass range under consideration, this effective mass is above the $\approx 13\,\text{keV}$ limit where damping of structure from dark matter particle free streaming could come into conflict with observation \cite{Seljak:2006fk}. Therefore, in this model, a relatively light ($m_s \sim$ a few keV) sterile neutrino can also function as cold dark matter. This is because the high dilution factor that is required to get the correct relic density, also serves to suppress the free-streaming scale. The total mass inside the sterile neutrino free streaming scale, $M_\text{FS} \equiv (4\pi/3)\,\lm_\text{FS}^3\, \rho_{m}$, can be calculated as
\be \label{eq:mfs}
\begin{split}
M_\text{FS} &\approx 4 \times 10^5 \, M_\odot \, \(\frac{20\,\text{keV}}{m_s^\text{cd}}\)^3 \(\frac{\Omega_{m}\,h^2}{0.14}\) \\
&\quad \times \[10 + \ln\(\frac{m_s^\text{cd}}{20\,\text{keV}} \cdot \frac{0.14}{\Omega_mh^2}\)\]^3,
\end{split}
\ee
where $\rho_{m}$ is the total energy density in nonrelativistic matter (baryons $+$ dark matter) at the present epoch, corresponding to closure fraction $\Omega_{m}$. $M_\text{FS}$ defines a mass scale for fluctuations, below which they would experience considerable damping via dark matter particle free streaming. Again, it must be emphasised that the effective thermally-adjusted particle mass $m_s^\text{cd}$ that sets this scale is not the sterile neutrino rest mass, but is scaled relative to it by the ratio of photon temperature to sterile neutrino temperature.

\begin{figure}[htb]
	\centering
	\subfloat{\label{fig:mfs2}\includegraphics[width=0.9\linewidth]{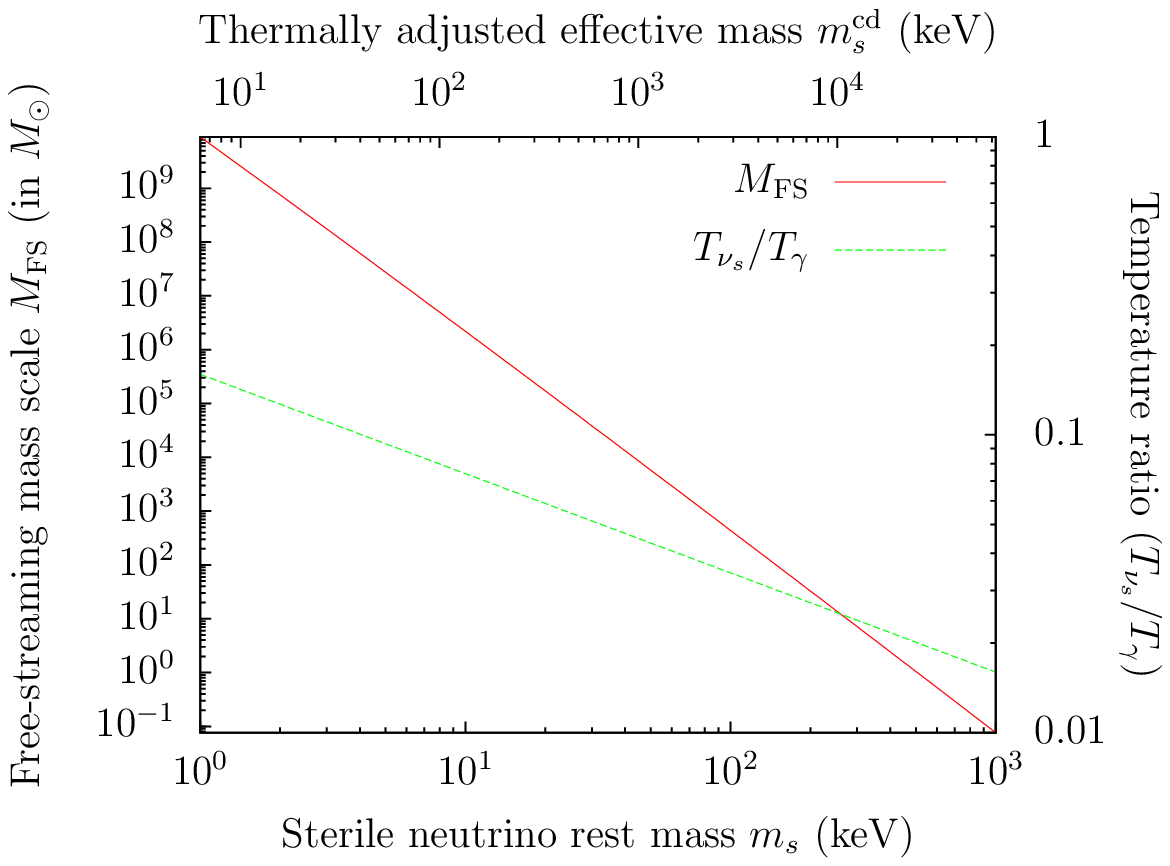}} \\
	\vskip 20pt
	\subfloat{\label{fig:mfsctr}\includegraphics[width=0.9\linewidth]{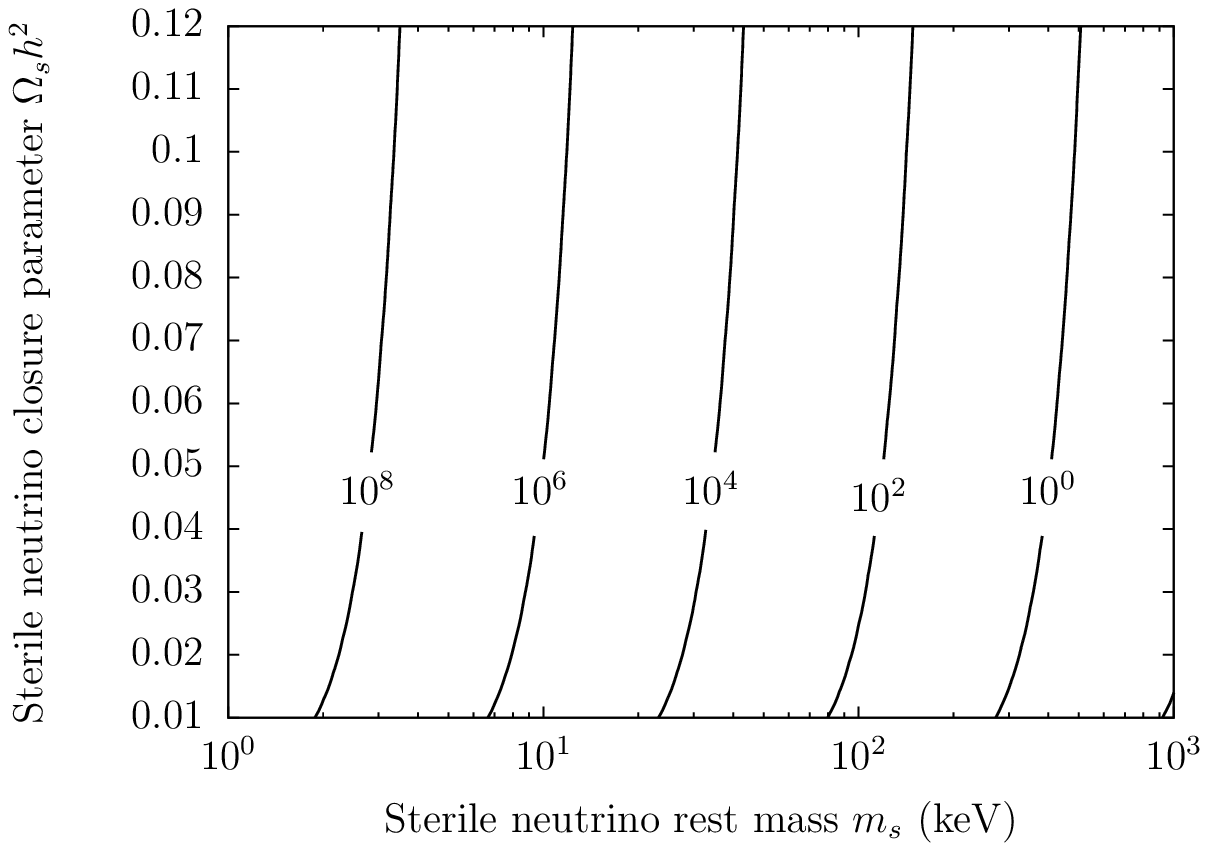}}
	\caption{Top: total mass inside the sterile neutrino free-streaming (i.e., collisionless damping) scale, $M_\text{FS}$, as a function of sterile neutrino rest mass $m_s$, for $\Omega_s\,h^2 = 0.12$. Also shown are the temperature ratio $T_{\nu_s}/T_\gamma$ (on the $y2$ axis) as a function of $m_s$, for $\Omega_s\,h^2 = 0.12$, and the corresponding thermally adjusted effective mass $m_s^\text{cd}$ (on the $x2$ axis). Bottom: contours of $M_\text{FS}$, in solar masses, plotted against sterile neutrino rest mass $m_s$ on the $x$ axis, and sterile neutrino closure density parameter $\Omega_s h^2$ on the $y$ axis. }
	\label{fig:mfs}
\end{figure}

Figure \ref{fig:mfs} depicts how the total mass inside the sterile neutrino free-streaming scale varies as a function of sterile neutrino rest mass, in this model. In Fig.~\ref{fig:mfs2} (top), we consider the case where the diluted-equilibrium sterile neutrinos are all of the dark matter, i.e., $\Omega_s\,h^2 = \Omega_\text{DM}\,h^2 \approx 0.12$, whereas in Fig.~\ref{fig:mfsctr} (bottom), we allow the closure density parameter $\Omega_s\,h^2$ to vary, in order to account for situations where these sterile neutrinos may not be all of the dark matter. Aside from the relatively unimportant logarithmic dependence, the variation of $M_\text{FS}$ with $m_s$ and $\Omega_s\,h^2$ can be quantified as $M_\text{FS} \propto m_s^{-4}\cdot \Omega_s\,h^2$.

For a broad range of sterile neutrino rest mass and diluton properties, the DESNDM model can produce what is effectively CDM, at least as far as the existence of a Lyman-$\alpha$ forest is concerned. However, in certtain ranges, the sterile neutrino character may not be entirely CDM-like, from the point of view of certain other aspects of structure-formation. For example, for sterile neutrino rest masses in the range $\sim 5$--$10$ keV in the DESNDM model, collisionless damping scales can be $\sim 10^7\,M_\odot$. These could fall in a range of interest for the core/cusp problem and other issues in dwarf galaxy morphology under current investigation. Further discussion follows in Sec.~\ref{sec:struct}.

\subsection{Matter-dominated epochs in the early Universe} \label{sec:matdom}

\begin{figure}[htb]
	\includegraphics[width=	0.45\textwidth]{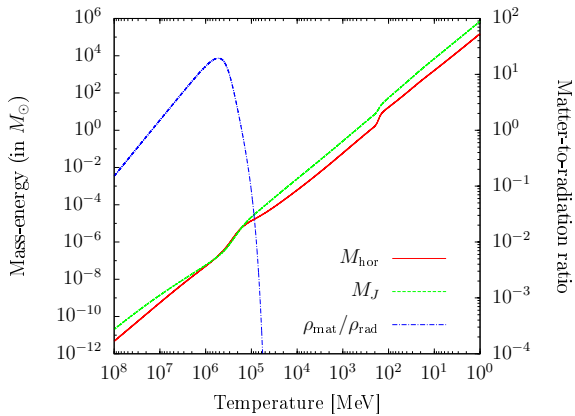}
	\caption{Curves showing the total horizon mass energy (solid, red), the Jeans mass (dashed, green), and the ratio of the energy density contributed by diluton rest mass to the total energy density in radiation (dot-dashed, blue, plotted on $y2$ axis), as a function of plasma temperature. The Jeans mass can be seen to drop relative to the horizon mass in the matter-dominated epochs. The features (bends) in the $M_J$ and $M_\text{hor}$ curves at $T \sim 100$ MeV are a consequence of the relatively sudden change in relativistic degrees of freedom across the QCD transition. The diluton rest mass and lifetime in this example are $m_H = 2.91\,\text{PeV}$ and $\tau_H = 0.7406\times {10}^{-11}\,\text{s}$, respectively.}
	\label{fig:zhormas}
\end{figure}

The presence of a heavy particle with a number density comparable to that of thermal particles raises the possibility of an epoch of matter domination in the early Universe. For example, a $\sim$ PeV rest-mass-scale diluton with a lifetime of order $10^{-11}\,\text{s}$ will linger around until the temperature of the plasma has dropped to about a $100$ GeV. This means that there will be a period of time where the total energy density of the Universe is dominated by the diluton rest mass. This is illustrated in Fig.~\ref{fig:zhormas} (dot-dashed, blue curve).

Because of this effect, the causal horizon proper length starts to increase, from a value of $d_\text{hor}(t) = 2\,t$ in radiation-dominated conditions, towards $d_\text{hor}(t) = 3\,t$ in matter-dominated conditions. In a radiation-dominated Universe, the total mass energy contained in the causal horizon, $M_\text{hor} = (4\pi/3)\, d_\text{hor}^3 \,\rho_\text{tot}$, is a factor of few smaller than the total Jeans mass, $M_J = (4\pi/3)\, \lambda_J^3\, \rho_\text{tot}$, which is the mass scale above which gravitation can overcome pressure support and cause fluctuations to grow in amplitude. Here, $\rho_\text{tot}$ is the total energy density, and the Jeans length is $\lambda_J = c_s\,m_\text{pl}/\sqrt{\rho_\text{tot}}$, with $c_s = 1/\sqrt{3}$ as the sound speed (even when the energy density is dominated by the diluton rest mass, it is the thermally coupled, relativistic particles in the plasma that provide the pressure support, and hence determine the sound speed).

An epoch of early matter domination can cause the Jeans mass to drop below the causal horizon mass scale, meaning perturbations that are in causal contact can now start to grow in amplitude. For our test case of a $2.91$ PeV rest-mass diluton with a lifetime $\tau_H = 0.7406 \times 10^{-11}$ s, this phase lasts for less than a decade in temperature (see Fig.~\ref{fig:zhormas}), i.e., a few Hubble times. For this particular choice of parameter values, the relevant mass scale of the fluctuations in this regime is $10^{-7}\text{--}10^{-5}\, M_\odot$. Given the small horizon mass scale, the limited interval of matter domination, and significant radiation content of the plasma, it is unlikely that any nonlinear regime fluctuations produced in this epoch can survive to later epochs with appreciable and constrainable amplitudes \cite{Heckler:1993qy,Dolgov:1993fk,Jedamzik:1994lr,Jedamzik:1994uq}.


\section{Observational and experimental handles} \label{sec:obs}

\begin{figure}[htb]
	\includegraphics[width=	0.45\textwidth]{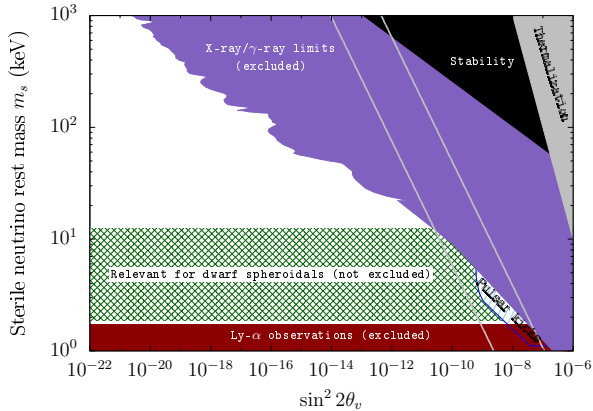}
\caption{Sterile neutrino rest mass and vacuum mixing parameter space, as constrained by x-ray/$\gamma$-ray observations (constraints from Refs.~\cite{Boyarsky:2007zr,Boyarsky:2008pd,Ng:2015th,Watson:2006rm,Boyarsky:2006yq,Boyarsky:2006kx} shown here in purple), Ly-$\alpha$ forest limits on collisionless damping (from Ref.~\cite{Seljak:2006fk}, shown here in dark red), as well as the requirements that dark matter is stable (black) and does not rethermalize after dilution (gray). Also shown are the regions of allowed parameter space that can produce some observable effects on dwarf-galaxy morphology (green, cross-hatched region corresponds to $M_\text{FS} \sim 10^6\text{--}10^9 M_\odot$ in the DESNDM model), and on pulsar kicks (sky-blue, diagonally hatched region reproduced from Ref.~\cite{Kusenko:2009lr}), respectively. The x-ray/$\gamma$-ray constraints shown here are premised on sterile neutrinos being all of the dark matter. The rest-mass ranges encompassed by the Ly-$\alpha$ and dwarf galaxy regions shown here are specific to the DESNDM model. Finally, the gray solid lines are contours of sterile neutrino closure fraction produced by the Dodelson-Widrow mechanism: $\Omega_\text{DW}$ = 0.26 (right) and 0.0055 (left).}
	\label{fig:constr}
\end{figure}

Broadly speaking, the parameter space of sterile neutrino rest mass and vacuum mixing angle can be constrained by dark matter stability considerations, various x-ray observations, and also through kinematic arguments, e.g., phase space considerations \cite{Tremaine:1979nr}, and bounds on dark matter collisionless damping from Ly-$\alpha$ forest observations. Some of these constraints are summarized in Fig.~\ref{fig:constr} and described in further detail in the subsequent subsections.

\subsection{Dark matter stability considerations} \label{sec:DMtau}

For any particle to be an acceptable dark matter candidate, it has to be stable against decay or annihilation processes over the lifetime of the Universe ($t_0 \sim 4 \times 10^{17}$ s). Massive sterile neutrinos are able to decay by virtue of the posited mixing with active species in vacuum. For a sterile neutrino in the keV--MeV 
 range, the predominant disappearance channel is a tree-level decay mediated by a $Z$ boson, into three active neutrinos, i.e., $\nu_s \rightarrow 3\nu$. For a sterile neutrino that mixes with a single active species, the decay rate is given by
\be
\Gamma_{\nu_s \rightarrow 3\nu} = \frac{G_F^2}{192 \pi^3}\, m_s^5 \, \sin^2{\theta_v}.
\ee

Requiring that the lifetime $\tau_s \approx 1/\Gamma_{\nu_s \rightarrow 3\nu} \gtrsim 10^{18}$ s leads to the following constraint on the sterile neutrino mass-mixing parameter space
\be \label{eq:DMtauconst}
\(\frac{m_s}{10\,\text{keV}}\)^5\(\frac{\sin^2{2\theta_v}}{10^{-10}}\) \lesssim 10^7.
\ee

If the sterile neutrino rest mass is greater than twice the electron rest mass, i.e., $m_s \gtrsim$ MeV, then that opens up an additional decay channel $\nu_s \rightarrow \nu e^+e^-$. The rate for this channel is one-third the decay rate into $3\nu$, resulting in a factor of $4/3$ enhancement in the overall decay rate, making the above constraint slightly more stringent at these relatively higher rest masses.

\subsection{X-ray observations} \label{sec:xray}

In addition to the aforementioned decays via tree-level weak processes, there is a radiative electromagnetic decay branch arising via one-loop interactions \cite{Pal:1982qy,Abazajian:2001lr,Abazajian:2001fk} which provides a photon with energy $m_s/2$. The decay rate for this electromagnetic branch is given by 
\be
\begin{split}
\Gamma_{\nu_s \rightarrow \nu\gamma} &= \sin^2{2\theta_v}\,\alpha\,G_F^2\,\(\frac{9m_s^5}{2048 \pi^4}\) \\
&\approx 6.8 \times 10^{-33} \text{ s}^{-1} \(\frac{\sin^2{2\theta_v}}{10^{-10}}\)\(\frac{m_s}{1\text{ keV}}\)^5.
\end{split}
\ee

Although this rate is $\mathcal{O}(\alpha)$ smaller compared to the $3\nu$ channel, the fact that it leaves an electromagnetic imprint makes it much more viable for indirect detection \cite{Abazajian:2001fk,Hansen:2002lr}. Consequently, this radiative decay channel has been used to give the currently most stringent constraints on many models for sterile neutrino dark matter. X-ray observations of the Milky Way~\cite{Boyarsky:2007zr,Boyarsky:2006fj,Yuksel:2008fr,Boyarsky:2008pd,Ng:2015th}, Andromeda (M31) and other local group galaxies~\cite{Horiuchi:2014zl,Watson:2012rt,Watson:2006rm,Boyarsky:2008gf}, dwarf spheroidals~\cite{Loewenstein:2009lr,Riemer-Sorensen:2009qy,Loewenstein:2010fk,Loewenstein:2012fk}, and galaxy clusters~\cite{Boyarsky:2008kx,Boyarsky:2006yq,Riemer-Sorensen:2007vn,Riemer-Sorensen:2015lr}, as well as measurements of the diffuse and unresolved cosmic x-ray backgrounds~\cite{Boyarsky:2006kx,Abazajian:2007ys}, have been used to constrain the parameter space of rest mass and vacuum mixing for sterile-neutrino dark matter models.

The model proposed here can evade these constraints, again effectively because of the dilution involved in creating their relic densities. Since the active-sterile mixing angle has no bearing on the relic density in this model, it can be made arbitrarily small. Since the decay-photon emissivity from the sterile neutrinos is proportional to the square of the appropriate vacuum mixing angle, all the above bounds could therefore be evaded.

\subsubsection{The $3.55$ keV x-ray line}

Recent analysis~\cite{Bulbul:2014lr,Boyarsky:2014fk,Boyarsky:2014qy} of x-ray emission from various sources has led to the detection of a previously unidentified monochromatic x-ray emission at a photon energy of around $3.55$ keV, possibly arising from electromagnetic decay of a $7.1$ keV rest-mass sterile neutrino into an active neutrino and a photon \cite{Abazajian:2014ys}. From the observed fluxes, and with the assumption that the sterile neutrinos constitute all of the dark matter, the inferred best-fit vacuum mixing angles are $\sin^2{2\theta_v} \approx 7 \times 10^{-11}$. While the existence of this line in terms of statistical significance, as well as its interpretation as having a dark matter origin are still up for debate \cite{Malyshev:2014fr,Lovell:2014zr,Anderson:2014mz,Jeltema:2014ly,Riemer-Sorensen:2014fp,Urban:2014db,Bulbul:2014xe,Boyarsky:2014uo,Jeltema:2014qq,Iakubovskyi:2014ek,Carlson:2015gf,Tamura:2015hl}, the possibility remains intriguing, and various sterile neutrino dark matter models can have their parameters tailored to fit this particular scenario~\cite{Abazajian:2014zl,Tsuyuki:2014uq,Bezrukov:2014nr,Roland:2014lr,Abada:2014fk,Lello:2015lr,Humbert:2015jk, Adulpravitchai:2015qv}. Future x-ray telescopes such as ASTRO-H and ATHENA, as well as microcalorimeter sounding rocket experiments such as Micro-X \cite{Boyarsky:2007lr,Figueroa-Feliciano:2015fk}, with their high energy resolution, could help settle the verdict on this case one way or the other \cite{Speckhard:2015qy}.

Some of the results presented in this paper, such as in Fig.~\ref{fig:dilcurves}, as well as the discussion in Secs.~\ref{sec:matdom} and \ref{sec:theta}, have used this posited $7.1$ keV rest-mass sterile neutrino as an example. However, much of the analysis is also applicable more generally, for a wide range of sterile neutrino parameters.

\subsubsection{Looking for heavier sterile neutrino dark matter}

We have shown that our model can dilute an initial thermal distribution down to the right relic density even for much heavier sterile neutrinos, whose electromagnetic decay branches would fall outside the purview of telescopes such as Chandra, XMM-Newton or Suzaku. However, some of this higher rest-mass range would lie in a suitable regime for other x-ray/$\gamma$-ray telescopes, e.g., Fermi-GBM, which can probe the $20\text{--}50$ keV rest-mass range for sterile neutrinos~\cite{Ng:2015th}, or NuSTAR, which is designed to see x-rays in the $3\mbox{--}79$ keV range~\cite{Harrison:2013fv,Riemer-Sorensen:2015lr}, corresponding to $m_s = 6\text{--}158$ keV, or INTEGRAL, looking at $18$ keV--$8$ MeV photon energies~\cite{Yuksel:2008fr,Boyarsky:2008pd}.

\subsection{Dependence on mixing angle} \label{sec:theta}

In our model, the dark matter relic density is set by assuming an equilibrium distribution of sterile neutrinos in the early Universe, followed by an epoch of out-of-equilibrium heavy particle decay, which engineers an appropriate amount of dilution. Therefore, our model does not rely on the active-sterile mixing angle, as far as setting the relic density is concerned. However, in certain regimes, the mixing angle can have other important consequences, and can therefore be constrained.

\subsubsection{Avoiding re-thermalization after dilution} \label{sec:retherm}

The DESNDM model rides on the assumption that the sterile neutrino decouples in the early Universe, followed by an epoch of entropy injection which cools the sterile sea relative to the plasma. However, if the sterile neutrinos were to re-thermalize after this, the entire purpose of dilution would be lost.

The sterile neutrino scattering rate in the plasma would have to be greater than the expansion rate at some epoch for them to re-thermalize, and as demonstrated in Sec. \ref{sec:prod}, this would not happen for small enough active-sterile vacuum mixing angles, as long as the lepton number is negligible. This puts an upper limit on the active-sterile vacuum mixing angle of $\sin^2{2\theta_v} \lesssim 10^{-6} \, (10\text{ keV}/m_s)$, for $m_s \sim$ keV--MeV, if the dilution model is to work. In practice, vacuum mixing angles in this range are already ruled out by x-ray observations and/or constraints on sterile neutrino lifetime, so the upper limits arrived at using thermalization arguments are merely of academic interest.

\subsubsection{Warm dark matter component produced by scattering-induced decoherence} \label{sec:DW}

As mentioned in Sec. \ref{sec:prod}, sterile neutrinos can be produced in the early Universe via scattering induced decoherence, even in the absence of a lepton number. So long as the sterile neutrinos mix with the active neutrinos, this process is unavoidable, and in the context of our model, it can produce an additional contribution to the sterile neutrino relic density. Also, if the dilution event were to happen prior to $T \sim 0.1\text{--}1$ GeV, the sterile neutrinos produced by scattering-induced decoherence would possess a higher average kinetic energy compared to their diluted-thermal-relic counterparts. This decoherently produced component will therefore contribute a warm tail to the overall sterile neutrino energy spectrum, leading to an increase in the effective dark matter collisionless damping scale, with likely implications for structure formation models.

For a sterile neutrino with mass $m_s$ and vacuum mixing angle $\theta_v$ with the active neutrinos, the contribution of this component to the closure density of the Universe in the zero-lepton number limit is given by \cite{Dodelson:1994rt,Kusenko:2009lr}
\be \label{eq:DW}
\Omega_\text{DW} \sim 0.2 \(\frac{\sin^2\theta_v}{3 \times 10^{-9}}\)\(\frac{m_s}{3\text{ keV}}\)^{1.8},
\ee
where \lq DW\rq\ is an acronym for Dodelson-Widrow. Clearly, one must have $\Omega_\text{DW} \leq \Omega_\text{DM} \approx 0.26$, and this puts an upper limit on the mixing angle as a function of sterile neutrino rest mass, in order to avoid overabundance of these steriles. For the particular case of the $m_s = 7.1$ keV sterile neutrino described in Sec.~\ref{sec:xray}, Eq.~(\ref{eq:DW}) becomes
\be \label{eq:DW7}
\Omega_\text{DW} \sim 0.94 \(\frac{\sin^2\theta_v}{3 \times 10^{-9}}\).
\ee

In scenarios where the $7.1$ keV sterile neutrino is all of the dark matter, i.e., $\Omega_s = \Omega_\text{DM}$, the inferred mixing angle from the observed flux of the x-ray line is $\sin^2{2\theta_v} \approx 7 \times 10^{-11}$. This implies $\sin^2{\theta_v} \approx 1.75 \times 10^{-11}$, which using Eq. (\ref{eq:DW7}) gives $\Omega_\text{DW} \approx 0.0055 \approx 0.021\, \Omega_\text{DM}$, for $\Omega_\text{DM} = 0.26$. Thus, if the $7.1$ keV sterile hinted at by x-ray observations were to be all of the dark matter, then about $2\%$ of its total number density (in the zero-lepton number limit) would be produced by the Dodelson-Widrow mechanism.

However, we can also look at cases where the sterile neutrino need not be all of the dark matter, i.e., $\Omega_s < \Omega_\text{DM}$. The inferred mixing angle from the observed x-ray line flux is then higher, and is given by 
\be \label{eq:thetaXray}
\sin^2{2\theta} \approx 7 \times 10^{-11}\,\(\Omega_\text{DM}/\Omega_s\).
\ee

An interesting limit to contemplate is the one where sterile neutrinos are produced only by the Dodelson-Widrow mechanism. One can then estimate the fraction of the total dark matter that would be constituted by these steriles. Solving Eqs. (\ref{eq:DW7}) and (\ref{eq:thetaXray}), with $\Omega_s = \Omega_\text{DW}$ and $\Omega_\text{DM} = 0.26$ gives $\Omega_s = \Omega_\text{DW} \approx 0.038 \approx 0.15\,\Omega_\text{DM}$. Therefore, the posited $7.1$ keV sterile neutrino, even in the absence of an appreciable lepton number or other nonstandard production scenarios such as DESNDM, could still account for roughly $15\%$ of the total dark matter in this purely quantum mechanical, Dodelson-Widrow limit.

\subsection{Kinematic constraints from small- and large-scale structure} \label{sec:struct}

As discussed in Sec.~\ref{sec:dmcd}, the collisionless damping mass/length sets the scale below which fluctuations can get damped by dark matter particle free-streaming. Observations of large-scale structure, e.g., the Ly-$\alpha$ forest and galaxy clustering, put upper bounds on the collisionless damping scale. These correspond to model-dependent lower limits on sterile neutrino rest mass~\cite{Abazajian:2006yg,Abazajian:2006yz,Seljak:2006fk,Boyarsky:2009fk,Boyarsky:2009jk,Viel:2005fv,Viel:2013eu,Merle:2014fk}. Warm dark matter is also known to flatten the cores of dark matter haloes in dwarf spheroidals, as well as decrease the expected number of low-mass satellites in larger dark matter haloes~\cite{Bode:2001zl}, although some recent work has argued that the deviation of warm dark matter density profiles, relative to the Navarro-Frenk-White (NFW) profile observed in CDM simulations, could be minimal~\cite{Schneider:2012ul}. Observations of stellar velocity dispersion profiles of dwarf galaxies can therefore be used to put constraints on the dark matter phase-space density, which again leads to model-dependent bounds on the rest mass of the candidate sterile neutrino~\cite{Strigari:2006rm}.

It has been suggested that observations of structure on small (i.e., dwarf galaxy) scales are inconsistent with the results of $\Lambda$CDM~\cite{Bode:2001zl,Boylan-Kolchin:2012nr} simulations. While it is not clear at this point whether the discrepancies could be resolved by incorporating baryonic feedback effects, solutions are also being sought via alterations to the standard CDM paradigm. For example, it has been argued that a resonantly enhanced sterile neutrino, that is slightly warm, but nevertheless not ruled out by present-day observations, could lie in the sweet spot for alleviating some of these inconsistencies~\cite{Abazajian:2014zl}. Eventually, 21-cm observations also may weigh in on these issues, either through flagging intergalactic medium heating from WIMP annihilation~\cite{Evoli:2014lr}, or other insights on large-scale structure~\cite{Sekiguchi:2014qf,Sitwell:2014vn,Shimabukuro:2014rm}. The latter indicate that perhaps 21-cm observations could probe structure down to length scales (comoving to the current epoch) of about $\sim 0.01$ Mpc, about an order of magnitude better than the scales probed by Lyman-$\alpha$ forest observations.

Our model would likely be confronted with some of these constraints towards the lighter end of sterile neutrino rest-mass range, i.e., for $m_s \lesssim$ a few keV, where the collisionless damping scales would be relevant for the issues discussed above.

\subsection{Laboratory constraints}

It is difficult to engineer direct laboratory probes of the candidate sterile neutrino dark matter particles and the candidate diluton particles we discuss here. This is because the couplings or rest masses of the particles involved can be out of reach for energies and sensitivities of existing or future experiments. 

Nevertheless, some laboratory probes can nibble around the edges of interesting parameter space for dark matter sterile neutrino candidates. For example, the KATRIN experiment and other direct beta decay endpoint experiments can target the contribution of heavy neutrino mass eigenvalues in the coherent sum entering into the projection of electron flavor neutrinos in this process \cite{de-Vega:2013tg,Barry:2014ai,Mertens:2015xe,Mertens:2015uo}.

Collider experiments, in principle, have much to say about beyond standard model particles and potentially about dark matter \cite{Alekhin:2015dq}. For example, if the dilutons were on the lighter end of the rest-mass range considered here, i.e., $m_H \sim$ TeV, then existing and future colliders could allow us to constrain their lifetimes \cite{Fuller:2011lr}. Dilutons produced in colliders could be detected if they were to subsequently decay inside a detector. The detection rate would be proportional to the production rate times the ratio of crossing time to the Lorentz dilated lifetime of the diluton.

Dilutons that are heavy sterile neutrinos also could be indirectly inferred via their impact on electroweak precision observables such as the invisible $Z$-decay width, the $W$-boson mass, and the charged-to-neutral current ratio for neutrino scattering~\cite{Akhmedov:2013xy}. Near-future collider experiments could potentially probe the effects of heavy sterile neutrinos on lepton-flavor-violating $Z$ decays~\cite{Abada:2015lq}, unfortunately their predicted sensitivities would not be high enough for sterile neutrinos that are sufficiently long-lived to be dilutons. TeV-scale sterile neutrinos could also influence the neutrinoless double beta decay rate through their contribution to the effective Majorana neutrino mass. However, this would require a significant amount of fine-tuning \cite{Lopez-Pavon:2015uq}, resulting in large active-sterile mixing and possibly rendering the sterile neutrino an unsuitable diluton candidate.

Of course, current and near-future colliders are not likely to have the energy reach required to probe the physics of dilutons heavier than a few TeV in rest mass. High-energy cosmic ray detectors and neutrino telescopes such as IceCube, on the other hand, could potentially be useful in probing this high-energy scale physics.

\subsection{Compact object constraints}

A significant fraction of the range of rest-mass and vacuum mixing parameters for viable sterile neutrino dark matter created through scattering-induced decoherence or resonant channels can also affect core collapse supernova physics \cite{Abazajian:2001lr}. Conversion of active-to-sterile neutrinos, and perhaps sometimes back again to active states, can affect energy and lepton number transport in the core, energy deposition in the mantle below the shock \cite{Kolb:1996sp, Abazajian:2001lr,Hidaka:2006yq,Fryer:2006rt,Hidaka:2007kx, Choubey:2007bs,Fuller:2009uq,Raffelt:2011ij,Warren:2014dp,Zhou:2015th}, and even proto-neutron star \lq\lq kicks\rq\rq\ associated with the neutrino burst~\cite{Kusenko:1999hc,Fuller:2003vn,Kusenko:2008lr,Kusenko:2009lr,Kishimoto:2011fk}. However, with a small enough mixing angle, and with sufficiently high rest mass (i.e., well above any resonant condition in the core), sterile neutrino dark matter candidates considered here in the DESNDM model can manage to avoid changing compact object physics.

\subsection{Differentiating between resonant production and DESNDM scenarios}

A relic density of sterile neutrinos comprising the dark matter could be produced by resonantly enhanced scattering-induced decoherence, as stated earlier. However, for sterile neutrinos with rest masses in our range of interest, this would require a lepton number that is several orders of magnitude bigger than the baryon number. If future lepton number constraints, e.g., from precision measurements of primordial helium and deuterium abundances, were to push the upper limit on the observationally inferred lepton number to below what would be required for resonant production, then it would force us to consider alternative models, if sterile neutrinos indeed were to be the dark matter. Additionally, sterile neutrinos produced resonantly would have warmer energy spectra, i.e., larger collisionless damping scales, compared to dilution generated sterile neutrinos of the same mass. Improved constraints on the collisionless damping scale from future Ly-$\alpha$ observations and possibly 21-cm observations, as well as an improved understanding of small- and large-scale structure formation through sophisticated simulations, could lead to certain models gaining favor over others.

\section{Conclusions} \label{sec:concl}

We have described a generic mechanism, the DESNDM model, whereby sterile neutrinos with rest masses in the range $\sim$ keV to $\sim$ MeV could acquire relic densities that allow them to be the dark matter. The key assumptions of this model are that: (1) the dark matter candidate sterile neutrino is in thermal and chemical equilibrium at very high temperature scale in the early Universe, (2) this particle decouples at very early epochs, (3) subsequent to this dark matter candidate sterile neutrino decoupling event there is prodigious entropy generation from the out-of-equilibrium decay of a different particle, the diluton, and (4) the diluton is presumed to be very massive and to possess nonrelativistic kinematics when it decays, but is also assumed to have been in thermal and chemical equilibrium at very early epochs, possessing relativistic kinematics at the time of its decoupling. 

We do not identify the diluton with a specific particle candidate. Instead, we consider the generic issues involved with out-of-equilibrium particle decay in the early Universe and attendant cosmological, observational, and laboratory constraints. This leads us to consider dilutons with rest masses in the $\sim$ TeV to $\sim$ EeV range, possessing rather long decay lifetimes. Candidate diluton particles might include, for example, heavy sterile neutrinos, different from the ones we might consider for the dark matter particle, and supersymmetric particles with R-parity-violating decays into standard model particles. The latter scenario would suggest a very high supersymmetry scale, and a novel and heterodox role for supersymmetric particles in the dark matter problem.

The DESNDM dilution mechanism for producing the sterile neutrino relic densities of interest necessarily results in a corresponding relic energy spectrum which, though thermal in \textit{shape}, can be quite cold compared to a standard energy spectrum characterized by a relic photon or relic active neutrino temperature. This allows our sterile neutrino dark matter to behave like CDM in many cases, even though the actual rest masses of the sterile neutrinos are modest. Additionally, the sterile neutrino population could acquire a \lq\lq warm,\rq\rq\ albeit subdominant component arising via scattering-induced decoherence, even in the zero lepton number limit, which could have some observable effects on structure formation on the small scales.

Interestingly, the DESNDM model for generating sterile neutrino relic densities can be, effectively, nearly independent of the vacuum mixing angle characterizing the mixing of this sterile neutrino with any of the active neutrino flavors. The model requires only that this mixing angle be smaller than that required to effect population of a sterile neutrino sea from the seas of active neutrinos in the very early Universe. This is unlike other models for producing a sterile neutrino relic dark matter density. For example, in scattering-induced decoherence and resonant enhancement of this process, the vacuum mixing angle is a key parameter, so that sterile neutrino rest mass, this mixing angle, and perhaps other parameters like lepton number, uniquely determine the relic density. This means that x-ray observational constraints can, in principle, definitively rule out ranges of sterile neutrino rest mass and vacuum mixing parameter space. 

In the DESNDM model the relic density is set by different physics. The existence of a nonzero vacuum mixing with active neutrinos will, of course, still guarantee a radiative decay channel for this sterile neutrino particle. However, in the DESNDM model a given sterile neutrino rest mass with a dark matter relic density need not have a mixing angle large enough to produce an x-ray flux sufficient for detection. As a consequence, the DESNDM mechanism can evade all x-ray bounds, and the detection of a dark matter sterile neutrino decay line would be a lucky, but not inevitable development.

That does not mean that there are no potential observational or experimental handles on sterile neutrino dark matter produced via the DESNDM mechanism. First, note that dilution can make dark matter sterile neutrinos with a wider range of rest masses than is possible in scattering-induced decoherence models. As discussed above, new experiments like NuSTAR can probe higher-energy x rays. At x-ray energies above $\sim 10$ keV, the expected x-ray backgrounds are lower than they are in the \lq\lq sweet spot\rq\rq\ of a few to $10$ keV for the XMM and Chandra x-ray telescopes. Second, as discussed above, large-scale structure simulations and observations may be able to produce finer probes of the dark matter character.

\begin{acknowledgments}

We would like to thank K.~Abazajian, M.~Boylan-Kolchin, M.~Drewes, E.~Figueroa-Feliciano, E.~Grohs, D.~Keres, S.~Mertens, P.~Smith, and A.~Vlasenko for valuable conversations. This work was supported in part by National Science Foundation grant PHY-1307372 at UCSD and by Department of Energy grant DE-SC0009937 at UCLA. The work of A.K. was also supported by the World Premier International Research Center Initiative (WPI), MEXT, Japan.

\end{acknowledgments}

\bibliography{allref}

\begin{thebibliography}{133}%
\makeatletter
\providecommand \@ifxundefined [1]{%
 \@ifx{#1\undefined}
}%
\providecommand \@ifnum [1]{%
 \ifnum #1\expandafter \@firstoftwo
 \else \expandafter \@secondoftwo
 \fi
}%
\providecommand \@ifx [1]{%
 \ifx #1\expandafter \@firstoftwo
 \else \expandafter \@secondoftwo
 \fi
}%
\providecommand \natexlab [1]{#1}%
\providecommand \enquote  [1]{``#1''}%
\providecommand \bibnamefont  [1]{#1}%
\providecommand \bibfnamefont [1]{#1}%
\providecommand \citenamefont [1]{#1}%
\providecommand \href@noop [0]{\@secondoftwo}%
\providecommand \href [0]{\begingroup \@sanitize@url \@href}%
\providecommand \@href[1]{\@@startlink{#1}\@@href}%
\providecommand \@@href[1]{\endgroup#1\@@endlink}%
\providecommand \@sanitize@url [0]{\catcode `\\12\catcode `\$12\catcode
  `\&12\catcode `\#12\catcode `\^12\catcode `\_12\catcode `\%12\relax}%
\providecommand \@@startlink[1]{}%
\providecommand \@@endlink[0]{}%
\providecommand \url  [0]{\begingroup\@sanitize@url \@url }%
\providecommand \@url [1]{\endgroup\@href {#1}{\urlprefix }}%
\providecommand \urlprefix  [0]{URL }%
\providecommand \Eprint [0]{\href }%
\providecommand \doibase [0]{http://dx.doi.org/}%
\providecommand \selectlanguage [0]{\@gobble}%
\providecommand \bibinfo  [0]{\@secondoftwo}%
\providecommand \bibfield  [0]{\@secondoftwo}%
\providecommand \translation [1]{[#1]}%
\providecommand \BibitemOpen [0]{}%
\providecommand \bibitemStop [0]{}%
\providecommand \bibitemNoStop [0]{.\EOS\space}%
\providecommand \EOS [0]{\spacefactor3000\relax}%
\providecommand \BibitemShut  [1]{\csname bibitem#1\endcsname}%
\let\auto@bib@innerbib\@empty
\bibitem [{\citenamefont {{Dodelson}}\ and\ \citenamefont
  {{Widrow}}(1994)}]{Dodelson:1994rt}%
  \BibitemOpen
  \bibfield  {author} {\bibinfo {author} {\bibfnamefont {S.}~\bibnamefont
  {{Dodelson}}}\ and\ \bibinfo {author} {\bibfnamefont {L.~M.}\ \bibnamefont
  {{Widrow}}},\ }\href {\doibase 10.1103/PhysRevLett.72.17} {\bibfield
  {journal} {\bibinfo  {journal} {Physical Review Letters}\ }\textbf {\bibinfo
  {volume} {72}},\ \bibinfo {pages} {17} (\bibinfo {year} {1994})},\ \Eprint
  {http://arxiv.org/abs/arXiv:hep-ph/9303287} {arXiv:hep-ph/9303287}
  \BibitemShut {NoStop}%
\bibitem [{\citenamefont {{Shi}}\ and\ \citenamefont
  {{Fuller}}(1999)}]{Shi:1999lq}%
  \BibitemOpen
  \bibfield  {author} {\bibinfo {author} {\bibfnamefont {X.}~\bibnamefont
  {{Shi}}}\ and\ \bibinfo {author} {\bibfnamefont {G.~M.}\ \bibnamefont
  {{Fuller}}},\ }\href {\doibase 10.1103/PhysRevLett.82.2832} {\bibfield
  {journal} {\bibinfo  {journal} {Physical Review Letters}\ }\textbf {\bibinfo
  {volume} {82}},\ \bibinfo {pages} {2832} (\bibinfo {year} {1999})},\ \Eprint
  {http://arxiv.org/abs/arXiv:astro-ph/9810076} {arXiv:astro-ph/9810076}
  \BibitemShut {NoStop}%
\bibitem [{\citenamefont {{Abazajian}}\ \emph
  {et~al.}(2001{\natexlab{a}})\citenamefont {{Abazajian}}, \citenamefont
  {{Fuller}},\ and\ \citenamefont {{Patel}}}]{Abazajian:2001lr}%
  \BibitemOpen
  \bibfield  {author} {\bibinfo {author} {\bibfnamefont {K.}~\bibnamefont
  {{Abazajian}}}, \bibinfo {author} {\bibfnamefont {G.~M.}\ \bibnamefont
  {{Fuller}}}, \ and\ \bibinfo {author} {\bibfnamefont {M.}~\bibnamefont
  {{Patel}}},\ }\href {\doibase 10.1103/PhysRevD.64.023501} {\bibfield
  {journal} {\bibinfo  {journal} {\prd}\ }\textbf {\bibinfo {volume} {64}},\
  \bibinfo {pages} {023501} (\bibinfo {year} {2001}{\natexlab{a}})},\ \Eprint
  {http://arxiv.org/abs/arXiv:astro-ph/0101524} {arXiv:astro-ph/0101524}
  \BibitemShut {NoStop}%
\bibitem [{\citenamefont {{Dolgov}}\ and\ \citenamefont
  {{Hansen}}(2002)}]{Dolgov:2002ve}%
  \BibitemOpen
  \bibfield  {author} {\bibinfo {author} {\bibfnamefont {A.~D.}\ \bibnamefont
  {{Dolgov}}}\ and\ \bibinfo {author} {\bibfnamefont {S.~H.}\ \bibnamefont
  {{Hansen}}},\ }\href {\doibase 10.1016/S0927-6505(01)00115-3} {\bibfield
  {journal} {\bibinfo  {journal} {Astroparticle Physics}\ }\textbf {\bibinfo
  {volume} {16}},\ \bibinfo {pages} {339} (\bibinfo {year} {2002})},\ \Eprint
  {http://arxiv.org/abs/arXiv:hep-ph/0009083} {arXiv:hep-ph/0009083}
  \BibitemShut {NoStop}%
\bibitem [{\citenamefont {{Abazajian}}\ and\ \citenamefont
  {{Fuller}}(2002)}]{Abazajian:2002bh}%
  \BibitemOpen
  \bibfield  {author} {\bibinfo {author} {\bibfnamefont {K.~N.}\ \bibnamefont
  {{Abazajian}}}\ and\ \bibinfo {author} {\bibfnamefont {G.~M.}\ \bibnamefont
  {{Fuller}}},\ }\href {\doibase 10.1103/PhysRevD.66.023526} {\bibfield
  {journal} {\bibinfo  {journal} {\prd}\ }\textbf {\bibinfo {volume} {66}},\
  \bibinfo {pages} {023526} (\bibinfo {year} {2002})},\ \Eprint
  {http://arxiv.org/abs/arXiv:astro-ph/0204293} {arXiv:astro-ph/0204293}
  \BibitemShut {NoStop}%
\bibitem [{\citenamefont {{Asaka}}\ and\ \citenamefont
  {{Shaposhnikov}}(2005)}]{Asaka:2005fx}%
  \BibitemOpen
  \bibfield  {author} {\bibinfo {author} {\bibfnamefont {T.}~\bibnamefont
  {{Asaka}}}\ and\ \bibinfo {author} {\bibfnamefont {M.}~\bibnamefont
  {{Shaposhnikov}}},\ }\href {\doibase 10.1016/j.physletb.2005.06.020}
  {\bibfield  {journal} {\bibinfo  {journal} {Physics Letters B}\ }\textbf
  {\bibinfo {volume} {620}},\ \bibinfo {pages} {17} (\bibinfo {year} {2005})},\
  \Eprint {http://arxiv.org/abs/hep-ph/0505013} {hep-ph/0505013} \BibitemShut
  {NoStop}%
\bibitem [{\citenamefont {{Asaka}}\ \emph {et~al.}(2005)\citenamefont
  {{Asaka}}, \citenamefont {{Blanchet}},\ and\ \citenamefont
  {{Shaposhnikov}}}]{Asaka:2005fj}%
  \BibitemOpen
  \bibfield  {author} {\bibinfo {author} {\bibfnamefont {T.}~\bibnamefont
  {{Asaka}}}, \bibinfo {author} {\bibfnamefont {S.}~\bibnamefont {{Blanchet}}},
  \ and\ \bibinfo {author} {\bibfnamefont {M.}~\bibnamefont {{Shaposhnikov}}},\
  }\href {\doibase 10.1016/j.physletb.2005.09.070} {\bibfield  {journal}
  {\bibinfo  {journal} {Physics Letters B}\ }\textbf {\bibinfo {volume}
  {631}},\ \bibinfo {pages} {151} (\bibinfo {year} {2005})},\ \Eprint
  {http://arxiv.org/abs/arXiv:hep-ph/0503065} {arXiv:hep-ph/0503065}
  \BibitemShut {NoStop}%
\bibitem [{\citenamefont {{Abazajian}}(2006{\natexlab{a}})}]{Abazajian:2006qf}%
  \BibitemOpen
  \bibfield  {author} {\bibinfo {author} {\bibfnamefont {K.}~\bibnamefont
  {{Abazajian}}},\ }\href {\doibase 10.1103/PhysRevD.73.063506} {\bibfield
  {journal} {\bibinfo  {journal} {\prd}\ }\textbf {\bibinfo {volume} {73}},\
  \bibinfo {pages} {063506} (\bibinfo {year} {2006}{\natexlab{a}})},\ \Eprint
  {http://arxiv.org/abs/arXiv:astro-ph/0511630} {arXiv:astro-ph/0511630}
  \BibitemShut {NoStop}%
\bibitem [{\citenamefont {{Asaka}}\ \emph {et~al.}(2006)\citenamefont
  {{Asaka}}, \citenamefont {{Shaposhnikov}},\ and\ \citenamefont
  {{Kusenko}}}]{Asaka:2006qy}%
  \BibitemOpen
  \bibfield  {author} {\bibinfo {author} {\bibfnamefont {T.}~\bibnamefont
  {{Asaka}}}, \bibinfo {author} {\bibfnamefont {M.}~\bibnamefont
  {{Shaposhnikov}}}, \ and\ \bibinfo {author} {\bibfnamefont {A.}~\bibnamefont
  {{Kusenko}}},\ }\href {\doibase 10.1016/j.physletb.2006.05.067} {\bibfield
  {journal} {\bibinfo  {journal} {Physics Letters B}\ }\textbf {\bibinfo
  {volume} {638}},\ \bibinfo {pages} {401} (\bibinfo {year} {2006})},\ \Eprint
  {http://arxiv.org/abs/hep-ph/0602150} {hep-ph/0602150} \BibitemShut {NoStop}%
\bibitem [{\citenamefont {Kusenko}(2006)}]{Kusenko:2006rh}%
  \BibitemOpen
  \bibfield  {author} {\bibinfo {author} {\bibfnamefont {A.}~\bibnamefont
  {Kusenko}},\ }\href {\doibase 10.1103/PhysRevLett.97.241301} {\bibfield
  {journal} {\bibinfo  {journal} {Phys.Rev.Lett.}\ }\textbf {\bibinfo {volume}
  {97}},\ \bibinfo {pages} {241301} (\bibinfo {year} {2006})},\ \Eprint
  {http://arxiv.org/abs/hep-ph/0609081} {arXiv:hep-ph/0609081 [hep-ph]}
  \BibitemShut {NoStop}%
\bibitem [{\citenamefont {{Shaposhnikov}}\ and\ \citenamefont
  {{Tkachev}}(2006)}]{Shaposhnikov:2006fj}%
  \BibitemOpen
  \bibfield  {author} {\bibinfo {author} {\bibfnamefont {M.}~\bibnamefont
  {{Shaposhnikov}}}\ and\ \bibinfo {author} {\bibfnamefont {I.}~\bibnamefont
  {{Tkachev}}},\ }\href {\doibase 10.1016/j.physletb.2006.06.063} {\bibfield
  {journal} {\bibinfo  {journal} {Physics Letters B}\ }\textbf {\bibinfo
  {volume} {639}},\ \bibinfo {pages} {414} (\bibinfo {year} {2006})},\ \Eprint
  {http://arxiv.org/abs/arXiv:hep-ph/0604236} {arXiv:hep-ph/0604236}
  \BibitemShut {NoStop}%
\bibitem [{\citenamefont {{Boyanovsky}}\ and\ \citenamefont
  {{Ho}}(2007)}]{Boyanovsky:2007lr}%
  \BibitemOpen
  \bibfield  {author} {\bibinfo {author} {\bibfnamefont {D.}~\bibnamefont
  {{Boyanovsky}}}\ and\ \bibinfo {author} {\bibfnamefont {C.-M.}\ \bibnamefont
  {{Ho}}},\ }\href {\doibase 10.1088/1126-6708/2007/07/030} {\bibfield
  {journal} {\bibinfo  {journal} {Journal of High Energy Physics}\ }\textbf
  {\bibinfo {volume} {7}},\ \bibinfo {pages} {30} (\bibinfo {year} {2007})},\
  \Eprint {http://arxiv.org/abs/arXiv:hep-ph/0612092} {arXiv:hep-ph/0612092}
  \BibitemShut {NoStop}%
\bibitem [{\citenamefont {{Boyanovsky}}(2007)}]{Boyanovsky:2007fk}%
  \BibitemOpen
  \bibfield  {author} {\bibinfo {author} {\bibfnamefont {D.}~\bibnamefont
  {{Boyanovsky}}},\ }\href {\doibase 10.1103/PhysRevD.76.103514} {\bibfield
  {journal} {\bibinfo  {journal} {\prd}\ }\textbf {\bibinfo {volume} {76}},\
  \bibinfo {pages} {103514} (\bibinfo {year} {2007})},\ \Eprint
  {http://arxiv.org/abs/0706.3167} {arXiv:0706.3167 [hep-ph]} \BibitemShut
  {NoStop}%
\bibitem [{\citenamefont {{Asaka}}\ \emph {et~al.}(2007)\citenamefont
  {{Asaka}}, \citenamefont {{Shaposhnikov}},\ and\ \citenamefont
  {{Laine}}}]{Asaka:2007fk}%
  \BibitemOpen
  \bibfield  {author} {\bibinfo {author} {\bibfnamefont {T.}~\bibnamefont
  {{Asaka}}}, \bibinfo {author} {\bibfnamefont {M.}~\bibnamefont
  {{Shaposhnikov}}}, \ and\ \bibinfo {author} {\bibfnamefont {M.}~\bibnamefont
  {{Laine}}},\ }\href {\doibase 10.1088/1126-6708/2007/01/091} {\bibfield
  {journal} {\bibinfo  {journal} {Journal of High Energy Physics}\ }\textbf
  {\bibinfo {volume} {1}},\ \bibinfo {eid} {091} (\bibinfo {year} {2007})},\
  \Eprint {http://arxiv.org/abs/hep-ph/0612182} {hep-ph/0612182} \BibitemShut
  {NoStop}%
\bibitem [{\citenamefont {{Shaposhnikov}}(2007)}]{Shaposhnikov:2007qy}%
  \BibitemOpen
  \bibfield  {author} {\bibinfo {author} {\bibfnamefont {M.}~\bibnamefont
  {{Shaposhnikov}}},\ }\href {\doibase 10.1016/j.nuclphysb.2006.11.003}
  {\bibfield  {journal} {\bibinfo  {journal} {Nuclear Physics B}\ }\textbf
  {\bibinfo {volume} {763}},\ \bibinfo {pages} {49} (\bibinfo {year} {2007})},\
  \Eprint {http://arxiv.org/abs/arXiv:hep-ph/0605047} {arXiv:hep-ph/0605047}
  \BibitemShut {NoStop}%
\bibitem [{\citenamefont {{Gorbunov}}\ and\ \citenamefont
  {{Shaposhnikov}}(2007)}]{Gorbunov:2007uq}%
  \BibitemOpen
  \bibfield  {author} {\bibinfo {author} {\bibfnamefont {D.}~\bibnamefont
  {{Gorbunov}}}\ and\ \bibinfo {author} {\bibfnamefont {M.}~\bibnamefont
  {{Shaposhnikov}}},\ }\href {\doibase 10.1088/1126-6708/2007/10/015}
  {\bibfield  {journal} {\bibinfo  {journal} {Journal of High Energy Physics}\
  }\textbf {\bibinfo {volume} {10}},\ \bibinfo {pages} {15} (\bibinfo {year}
  {2007})},\ \Eprint {http://arxiv.org/abs/0705.1729} {arXiv:0705.1729
  [hep-ph]} \BibitemShut {NoStop}%
\bibitem [{\citenamefont {{Kishimoto}}\ and\ \citenamefont
  {{Fuller}}(2008)}]{Kishimoto:2008pd}%
  \BibitemOpen
  \bibfield  {author} {\bibinfo {author} {\bibfnamefont {C.~T.}\ \bibnamefont
  {{Kishimoto}}}\ and\ \bibinfo {author} {\bibfnamefont {G.~M.}\ \bibnamefont
  {{Fuller}}},\ }\href {\doibase 10.1103/PhysRevD.78.023524} {\bibfield
  {journal} {\bibinfo  {journal} {\prd}\ }\textbf {\bibinfo {volume} {78}},\
  \bibinfo {pages} {023524} (\bibinfo {year} {2008})},\ \Eprint
  {http://arxiv.org/abs/0802.3377} {arXiv:0802.3377} \BibitemShut {NoStop}%
\bibitem [{\citenamefont {{Laine}}\ and\ \citenamefont
  {{Shaposhnikov}}(2008)}]{Laine:2008kx}%
  \BibitemOpen
  \bibfield  {author} {\bibinfo {author} {\bibfnamefont {M.}~\bibnamefont
  {{Laine}}}\ and\ \bibinfo {author} {\bibfnamefont {M.}~\bibnamefont
  {{Shaposhnikov}}},\ }\href {\doibase 10.1088/1475-7516/2008/06/031}
  {\bibfield  {journal} {\bibinfo  {journal} {JCAP}\ }\textbf {\bibinfo
  {volume} {6}},\ \bibinfo {pages} {31} (\bibinfo {year} {2008})},\ \Eprint
  {http://arxiv.org/abs/0804.4543} {arXiv:0804.4543 [hep-ph]} \BibitemShut
  {NoStop}%
\bibitem [{\citenamefont {{Petraki}}(2008)}]{Petraki:2008yq}%
  \BibitemOpen
  \bibfield  {author} {\bibinfo {author} {\bibfnamefont {K.}~\bibnamefont
  {{Petraki}}},\ }\href {\doibase 10.1103/PhysRevD.77.105004} {\bibfield
  {journal} {\bibinfo  {journal} {\prd}\ }\textbf {\bibinfo {volume} {77}},\
  \bibinfo {pages} {105004} (\bibinfo {year} {2008})},\ \Eprint
  {http://arxiv.org/abs/0801.3470} {arXiv:0801.3470 [hep-ph]} \BibitemShut
  {NoStop}%
\bibitem [{\citenamefont {{Petraki}}\ and\ \citenamefont
  {{Kusenko}}(2008)}]{Petraki:2008vn}%
  \BibitemOpen
  \bibfield  {author} {\bibinfo {author} {\bibfnamefont {K.}~\bibnamefont
  {{Petraki}}}\ and\ \bibinfo {author} {\bibfnamefont {A.}~\bibnamefont
  {{Kusenko}}},\ }\href {\doibase 10.1103/PhysRevD.77.065014} {\bibfield
  {journal} {\bibinfo  {journal} {\prd}\ }\textbf {\bibinfo {volume} {77}},\
  \bibinfo {pages} {065014} (\bibinfo {year} {2008})},\ \Eprint
  {http://arxiv.org/abs/0711.4646} {arXiv:0711.4646 [hep-ph]} \BibitemShut
  {NoStop}%
\bibitem [{\citenamefont {{Kusenko}}(2009)}]{Kusenko:2009lr}%
  \BibitemOpen
  \bibfield  {author} {\bibinfo {author} {\bibfnamefont {A.}~\bibnamefont
  {{Kusenko}}},\ }\href {\doibase 10.1016/j.physrep.2009.07.004} {\bibfield
  {journal} {\bibinfo  {journal} {Physics Reports}\ }\textbf {\bibinfo {volume}
  {481}},\ \bibinfo {pages} {1} (\bibinfo {year} {2009})},\ \Eprint
  {http://arxiv.org/abs/0906.2968} {arXiv:0906.2968 [hep-ph]} \BibitemShut
  {NoStop}%
\bibitem [{\citenamefont {{Kusenko}}\ \emph {et~al.}(2010)\citenamefont
  {{Kusenko}}, \citenamefont {{Takahashi}},\ and\ \citenamefont
  {{Yanagida}}}]{Kusenko:2010uq}%
  \BibitemOpen
  \bibfield  {author} {\bibinfo {author} {\bibfnamefont {A.}~\bibnamefont
  {{Kusenko}}}, \bibinfo {author} {\bibfnamefont {F.}~\bibnamefont
  {{Takahashi}}}, \ and\ \bibinfo {author} {\bibfnamefont {T.~T.}\ \bibnamefont
  {{Yanagida}}},\ }\href {\doibase 10.1016/j.physletb.2010.08.031} {\bibfield
  {journal} {\bibinfo  {journal} {Physics Letters B}\ }\textbf {\bibinfo
  {volume} {693}},\ \bibinfo {pages} {144} (\bibinfo {year} {2010})},\ \Eprint
  {http://arxiv.org/abs/1006.1731} {arXiv:1006.1731 [hep-ph]} \BibitemShut
  {NoStop}%
\bibitem [{\citenamefont {{Bezrukov}}\ \emph {et~al.}(2010)\citenamefont
  {{Bezrukov}}, \citenamefont {{Hettmansperger}},\ and\ \citenamefont
  {{Lindner}}}]{Bezrukov:2010rm}%
  \BibitemOpen
  \bibfield  {author} {\bibinfo {author} {\bibfnamefont {F.}~\bibnamefont
  {{Bezrukov}}}, \bibinfo {author} {\bibfnamefont {H.}~\bibnamefont
  {{Hettmansperger}}}, \ and\ \bibinfo {author} {\bibfnamefont
  {M.}~\bibnamefont {{Lindner}}},\ }\href {\doibase 10.1103/PhysRevD.81.085032}
  {\bibfield  {journal} {\bibinfo  {journal} {\prd}\ }\textbf {\bibinfo
  {volume} {81}},\ \bibinfo {eid} {085032} (\bibinfo {year} {2010})},\ \Eprint
  {http://arxiv.org/abs/0912.4415} {arXiv:0912.4415 [hep-ph]} \BibitemShut
  {NoStop}%
\bibitem [{\citenamefont {{Nemev{\v s}ek}}\ \emph {et~al.}(2012)\citenamefont
  {{Nemev{\v s}ek}}, \citenamefont {{Senjanovi{\'c}}},\ and\ \citenamefont
  {{Zhang}}}]{Nemevsek:2012zl}%
  \BibitemOpen
  \bibfield  {author} {\bibinfo {author} {\bibfnamefont {M.}~\bibnamefont
  {{Nemev{\v s}ek}}}, \bibinfo {author} {\bibfnamefont {G.}~\bibnamefont
  {{Senjanovi{\'c}}}}, \ and\ \bibinfo {author} {\bibfnamefont
  {Y.}~\bibnamefont {{Zhang}}},\ }\href {\doibase
  10.1088/1475-7516/2012/07/006} {\bibfield  {journal} {\bibinfo  {journal}
  {Journal of Cosmology and Astroparticle Physics}\ }\textbf {\bibinfo {volume}
  {7}},\ \bibinfo {eid} {006} (\bibinfo {year} {2012})},\ \Eprint
  {http://arxiv.org/abs/1205.0844} {arXiv:1205.0844 [hep-ph]} \BibitemShut
  {NoStop}%
\bibitem [{\citenamefont {{Canetti}}\ \emph
  {et~al.}(2013{\natexlab{a}})\citenamefont {{Canetti}}, \citenamefont
  {{Drewes}},\ and\ \citenamefont {{Shaposhnikov}}}]{Canetti:2013qy}%
  \BibitemOpen
  \bibfield  {author} {\bibinfo {author} {\bibfnamefont {L.}~\bibnamefont
  {{Canetti}}}, \bibinfo {author} {\bibfnamefont {M.}~\bibnamefont {{Drewes}}},
  \ and\ \bibinfo {author} {\bibfnamefont {M.}~\bibnamefont {{Shaposhnikov}}},\
  }\href {\doibase 10.1103/PhysRevLett.110.061801} {\bibfield  {journal}
  {\bibinfo  {journal} {Physical Review Letters}\ }\textbf {\bibinfo {volume}
  {110}},\ \bibinfo {eid} {061801} (\bibinfo {year} {2013}{\natexlab{a}})},\
  \Eprint {http://arxiv.org/abs/1204.3902} {arXiv:1204.3902 [hep-ph]}
  \BibitemShut {NoStop}%
\bibitem [{\citenamefont {{Bezrukov}}\ \emph {et~al.}(2013)\citenamefont
  {{Bezrukov}}, \citenamefont {{Kartavtsev}},\ and\ \citenamefont
  {{Lindner}}}]{Bezrukov:2013qv}%
  \BibitemOpen
  \bibfield  {author} {\bibinfo {author} {\bibfnamefont {F.}~\bibnamefont
  {{Bezrukov}}}, \bibinfo {author} {\bibfnamefont {A.}~\bibnamefont
  {{Kartavtsev}}}, \ and\ \bibinfo {author} {\bibfnamefont {M.}~\bibnamefont
  {{Lindner}}},\ }\href {\doibase 10.1088/0954-3899/40/9/095202} {\bibfield
  {journal} {\bibinfo  {journal} {Journal of Physics G Nuclear Physics}\
  }\textbf {\bibinfo {volume} {40}},\ \bibinfo {eid} {095202} (\bibinfo {year}
  {2013})},\ \Eprint {http://arxiv.org/abs/1204.5477} {arXiv:1204.5477
  [hep-ph]} \BibitemShut {NoStop}%
\bibitem [{\citenamefont {{Canetti}}\ \emph
  {et~al.}(2013{\natexlab{b}})\citenamefont {{Canetti}}, \citenamefont
  {{Drewes}}, \citenamefont {{Frossard}},\ and\ \citenamefont
  {{Shaposhnikov}}}]{Canetti:2013lr}%
  \BibitemOpen
  \bibfield  {author} {\bibinfo {author} {\bibfnamefont {L.}~\bibnamefont
  {{Canetti}}}, \bibinfo {author} {\bibfnamefont {M.}~\bibnamefont {{Drewes}}},
  \bibinfo {author} {\bibfnamefont {T.}~\bibnamefont {{Frossard}}}, \ and\
  \bibinfo {author} {\bibfnamefont {M.}~\bibnamefont {{Shaposhnikov}}},\ }\href
  {\doibase 10.1103/PhysRevD.87.093006} {\bibfield  {journal} {\bibinfo
  {journal} {\prd}\ }\textbf {\bibinfo {volume} {87}},\ \bibinfo {eid} {093006}
  (\bibinfo {year} {2013}{\natexlab{b}})},\ \Eprint
  {http://arxiv.org/abs/1208.4607} {arXiv:1208.4607 [hep-ph]} \BibitemShut
  {NoStop}%
\bibitem [{\citenamefont {{Merle}}(2013)}]{Merle:2013ty}%
  \BibitemOpen
  \bibfield  {author} {\bibinfo {author} {\bibfnamefont {A.}~\bibnamefont
  {{Merle}}},\ }\href {\doibase 10.1142/S0218271813300206} {\bibfield
  {journal} {\bibinfo  {journal} {International Journal of Modern Physics D}\
  }\textbf {\bibinfo {volume} {22}},\ \bibinfo {eid} {1330020} (\bibinfo {year}
  {2013})},\ \Eprint {http://arxiv.org/abs/1302.2625} {arXiv:1302.2625
  [hep-ph]} \BibitemShut {NoStop}%
\bibitem [{\citenamefont {{Abazajian}}(2014{\natexlab{a}})}]{Abazajian:2014zl}%
  \BibitemOpen
  \bibfield  {author} {\bibinfo {author} {\bibfnamefont {K.~N.}\ \bibnamefont
  {{Abazajian}}},\ }\href {\doibase 10.1103/PhysRevLett.112.161303} {\bibfield
  {journal} {\bibinfo  {journal} {Physical Review Letters}\ }\textbf {\bibinfo
  {volume} {112}},\ \bibinfo {eid} {161303} (\bibinfo {year}
  {2014}{\natexlab{a}})},\ \Eprint {http://arxiv.org/abs/1403.0954}
  {arXiv:1403.0954} \BibitemShut {NoStop}%
\bibitem [{\citenamefont {{Tsuyuki}}(2014)}]{Tsuyuki:2014uq}%
  \BibitemOpen
  \bibfield  {author} {\bibinfo {author} {\bibfnamefont {T.}~\bibnamefont
  {{Tsuyuki}}},\ }\href {\doibase 10.1103/PhysRevD.90.013007} {\bibfield
  {journal} {\bibinfo  {journal} {\prd}\ }\textbf {\bibinfo {volume} {90}},\
  \bibinfo {eid} {013007} (\bibinfo {year} {2014})},\ \Eprint
  {http://arxiv.org/abs/1403.5053} {arXiv:1403.5053 [hep-ph]} \BibitemShut
  {NoStop}%
\bibitem [{\citenamefont {{Merle}}\ \emph {et~al.}(2014)\citenamefont
  {{Merle}}, \citenamefont {{Niro}},\ and\ \citenamefont
  {{Schmidt}}}]{Merle:2014qy}%
  \BibitemOpen
  \bibfield  {author} {\bibinfo {author} {\bibfnamefont {A.}~\bibnamefont
  {{Merle}}}, \bibinfo {author} {\bibfnamefont {V.}~\bibnamefont {{Niro}}}, \
  and\ \bibinfo {author} {\bibfnamefont {D.}~\bibnamefont {{Schmidt}}},\ }\href
  {\doibase 10.1088/1475-7516/2014/03/028} {\bibfield  {journal} {\bibinfo
  {journal} {Journal of Cosmology and Astroparticle Physics}\ }\textbf
  {\bibinfo {volume} {3}},\ \bibinfo {eid} {028} (\bibinfo {year} {2014})},\
  \Eprint {http://arxiv.org/abs/1306.3996} {arXiv:1306.3996 [hep-ph]}
  \BibitemShut {NoStop}%
\bibitem [{\citenamefont {{Bezrukov}}\ and\ \citenamefont
  {{Gorbunov}}(2014)}]{Bezrukov:2014nr}%
  \BibitemOpen
  \bibfield  {author} {\bibinfo {author} {\bibfnamefont {F.~L.}\ \bibnamefont
  {{Bezrukov}}}\ and\ \bibinfo {author} {\bibfnamefont {D.~S.}\ \bibnamefont
  {{Gorbunov}}},\ }\href {\doibase 10.1016/j.physletb.2014.07.060} {\bibfield
  {journal} {\bibinfo  {journal} {Physics Letters B}\ }\textbf {\bibinfo
  {volume} {736}},\ \bibinfo {pages} {494} (\bibinfo {year} {2014})},\ \Eprint
  {http://arxiv.org/abs/1403.4638} {arXiv:1403.4638 [hep-ph]} \BibitemShut
  {NoStop}%
\bibitem [{\citenamefont {{Roland}}\ \emph {et~al.}(2014)\citenamefont
  {{Roland}}, \citenamefont {{Shakya}},\ and\ \citenamefont
  {{Wells}}}]{Roland:2014lr}%
  \BibitemOpen
  \bibfield  {author} {\bibinfo {author} {\bibfnamefont {S.~B.}\ \bibnamefont
  {{Roland}}}, \bibinfo {author} {\bibfnamefont {B.}~\bibnamefont {{Shakya}}},
  \ and\ \bibinfo {author} {\bibfnamefont {J.~D.}\ \bibnamefont {{Wells}}},\
  }\href@noop {} {\bibfield  {journal} {\bibinfo  {journal} {ArXiv e-prints}\ }
  (\bibinfo {year} {2014})},\ \Eprint {http://arxiv.org/abs/1412.4791}
  {arXiv:1412.4791 [hep-ph]} \BibitemShut {NoStop}%
\bibitem [{\citenamefont {{Abada}}\ \emph {et~al.}(2014)\citenamefont
  {{Abada}}, \citenamefont {{Arcadi}},\ and\ \citenamefont
  {{Lucente}}}]{Abada:2014fk}%
  \BibitemOpen
  \bibfield  {author} {\bibinfo {author} {\bibfnamefont {A.}~\bibnamefont
  {{Abada}}}, \bibinfo {author} {\bibfnamefont {G.}~\bibnamefont {{Arcadi}}}, \
  and\ \bibinfo {author} {\bibfnamefont {M.}~\bibnamefont {{Lucente}}},\ }\href
  {\doibase 10.1088/1475-7516/2014/10/001} {\bibfield  {journal} {\bibinfo
  {journal} {Journal of Cosmology and Astroparticle Physics}\ }\textbf
  {\bibinfo {volume} {10}},\ \bibinfo {eid} {001} (\bibinfo {year} {2014})},\
  \Eprint {http://arxiv.org/abs/1406.6556} {arXiv:1406.6556 [hep-ph]}
  \BibitemShut {NoStop}%
\bibitem [{\citenamefont {{Lello}}\ and\ \citenamefont
  {{Boyanovsky}}(2015)}]{Lello:2015lr}%
  \BibitemOpen
  \bibfield  {author} {\bibinfo {author} {\bibfnamefont {L.}~\bibnamefont
  {{Lello}}}\ and\ \bibinfo {author} {\bibfnamefont {D.}~\bibnamefont
  {{Boyanovsky}}},\ }\href {\doibase 10.1103/PhysRevD.91.063502} {\bibfield
  {journal} {\bibinfo  {journal} {\prd}\ }\textbf {\bibinfo {volume} {91}},\
  \bibinfo {eid} {063502} (\bibinfo {year} {2015})},\ \Eprint
  {http://arxiv.org/abs/1411.2690} {arXiv:1411.2690} \BibitemShut {NoStop}%
\bibitem [{\citenamefont {{Humbert}}\ \emph
  {et~al.}(2015{\natexlab{a}})\citenamefont {{Humbert}}, \citenamefont
  {{Lindner}},\ and\ \citenamefont {{Smirnov}}}]{Humbert:2015jk}%
  \BibitemOpen
  \bibfield  {author} {\bibinfo {author} {\bibfnamefont {P.}~\bibnamefont
  {{Humbert}}}, \bibinfo {author} {\bibfnamefont {M.}~\bibnamefont
  {{Lindner}}}, \ and\ \bibinfo {author} {\bibfnamefont {J.}~\bibnamefont
  {{Smirnov}}},\ }\href {\doibase 10.1007/JHEP06(2015)035} {\bibfield
  {journal} {\bibinfo  {journal} {Journal of High Energy Physics}\ }\textbf
  {\bibinfo {volume} {6}},\ \bibinfo {pages} {35} (\bibinfo {year}
  {2015}{\natexlab{a}})},\ \Eprint {http://arxiv.org/abs/1503.03066}
  {arXiv:1503.03066 [hep-ph]} \BibitemShut {NoStop}%
\bibitem [{\citenamefont {{Humbert}}\ \emph
  {et~al.}(2015{\natexlab{b}})\citenamefont {{Humbert}}, \citenamefont
  {{Lindner}}, \citenamefont {{Patra}},\ and\ \citenamefont
  {{Smirnov}}}]{Humbert:2015xy}%
  \BibitemOpen
  \bibfield  {author} {\bibinfo {author} {\bibfnamefont {P.}~\bibnamefont
  {{Humbert}}}, \bibinfo {author} {\bibfnamefont {M.}~\bibnamefont
  {{Lindner}}}, \bibinfo {author} {\bibfnamefont {S.}~\bibnamefont {{Patra}}},
  \ and\ \bibinfo {author} {\bibfnamefont {J.}~\bibnamefont {{Smirnov}}},\
  }\href@noop {} {\bibfield  {journal} {\bibinfo  {journal} {ArXiv e-prints}\ }
  (\bibinfo {year} {2015}{\natexlab{b}})},\ \Eprint
  {http://arxiv.org/abs/1505.07453} {arXiv:1505.07453 [hep-ph]} \BibitemShut
  {NoStop}%
\bibitem [{\citenamefont {{Adulpravitchai}}\ and\ \citenamefont
  {{Schmidt}}(2015)}]{Adulpravitchai:2015qv}%
  \BibitemOpen
  \bibfield  {author} {\bibinfo {author} {\bibfnamefont {A.}~\bibnamefont
  {{Adulpravitchai}}}\ and\ \bibinfo {author} {\bibfnamefont {M.~A.}\
  \bibnamefont {{Schmidt}}},\ }\href {\doibase 10.1007/JHEP01(2015)006}
  {\bibfield  {journal} {\bibinfo  {journal} {Journal of High Energy Physics}\
  }\textbf {\bibinfo {volume} {1}},\ \bibinfo {pages} {6} (\bibinfo {year}
  {2015})},\ \Eprint {http://arxiv.org/abs/1409.4330} {arXiv:1409.4330
  [hep-ph]} \BibitemShut {NoStop}%
\bibitem [{\citenamefont {{Abazajian}}\ \emph
  {et~al.}(2001{\natexlab{b}})\citenamefont {{Abazajian}}, \citenamefont
  {{Fuller}},\ and\ \citenamefont {{Tucker}}}]{Abazajian:2001fk}%
  \BibitemOpen
  \bibfield  {author} {\bibinfo {author} {\bibfnamefont {K.}~\bibnamefont
  {{Abazajian}}}, \bibinfo {author} {\bibfnamefont {G.~M.}\ \bibnamefont
  {{Fuller}}}, \ and\ \bibinfo {author} {\bibfnamefont {W.~H.}\ \bibnamefont
  {{Tucker}}},\ }\href {\doibase 10.1086/323867} {\bibfield  {journal}
  {\bibinfo  {journal} {\apj}\ }\textbf {\bibinfo {volume} {562}},\ \bibinfo
  {pages} {593} (\bibinfo {year} {2001}{\natexlab{b}})},\ \Eprint
  {http://arxiv.org/abs/arXiv:astro-ph/0106002} {arXiv:astro-ph/0106002}
  \BibitemShut {NoStop}%
\bibitem [{\citenamefont {{Hansen}}\ \emph {et~al.}(2002)\citenamefont
  {{Hansen}}, \citenamefont {{Lesgourgues}}, \citenamefont {{Pastor}},\ and\
  \citenamefont {{Silk}}}]{Hansen:2002lr}%
  \BibitemOpen
  \bibfield  {author} {\bibinfo {author} {\bibfnamefont {S.~H.}\ \bibnamefont
  {{Hansen}}}, \bibinfo {author} {\bibfnamefont {J.}~\bibnamefont
  {{Lesgourgues}}}, \bibinfo {author} {\bibfnamefont {S.}~\bibnamefont
  {{Pastor}}}, \ and\ \bibinfo {author} {\bibfnamefont {J.}~\bibnamefont
  {{Silk}}},\ }\href {\doibase 10.1046/j.1365-8711.2002.05410.x} {\bibfield
  {journal} {\bibinfo  {journal} {Mon. Not. Royal Astron. Soc.}\ }\textbf
  {\bibinfo {volume} {333}},\ \bibinfo {pages} {544} (\bibinfo {year}
  {2002})},\ \Eprint {http://arxiv.org/abs/astro-ph/0106108} {astro-ph/0106108}
  \BibitemShut {NoStop}%
\bibitem [{\citenamefont {{Abazajian}}(2009)}]{Abazajian:2009pd}%
  \BibitemOpen
  \bibfield  {author} {\bibinfo {author} {\bibfnamefont {K.}~\bibnamefont
  {{Abazajian}}},\ }in\ \href@noop {} {\emph {\bibinfo {booktitle} {astro2010:
  The Astronomy and Astrophysics Decadal Survey}}},\ \bibinfo {series}
  {Astronomy}, Vol.\ \bibinfo {volume} {2010}\ (\bibinfo {year} {2009})\
  p.~\bibinfo {pages} {1},\ \Eprint {http://arxiv.org/abs/0903.2040}
  {arXiv:0903.2040 [astro-ph.CO]} \BibitemShut {NoStop}%
\bibitem [{\citenamefont {{Merle}}\ and\ \citenamefont
  {{Schneider}}(2014)}]{Merle:2014fk}%
  \BibitemOpen
  \bibfield  {author} {\bibinfo {author} {\bibfnamefont {A.}~\bibnamefont
  {{Merle}}}\ and\ \bibinfo {author} {\bibfnamefont {A.}~\bibnamefont
  {{Schneider}}},\ }\href@noop {} {\bibfield  {journal} {\bibinfo  {journal}
  {ArXiv e-prints}\ } (\bibinfo {year} {2014})},\ \Eprint
  {http://arxiv.org/abs/1409.6311} {arXiv:1409.6311 [hep-ph]} \BibitemShut
  {NoStop}%
\bibitem [{\citenamefont {{Pal}}\ and\ \citenamefont
  {{Wolfenstein}}(1982)}]{Pal:1982qy}%
  \BibitemOpen
  \bibfield  {author} {\bibinfo {author} {\bibfnamefont {P.~B.}\ \bibnamefont
  {{Pal}}}\ and\ \bibinfo {author} {\bibfnamefont {L.}~\bibnamefont
  {{Wolfenstein}}},\ }\href {\doibase 10.1103/PhysRevD.25.766} {\bibfield
  {journal} {\bibinfo  {journal} {\prd}\ }\textbf {\bibinfo {volume} {25}},\
  \bibinfo {pages} {766} (\bibinfo {year} {1982})}\BibitemShut {NoStop}%
\bibitem [{\citenamefont {{Fuller}}\ \emph {et~al.}(2011)\citenamefont
  {{Fuller}}, \citenamefont {{Kishimoto}},\ and\ \citenamefont
  {{Kusenko}}}]{Fuller:2011lr}%
  \BibitemOpen
  \bibfield  {author} {\bibinfo {author} {\bibfnamefont {G.~M.}\ \bibnamefont
  {{Fuller}}}, \bibinfo {author} {\bibfnamefont {C.~T.}\ \bibnamefont
  {{Kishimoto}}}, \ and\ \bibinfo {author} {\bibfnamefont {A.}~\bibnamefont
  {{Kusenko}}},\ }\href@noop {} {\bibfield  {journal} {\bibinfo  {journal}
  {ArXiv e-prints}\ } (\bibinfo {year} {2011})},\ \Eprint
  {http://arxiv.org/abs/1110.6479} {arXiv:1110.6479 [astro-ph.CO]} \BibitemShut
  {NoStop}%
\bibitem [{\citenamefont {{Mohapatra}}(2015)}]{Mohapatra:2015lr}%
  \BibitemOpen
  \bibfield  {author} {\bibinfo {author} {\bibfnamefont {R.~N.}\ \bibnamefont
  {{Mohapatra}}},\ }\href@noop {} {\bibfield  {journal} {\bibinfo  {journal}
  {ArXiv e-prints}\ } (\bibinfo {year} {2015})},\ \Eprint
  {http://arxiv.org/abs/1503.06478} {arXiv:1503.06478 [hep-ph]} \BibitemShut
  {NoStop}%
\bibitem [{\citenamefont {{Scherrer}}\ and\ \citenamefont
  {{Turner}}(1985)}]{Scherrer:1985qv}%
  \BibitemOpen
  \bibfield  {author} {\bibinfo {author} {\bibfnamefont {R.~J.}\ \bibnamefont
  {{Scherrer}}}\ and\ \bibinfo {author} {\bibfnamefont {M.~S.}\ \bibnamefont
  {{Turner}}},\ }\href {\doibase 10.1103/PhysRevD.31.681} {\bibfield  {journal}
  {\bibinfo  {journal} {\prd}\ }\textbf {\bibinfo {volume} {31}},\ \bibinfo
  {pages} {681} (\bibinfo {year} {1985})}\BibitemShut {NoStop}%
\bibitem [{\citenamefont {{Dolgov}}\ \emph {et~al.}(2000)\citenamefont
  {{Dolgov}}, \citenamefont {{Hansen}}, \citenamefont {{Raffelt}},\ and\
  \citenamefont {{Semikoz}}}]{Dolgov:2000fr}%
  \BibitemOpen
  \bibfield  {author} {\bibinfo {author} {\bibfnamefont {A.~D.}\ \bibnamefont
  {{Dolgov}}}, \bibinfo {author} {\bibfnamefont {S.~H.}\ \bibnamefont
  {{Hansen}}}, \bibinfo {author} {\bibfnamefont {G.}~\bibnamefont {{Raffelt}}},
  \ and\ \bibinfo {author} {\bibfnamefont {D.~V.}\ \bibnamefont {{Semikoz}}},\
  }\href {\doibase 10.1016/S0550-3213(00)00566-6} {\bibfield  {journal}
  {\bibinfo  {journal} {Nuclear Physics B}\ }\textbf {\bibinfo {volume}
  {590}},\ \bibinfo {pages} {562} (\bibinfo {year} {2000})},\ \Eprint
  {http://arxiv.org/abs/arXiv:hep-ph/0008138} {arXiv:hep-ph/0008138}
  \BibitemShut {NoStop}%
\bibitem [{\citenamefont {{Menestrina}}\ and\ \citenamefont
  {{Scherrer}}(2012)}]{Menestrina:2012fj}%
  \BibitemOpen
  \bibfield  {author} {\bibinfo {author} {\bibfnamefont {J.~L.}\ \bibnamefont
  {{Menestrina}}}\ and\ \bibinfo {author} {\bibfnamefont {R.~J.}\ \bibnamefont
  {{Scherrer}}},\ }\href {\doibase 10.1103/PhysRevD.85.047301} {\bibfield
  {journal} {\bibinfo  {journal} {\prd}\ }\textbf {\bibinfo {volume} {85}},\
  \bibinfo {eid} {047301} (\bibinfo {year} {2012})},\ \Eprint
  {http://arxiv.org/abs/1111.0605} {arXiv:1111.0605 [astro-ph.CO]} \BibitemShut
  {NoStop}%
\bibitem [{\citenamefont {{Grohs}}\ \emph {et~al.}(2015)\citenamefont
  {{Grohs}}, \citenamefont {{Fuller}}, \citenamefont {{Kishimoto}},\ and\
  \citenamefont {{Paris}}}]{Grohs:2015lr}%
  \BibitemOpen
  \bibfield  {author} {\bibinfo {author} {\bibfnamefont {E.}~\bibnamefont
  {{Grohs}}}, \bibinfo {author} {\bibfnamefont {G.~M.}\ \bibnamefont
  {{Fuller}}}, \bibinfo {author} {\bibfnamefont {C.~T.}\ \bibnamefont
  {{Kishimoto}}}, \ and\ \bibinfo {author} {\bibfnamefont {M.~W.}\ \bibnamefont
  {{Paris}}},\ }\href {\doibase 10.1088/1475-7516/2015/05/017} {\bibfield
  {journal} {\bibinfo  {journal} {Journal of Cosmology and Astroparticle
  Physics}\ }\textbf {\bibinfo {volume} {5}},\ \bibinfo {eid} {017} (\bibinfo
  {year} {2015})},\ \Eprint {http://arxiv.org/abs/1502.02718}
  {arXiv:1502.02718} \BibitemShut {NoStop}%
\bibitem [{\citenamefont {{Pontecorvo}}(1968)}]{Pontecorvo:1968fk}%
  \BibitemOpen
  \bibfield  {author} {\bibinfo {author} {\bibfnamefont {B.}~\bibnamefont
  {{Pontecorvo}}},\ }\href@noop {} {\bibfield  {journal} {\bibinfo  {journal}
  {Soviet Journal of Experimental and Theoretical Physics}\ }\textbf {\bibinfo
  {volume} {26}},\ \bibinfo {pages} {984} (\bibinfo {year} {1968})}\BibitemShut
  {NoStop}%
\bibitem [{\citenamefont {Langacker}(1981)}]{Langacker:1980js}%
  \BibitemOpen
  \bibfield  {author} {\bibinfo {author} {\bibfnamefont {P.}~\bibnamefont
  {Langacker}},\ }\href {\doibase 10.1016/0370-1573(81)90059-4} {\bibfield
  {journal} {\bibinfo  {journal} {Phys.Rept.}\ }\textbf {\bibinfo {volume}
  {72}},\ \bibinfo {pages} {185} (\bibinfo {year} {1981})}\BibitemShut
  {NoStop}%
\bibitem [{\citenamefont {Langacker}(2009)}]{Langacker:2008yv}%
  \BibitemOpen
  \bibfield  {author} {\bibinfo {author} {\bibfnamefont {P.}~\bibnamefont
  {Langacker}},\ }\href {\doibase 10.1103/RevModPhys.81.1199} {\bibfield
  {journal} {\bibinfo  {journal} {Rev.Mod.Phys.}\ }\textbf {\bibinfo {volume}
  {81}},\ \bibinfo {pages} {1199} (\bibinfo {year} {2009})},\ \Eprint
  {http://arxiv.org/abs/0801.1345} {arXiv:0801.1345 [hep-ph]} \BibitemShut
  {NoStop}%
\bibitem [{\citenamefont {{Kolb}}\ and\ \citenamefont
  {{Turner}}(1990)}]{Kolb:1990bs}%
  \BibitemOpen
  \bibfield  {author} {\bibinfo {author} {\bibfnamefont {E.~W.}\ \bibnamefont
  {{Kolb}}}\ and\ \bibinfo {author} {\bibfnamefont {M.~S.}\ \bibnamefont
  {{Turner}}},\ }\href@noop {} {\emph {\bibinfo {title} {Front.~Phys.,
  Vol.~69,}}}\ (\bibinfo  {publisher} {Addison-Wesley},\ \bibinfo {year}
  {1990})\BibitemShut {NoStop}%
\bibitem [{\citenamefont {{Seljak}}\ \emph {et~al.}(2006)\citenamefont
  {{Seljak}}, \citenamefont {{Makarov}}, \citenamefont {{McDonald}},\ and\
  \citenamefont {{Trac}}}]{Seljak:2006fk}%
  \BibitemOpen
  \bibfield  {author} {\bibinfo {author} {\bibfnamefont {U.}~\bibnamefont
  {{Seljak}}}, \bibinfo {author} {\bibfnamefont {A.}~\bibnamefont {{Makarov}}},
  \bibinfo {author} {\bibfnamefont {P.}~\bibnamefont {{McDonald}}}, \ and\
  \bibinfo {author} {\bibfnamefont {H.}~\bibnamefont {{Trac}}},\ }\href
  {\doibase 10.1103/PhysRevLett.97.191303} {\bibfield  {journal} {\bibinfo
  {journal} {Physical Review Letters}\ }\textbf {\bibinfo {volume} {97}},\
  \bibinfo {eid} {191303} (\bibinfo {year} {2006})},\ \Eprint
  {http://arxiv.org/abs/arXiv:astro-ph/0602430} {arXiv:astro-ph/0602430}
  \BibitemShut {NoStop}%
\bibitem [{\citenamefont {{Heckler}}\ and\ \citenamefont
  {{Hogan}}(1993)}]{Heckler:1993qy}%
  \BibitemOpen
  \bibfield  {author} {\bibinfo {author} {\bibfnamefont {A.}~\bibnamefont
  {{Heckler}}}\ and\ \bibinfo {author} {\bibfnamefont {C.~J.}\ \bibnamefont
  {{Hogan}}},\ }\href {\doibase 10.1103/PhysRevD.47.4256} {\bibfield  {journal}
  {\bibinfo  {journal} {\prd}\ }\textbf {\bibinfo {volume} {47}},\ \bibinfo
  {pages} {4256} (\bibinfo {year} {1993})}\BibitemShut {NoStop}%
\bibitem [{\citenamefont {{Dolgov}}\ and\ \citenamefont
  {{Silk}}(1993)}]{Dolgov:1993fk}%
  \BibitemOpen
  \bibfield  {author} {\bibinfo {author} {\bibfnamefont {A.}~\bibnamefont
  {{Dolgov}}}\ and\ \bibinfo {author} {\bibfnamefont {J.}~\bibnamefont
  {{Silk}}},\ }\href {\doibase 10.1103/PhysRevD.47.4244} {\bibfield  {journal}
  {\bibinfo  {journal} {\prd}\ }\textbf {\bibinfo {volume} {47}},\ \bibinfo
  {pages} {4244} (\bibinfo {year} {1993})}\BibitemShut {NoStop}%
\bibitem [{\citenamefont {{Jedamzik}}\ and\ \citenamefont
  {{Fuller}}(1994)}]{Jedamzik:1994lr}%
  \BibitemOpen
  \bibfield  {author} {\bibinfo {author} {\bibfnamefont {K.}~\bibnamefont
  {{Jedamzik}}}\ and\ \bibinfo {author} {\bibfnamefont {G.~M.}\ \bibnamefont
  {{Fuller}}},\ }\href {\doibase 10.1086/173788} {\bibfield  {journal}
  {\bibinfo  {journal} {\apj}\ }\textbf {\bibinfo {volume} {423}},\ \bibinfo
  {pages} {33} (\bibinfo {year} {1994})},\ \Eprint
  {http://arxiv.org/abs/astro-ph/9312063} {astro-ph/9312063} \BibitemShut
  {NoStop}%
\bibitem [{\citenamefont {{Jedamzik}}\ \emph {et~al.}(1994)\citenamefont
  {{Jedamzik}}, \citenamefont {{Fuller}},\ and\ \citenamefont
  {{Mathews}}}]{Jedamzik:1994uq}%
  \BibitemOpen
  \bibfield  {author} {\bibinfo {author} {\bibfnamefont {K.}~\bibnamefont
  {{Jedamzik}}}, \bibinfo {author} {\bibfnamefont {G.~M.}\ \bibnamefont
  {{Fuller}}}, \ and\ \bibinfo {author} {\bibfnamefont {G.~J.}\ \bibnamefont
  {{Mathews}}},\ }\href {\doibase 10.1086/173789} {\bibfield  {journal}
  {\bibinfo  {journal} {\apj}\ }\textbf {\bibinfo {volume} {423}},\ \bibinfo
  {pages} {50} (\bibinfo {year} {1994})},\ \Eprint
  {http://arxiv.org/abs/astro-ph/9312065} {astro-ph/9312065} \BibitemShut
  {NoStop}%
\bibitem [{\citenamefont {{Tremaine}}\ and\ \citenamefont
  {{Gunn}}(1979)}]{Tremaine:1979nr}%
  \BibitemOpen
  \bibfield  {author} {\bibinfo {author} {\bibfnamefont {S.}~\bibnamefont
  {{Tremaine}}}\ and\ \bibinfo {author} {\bibfnamefont {J.~E.}\ \bibnamefont
  {{Gunn}}},\ }\href {\doibase 10.1103/PhysRevLett.42.407} {\bibfield
  {journal} {\bibinfo  {journal} {Physical Review Letters}\ }\textbf {\bibinfo
  {volume} {42}},\ \bibinfo {pages} {407} (\bibinfo {year} {1979})}\BibitemShut
  {NoStop}%
\bibitem [{\citenamefont {{Boyarsky}}\ \emph
  {et~al.}(2007{\natexlab{a}})\citenamefont {{Boyarsky}}, \citenamefont
  {{Nevalainen}},\ and\ \citenamefont {{Ruchayskiy}}}]{Boyarsky:2007zr}%
  \BibitemOpen
  \bibfield  {author} {\bibinfo {author} {\bibfnamefont {A.}~\bibnamefont
  {{Boyarsky}}}, \bibinfo {author} {\bibfnamefont {J.}~\bibnamefont
  {{Nevalainen}}}, \ and\ \bibinfo {author} {\bibfnamefont {O.}~\bibnamefont
  {{Ruchayskiy}}},\ }\href {\doibase 10.1051/0004-6361:20066774} {\bibfield
  {journal} {\bibinfo  {journal} {Astron. Astrophys.}\ }\textbf {\bibinfo
  {volume} {471}},\ \bibinfo {pages} {51} (\bibinfo {year}
  {2007}{\natexlab{a}})},\ \Eprint {http://arxiv.org/abs/astro-ph/0610961}
  {astro-ph/0610961} \BibitemShut {NoStop}%
\bibitem [{\citenamefont {{Boyarsky}}\ \emph
  {et~al.}(2008{\natexlab{a}})\citenamefont {{Boyarsky}}, \citenamefont
  {{Malyshev}}, \citenamefont {{Neronov}},\ and\ \citenamefont
  {{Ruchayskiy}}}]{Boyarsky:2008pd}%
  \BibitemOpen
  \bibfield  {author} {\bibinfo {author} {\bibfnamefont {A.}~\bibnamefont
  {{Boyarsky}}}, \bibinfo {author} {\bibfnamefont {D.}~\bibnamefont
  {{Malyshev}}}, \bibinfo {author} {\bibfnamefont {A.}~\bibnamefont
  {{Neronov}}}, \ and\ \bibinfo {author} {\bibfnamefont {O.}~\bibnamefont
  {{Ruchayskiy}}},\ }\href {\doibase 10.1111/j.1365-2966.2008.13003.x}
  {\bibfield  {journal} {\bibinfo  {journal} {Mon. Not. Royal Astron. Soc.}\
  }\textbf {\bibinfo {volume} {387}},\ \bibinfo {pages} {1345} (\bibinfo {year}
  {2008}{\natexlab{a}})},\ \Eprint {http://arxiv.org/abs/0710.4922}
  {arXiv:0710.4922} \BibitemShut {NoStop}%
\bibitem [{\citenamefont {{Ng}}\ \emph {et~al.}(2015)\citenamefont {{Ng}},
  \citenamefont {{Horiuchi}}, \citenamefont {{Gaskins}}, \citenamefont
  {{Smith}},\ and\ \citenamefont {{Preece}}}]{Ng:2015th}%
  \BibitemOpen
  \bibfield  {author} {\bibinfo {author} {\bibfnamefont {K.~C.~Y.}\
  \bibnamefont {{Ng}}}, \bibinfo {author} {\bibfnamefont {S.}~\bibnamefont
  {{Horiuchi}}}, \bibinfo {author} {\bibfnamefont {J.~M.}\ \bibnamefont
  {{Gaskins}}}, \bibinfo {author} {\bibfnamefont {M.}~\bibnamefont {{Smith}}},
  \ and\ \bibinfo {author} {\bibfnamefont {R.}~\bibnamefont {{Preece}}},\
  }\href@noop {} {\bibfield  {journal} {\bibinfo  {journal} {ArXiv e-prints}\ }
  (\bibinfo {year} {2015})},\ \Eprint {http://arxiv.org/abs/1504.04027}
  {arXiv:1504.04027} \BibitemShut {NoStop}%
\bibitem [{\citenamefont {{Watson}}\ \emph {et~al.}(2006)\citenamefont
  {{Watson}}, \citenamefont {{Beacom}}, \citenamefont {{Y{\"u}ksel}},\ and\
  \citenamefont {{Walker}}}]{Watson:2006rm}%
  \BibitemOpen
  \bibfield  {author} {\bibinfo {author} {\bibfnamefont {C.~R.}\ \bibnamefont
  {{Watson}}}, \bibinfo {author} {\bibfnamefont {J.~F.}\ \bibnamefont
  {{Beacom}}}, \bibinfo {author} {\bibfnamefont {H.}~\bibnamefont
  {{Y{\"u}ksel}}}, \ and\ \bibinfo {author} {\bibfnamefont {T.~P.}\
  \bibnamefont {{Walker}}},\ }\href {\doibase 10.1103/PhysRevD.74.033009}
  {\bibfield  {journal} {\bibinfo  {journal} {\prd}\ }\textbf {\bibinfo
  {volume} {74}},\ \bibinfo {eid} {033009} (\bibinfo {year} {2006})},\ \Eprint
  {http://arxiv.org/abs/astro-ph/0605424} {astro-ph/0605424} \BibitemShut
  {NoStop}%
\bibitem [{\citenamefont {{Boyarsky}}\ \emph
  {et~al.}(2006{\natexlab{a}})\citenamefont {{Boyarsky}}, \citenamefont
  {{Neronov}}, \citenamefont {{Ruchayskiy}},\ and\ \citenamefont
  {{Shaposhnikov}}}]{Boyarsky:2006yq}%
  \BibitemOpen
  \bibfield  {author} {\bibinfo {author} {\bibfnamefont {A.}~\bibnamefont
  {{Boyarsky}}}, \bibinfo {author} {\bibfnamefont {A.}~\bibnamefont
  {{Neronov}}}, \bibinfo {author} {\bibfnamefont {O.}~\bibnamefont
  {{Ruchayskiy}}}, \ and\ \bibinfo {author} {\bibfnamefont {M.}~\bibnamefont
  {{Shaposhnikov}}},\ }\href {\doibase 10.1103/PhysRevD.74.103506} {\bibfield
  {journal} {\bibinfo  {journal} {\prd}\ }\textbf {\bibinfo {volume} {74}},\
  \bibinfo {eid} {103506} (\bibinfo {year} {2006}{\natexlab{a}})},\ \Eprint
  {http://arxiv.org/abs/astro-ph/0603368} {astro-ph/0603368} \BibitemShut
  {NoStop}%
\bibitem [{\citenamefont {{Boyarsky}}\ \emph
  {et~al.}(2006{\natexlab{b}})\citenamefont {{Boyarsky}}, \citenamefont
  {{Neronov}}, \citenamefont {{Ruchayskiy}},\ and\ \citenamefont
  {{Shaposhnikov}}}]{Boyarsky:2006kx}%
  \BibitemOpen
  \bibfield  {author} {\bibinfo {author} {\bibfnamefont {A.}~\bibnamefont
  {{Boyarsky}}}, \bibinfo {author} {\bibfnamefont {A.}~\bibnamefont
  {{Neronov}}}, \bibinfo {author} {\bibfnamefont {O.}~\bibnamefont
  {{Ruchayskiy}}}, \ and\ \bibinfo {author} {\bibfnamefont {M.}~\bibnamefont
  {{Shaposhnikov}}},\ }\href {\doibase 10.1111/j.1365-2966.2006.10458.x}
  {\bibfield  {journal} {\bibinfo  {journal} {Mon. Not. Royal Astron. Soc.}\
  }\textbf {\bibinfo {volume} {370}},\ \bibinfo {pages} {213} (\bibinfo {year}
  {2006}{\natexlab{b}})},\ \Eprint {http://arxiv.org/abs/astro-ph/0512509}
  {astro-ph/0512509} \BibitemShut {NoStop}%
\bibitem [{\citenamefont {{Boyarsky}}\ \emph
  {et~al.}(2006{\natexlab{c}})\citenamefont {{Boyarsky}}, \citenamefont
  {{Neronov}}, \citenamefont {{Ruchayskiy}}, \citenamefont {{Shaposhnikov}},\
  and\ \citenamefont {{Tkachev}}}]{Boyarsky:2006fj}%
  \BibitemOpen
  \bibfield  {author} {\bibinfo {author} {\bibfnamefont {A.}~\bibnamefont
  {{Boyarsky}}}, \bibinfo {author} {\bibfnamefont {A.}~\bibnamefont
  {{Neronov}}}, \bibinfo {author} {\bibfnamefont {O.}~\bibnamefont
  {{Ruchayskiy}}}, \bibinfo {author} {\bibfnamefont {M.}~\bibnamefont
  {{Shaposhnikov}}}, \ and\ \bibinfo {author} {\bibfnamefont {I.}~\bibnamefont
  {{Tkachev}}},\ }\href {\doibase 10.1103/PhysRevLett.97.261302} {\bibfield
  {journal} {\bibinfo  {journal} {Physical Review Letters}\ }\textbf {\bibinfo
  {volume} {97}},\ \bibinfo {eid} {261302} (\bibinfo {year}
  {2006}{\natexlab{c}})},\ \Eprint {http://arxiv.org/abs/astro-ph/0603660}
  {astro-ph/0603660} \BibitemShut {NoStop}%
\bibitem [{\citenamefont {{Y{\"u}ksel}}\ \emph {et~al.}(2008)\citenamefont
  {{Y{\"u}ksel}}, \citenamefont {{Beacom}},\ and\ \citenamefont
  {{Watson}}}]{Yuksel:2008fr}%
  \BibitemOpen
  \bibfield  {author} {\bibinfo {author} {\bibfnamefont {H.}~\bibnamefont
  {{Y{\"u}ksel}}}, \bibinfo {author} {\bibfnamefont {J.~F.}\ \bibnamefont
  {{Beacom}}}, \ and\ \bibinfo {author} {\bibfnamefont {C.~R.}\ \bibnamefont
  {{Watson}}},\ }\href {\doibase 10.1103/PhysRevLett.101.121301} {\bibfield
  {journal} {\bibinfo  {journal} {Physical Review Letters}\ }\textbf {\bibinfo
  {volume} {101}},\ \bibinfo {eid} {121301} (\bibinfo {year} {2008})},\ \Eprint
  {http://arxiv.org/abs/0706.4084} {arXiv:0706.4084} \BibitemShut {NoStop}%
\bibitem [{\citenamefont {{Horiuchi}}\ \emph {et~al.}(2014)\citenamefont
  {{Horiuchi}}, \citenamefont {{Humphrey}}, \citenamefont {{O{\~n}orbe}},
  \citenamefont {{Abazajian}}, \citenamefont {{Kaplinghat}},\ and\
  \citenamefont {{Garrison-Kimmel}}}]{Horiuchi:2014zl}%
  \BibitemOpen
  \bibfield  {author} {\bibinfo {author} {\bibfnamefont {S.}~\bibnamefont
  {{Horiuchi}}}, \bibinfo {author} {\bibfnamefont {P.~J.}\ \bibnamefont
  {{Humphrey}}}, \bibinfo {author} {\bibfnamefont {J.}~\bibnamefont
  {{O{\~n}orbe}}}, \bibinfo {author} {\bibfnamefont {K.~N.}\ \bibnamefont
  {{Abazajian}}}, \bibinfo {author} {\bibfnamefont {M.}~\bibnamefont
  {{Kaplinghat}}}, \ and\ \bibinfo {author} {\bibfnamefont {S.}~\bibnamefont
  {{Garrison-Kimmel}}},\ }\href {\doibase 10.1103/PhysRevD.89.025017}
  {\bibfield  {journal} {\bibinfo  {journal} {\prd}\ }\textbf {\bibinfo
  {volume} {89}},\ \bibinfo {eid} {025017} (\bibinfo {year} {2014})},\ \Eprint
  {http://arxiv.org/abs/1311.0282} {arXiv:1311.0282} \BibitemShut {NoStop}%
\bibitem [{\citenamefont {{Watson}}\ \emph {et~al.}(2012)\citenamefont
  {{Watson}}, \citenamefont {{Li}},\ and\ \citenamefont
  {{Polley}}}]{Watson:2012rt}%
  \BibitemOpen
  \bibfield  {author} {\bibinfo {author} {\bibfnamefont {C.~R.}\ \bibnamefont
  {{Watson}}}, \bibinfo {author} {\bibfnamefont {Z.}~\bibnamefont {{Li}}}, \
  and\ \bibinfo {author} {\bibfnamefont {N.~K.}\ \bibnamefont {{Polley}}},\
  }\href {\doibase 10.1088/1475-7516/2012/03/018} {\bibfield  {journal}
  {\bibinfo  {journal} {Journal of Cosmology and Astroparticle Physics}\
  }\textbf {\bibinfo {volume} {3}},\ \bibinfo {eid} {018} (\bibinfo {year}
  {2012})},\ \Eprint {http://arxiv.org/abs/1111.4217} {arXiv:1111.4217
  [astro-ph.CO]} \BibitemShut {NoStop}%
\bibitem [{\citenamefont {{Boyarsky}}\ \emph
  {et~al.}(2008{\natexlab{b}})\citenamefont {{Boyarsky}}, \citenamefont
  {{Iakubovskyi}}, \citenamefont {{Ruchayskiy}},\ and\ \citenamefont
  {{Savchenko}}}]{Boyarsky:2008gf}%
  \BibitemOpen
  \bibfield  {author} {\bibinfo {author} {\bibfnamefont {A.}~\bibnamefont
  {{Boyarsky}}}, \bibinfo {author} {\bibfnamefont {D.}~\bibnamefont
  {{Iakubovskyi}}}, \bibinfo {author} {\bibfnamefont {O.}~\bibnamefont
  {{Ruchayskiy}}}, \ and\ \bibinfo {author} {\bibfnamefont {V.}~\bibnamefont
  {{Savchenko}}},\ }\href {\doibase 10.1111/j.1365-2966.2008.13266.x}
  {\bibfield  {journal} {\bibinfo  {journal} {Mon. Not. Royal Astron. Soc.}\
  }\textbf {\bibinfo {volume} {387}},\ \bibinfo {pages} {1361} (\bibinfo {year}
  {2008}{\natexlab{b}})},\ \Eprint {http://arxiv.org/abs/0709.2301}
  {arXiv:0709.2301} \BibitemShut {NoStop}%
\bibitem [{\citenamefont {{Loewenstein}}\ \emph {et~al.}(2009)\citenamefont
  {{Loewenstein}}, \citenamefont {{Kusenko}},\ and\ \citenamefont
  {{Biermann}}}]{Loewenstein:2009lr}%
  \BibitemOpen
  \bibfield  {author} {\bibinfo {author} {\bibfnamefont {M.}~\bibnamefont
  {{Loewenstein}}}, \bibinfo {author} {\bibfnamefont {A.}~\bibnamefont
  {{Kusenko}}}, \ and\ \bibinfo {author} {\bibfnamefont {P.~L.}\ \bibnamefont
  {{Biermann}}},\ }\href {\doibase 10.1088/0004-637X/700/1/426} {\bibfield
  {journal} {\bibinfo  {journal} {\apj}\ }\textbf {\bibinfo {volume} {700}},\
  \bibinfo {pages} {426} (\bibinfo {year} {2009})},\ \Eprint
  {http://arxiv.org/abs/0812.2710} {arXiv:0812.2710} \BibitemShut {NoStop}%
\bibitem [{\citenamefont {{Riemer-S{\o}rensen}}\ and\ \citenamefont
  {{Hansen}}(2009)}]{Riemer-Sorensen:2009qy}%
  \BibitemOpen
  \bibfield  {author} {\bibinfo {author} {\bibfnamefont {S.}~\bibnamefont
  {{Riemer-S{\o}rensen}}}\ and\ \bibinfo {author} {\bibfnamefont {S.~H.}\
  \bibnamefont {{Hansen}}},\ }\href {\doibase 10.1051/0004-6361/200912430}
  {\bibfield  {journal} {\bibinfo  {journal} {Astron. Astrophys.}\ }\textbf
  {\bibinfo {volume} {500}},\ \bibinfo {pages} {L37} (\bibinfo {year}
  {2009})}\BibitemShut {NoStop}%
\bibitem [{\citenamefont {{Loewenstein}}\ and\ \citenamefont
  {{Kusenko}}(2010)}]{Loewenstein:2010fk}%
  \BibitemOpen
  \bibfield  {author} {\bibinfo {author} {\bibfnamefont {M.}~\bibnamefont
  {{Loewenstein}}}\ and\ \bibinfo {author} {\bibfnamefont {A.}~\bibnamefont
  {{Kusenko}}},\ }\href {\doibase 10.1088/0004-637X/714/1/652} {\bibfield
  {journal} {\bibinfo  {journal} {\apj}\ }\textbf {\bibinfo {volume} {714}},\
  \bibinfo {pages} {652} (\bibinfo {year} {2010})},\ \Eprint
  {http://arxiv.org/abs/0912.0552} {arXiv:0912.0552 [astro-ph.HE]} \BibitemShut
  {NoStop}%
\bibitem [{\citenamefont {{Loewenstein}}\ and\ \citenamefont
  {{Kusenko}}(2012)}]{Loewenstein:2012fk}%
  \BibitemOpen
  \bibfield  {author} {\bibinfo {author} {\bibfnamefont {M.}~\bibnamefont
  {{Loewenstein}}}\ and\ \bibinfo {author} {\bibfnamefont {A.}~\bibnamefont
  {{Kusenko}}},\ }\href {\doibase 10.1088/0004-637X/751/2/82} {\bibfield
  {journal} {\bibinfo  {journal} {\apj}\ }\textbf {\bibinfo {volume} {751}},\
  \bibinfo {eid} {82} (\bibinfo {year} {2012})},\ \Eprint
  {http://arxiv.org/abs/1203.5229} {arXiv:1203.5229 [astro-ph.CO]} \BibitemShut
  {NoStop}%
\bibitem [{\citenamefont {{Boyarsky}}\ \emph
  {et~al.}(2008{\natexlab{c}})\citenamefont {{Boyarsky}}, \citenamefont
  {{Ruchayskiy}},\ and\ \citenamefont {{Markevitch}}}]{Boyarsky:2008kx}%
  \BibitemOpen
  \bibfield  {author} {\bibinfo {author} {\bibfnamefont {A.}~\bibnamefont
  {{Boyarsky}}}, \bibinfo {author} {\bibfnamefont {O.}~\bibnamefont
  {{Ruchayskiy}}}, \ and\ \bibinfo {author} {\bibfnamefont {M.}~\bibnamefont
  {{Markevitch}}},\ }\href {\doibase 10.1086/524397} {\bibfield  {journal}
  {\bibinfo  {journal} {\apj}\ }\textbf {\bibinfo {volume} {673}},\ \bibinfo
  {pages} {752} (\bibinfo {year} {2008}{\natexlab{c}})},\ \Eprint
  {http://arxiv.org/abs/astro-ph/0611168} {astro-ph/0611168} \BibitemShut
  {NoStop}%
\bibitem [{\citenamefont {{Riemer-Sorensen}}\ \emph {et~al.}(2007)\citenamefont
  {{Riemer-Sorensen}}, \citenamefont {{Pedersen}}, \citenamefont {{Hansen}},\
  and\ \citenamefont {{Dahle}}}]{Riemer-Sorensen:2007vn}%
  \BibitemOpen
  \bibfield  {author} {\bibinfo {author} {\bibfnamefont {S.}~\bibnamefont
  {{Riemer-Sorensen}}}, \bibinfo {author} {\bibfnamefont {K.}~\bibnamefont
  {{Pedersen}}}, \bibinfo {author} {\bibfnamefont {S.~H.}\ \bibnamefont
  {{Hansen}}}, \ and\ \bibinfo {author} {\bibfnamefont {H.}~\bibnamefont
  {{Dahle}}},\ }\href {\doibase 10.1103/PhysRevD.76.043524} {\bibfield
  {journal} {\bibinfo  {journal} {\prd}\ }\textbf {\bibinfo {volume} {76}},\
  \bibinfo {eid} {043524} (\bibinfo {year} {2007})},\ \Eprint
  {http://arxiv.org/abs/astro-ph/0610034} {astro-ph/0610034} \BibitemShut
  {NoStop}%
\bibitem [{\citenamefont {{Riemer-S{\o}rensen}}\ \emph
  {et~al.}(2015)\citenamefont {{Riemer-S{\o}rensen}}, \citenamefont {{Wik}},
  \citenamefont {{Madejski}}, \citenamefont {{Molendi}}, \citenamefont
  {{Gastaldello}}, \citenamefont {{Harrison}}, \citenamefont {{Craig}},
  \citenamefont {{Hailey}}, \citenamefont {{Boggs}}, \citenamefont
  {{Christensen}}, \citenamefont {{Stern}}, \citenamefont {{Zhang}},\ and\
  \citenamefont {{Hornstrup}}}]{Riemer-Sorensen:2015lr}%
  \BibitemOpen
  \bibfield  {author} {\bibinfo {author} {\bibfnamefont {S.}~\bibnamefont
  {{Riemer-S{\o}rensen}}}, \bibinfo {author} {\bibfnamefont {D.}~\bibnamefont
  {{Wik}}}, \bibinfo {author} {\bibfnamefont {G.}~\bibnamefont {{Madejski}}},
  \bibinfo {author} {\bibfnamefont {S.}~\bibnamefont {{Molendi}}}, \bibinfo
  {author} {\bibfnamefont {F.}~\bibnamefont {{Gastaldello}}}, \bibinfo {author}
  {\bibfnamefont {F.~A.}\ \bibnamefont {{Harrison}}}, \bibinfo {author}
  {\bibfnamefont {W.~W.}\ \bibnamefont {{Craig}}}, \bibinfo {author}
  {\bibfnamefont {C.~J.}\ \bibnamefont {{Hailey}}}, \bibinfo {author}
  {\bibfnamefont {S.~E.}\ \bibnamefont {{Boggs}}}, \bibinfo {author}
  {\bibfnamefont {F.~E.}\ \bibnamefont {{Christensen}}}, \bibinfo {author}
  {\bibfnamefont {D.}~\bibnamefont {{Stern}}}, \bibinfo {author} {\bibfnamefont
  {W.~W.}\ \bibnamefont {{Zhang}}}, \ and\ \bibinfo {author} {\bibfnamefont
  {A.}~\bibnamefont {{Hornstrup}}},\ }\href@noop {} {\bibfield  {journal}
  {\bibinfo  {journal} {ArXiv e-prints}\ } (\bibinfo {year} {2015})},\ \Eprint
  {http://arxiv.org/abs/1507.01378} {arXiv:1507.01378} \BibitemShut {NoStop}%
\bibitem [{\citenamefont {{Abazajian}}\ \emph {et~al.}(2007)\citenamefont
  {{Abazajian}}, \citenamefont {{Markevitch}}, \citenamefont {{Koushiappas}},\
  and\ \citenamefont {{Hickox}}}]{Abazajian:2007ys}%
  \BibitemOpen
  \bibfield  {author} {\bibinfo {author} {\bibfnamefont {K.~N.}\ \bibnamefont
  {{Abazajian}}}, \bibinfo {author} {\bibfnamefont {M.}~\bibnamefont
  {{Markevitch}}}, \bibinfo {author} {\bibfnamefont {S.~M.}\ \bibnamefont
  {{Koushiappas}}}, \ and\ \bibinfo {author} {\bibfnamefont {R.~C.}\
  \bibnamefont {{Hickox}}},\ }\href {\doibase 10.1103/PhysRevD.75.063511}
  {\bibfield  {journal} {\bibinfo  {journal} {\prd}\ }\textbf {\bibinfo
  {volume} {75}},\ \bibinfo {eid} {063511} (\bibinfo {year} {2007})},\ \Eprint
  {http://arxiv.org/abs/astro-ph/0611144} {astro-ph/0611144} \BibitemShut
  {NoStop}%
\bibitem [{\citenamefont {{Bulbul}}\ \emph
  {et~al.}(2014{\natexlab{a}})\citenamefont {{Bulbul}}, \citenamefont
  {{Markevitch}}, \citenamefont {{Foster}}, \citenamefont {{Smith}},
  \citenamefont {{Loewenstein}},\ and\ \citenamefont
  {{Randall}}}]{Bulbul:2014lr}%
  \BibitemOpen
  \bibfield  {author} {\bibinfo {author} {\bibfnamefont {E.}~\bibnamefont
  {{Bulbul}}}, \bibinfo {author} {\bibfnamefont {M.}~\bibnamefont
  {{Markevitch}}}, \bibinfo {author} {\bibfnamefont {A.}~\bibnamefont
  {{Foster}}}, \bibinfo {author} {\bibfnamefont {R.~K.}\ \bibnamefont
  {{Smith}}}, \bibinfo {author} {\bibfnamefont {M.}~\bibnamefont
  {{Loewenstein}}}, \ and\ \bibinfo {author} {\bibfnamefont {S.~W.}\
  \bibnamefont {{Randall}}},\ }\href {\doibase 10.1088/0004-637X/789/1/13}
  {\bibfield  {journal} {\bibinfo  {journal} {\apj}\ }\textbf {\bibinfo
  {volume} {789}},\ \bibinfo {eid} {13} (\bibinfo {year}
  {2014}{\natexlab{a}})},\ \Eprint {http://arxiv.org/abs/1402.2301}
  {arXiv:1402.2301} \BibitemShut {NoStop}%
\bibitem [{\citenamefont {{Boyarsky}}\ \emph
  {et~al.}(2014{\natexlab{a}})\citenamefont {{Boyarsky}}, \citenamefont
  {{Ruchayskiy}}, \citenamefont {{Iakubovskyi}},\ and\ \citenamefont
  {{Franse}}}]{Boyarsky:2014fk}%
  \BibitemOpen
  \bibfield  {author} {\bibinfo {author} {\bibfnamefont {A.}~\bibnamefont
  {{Boyarsky}}}, \bibinfo {author} {\bibfnamefont {O.}~\bibnamefont
  {{Ruchayskiy}}}, \bibinfo {author} {\bibfnamefont {D.}~\bibnamefont
  {{Iakubovskyi}}}, \ and\ \bibinfo {author} {\bibfnamefont {J.}~\bibnamefont
  {{Franse}}},\ }\href {\doibase 10.1103/PhysRevLett.113.251301} {\bibfield
  {journal} {\bibinfo  {journal} {Physical Review Letters}\ }\textbf {\bibinfo
  {volume} {113}},\ \bibinfo {eid} {251301} (\bibinfo {year}
  {2014}{\natexlab{a}})},\ \Eprint {http://arxiv.org/abs/1402.4119}
  {arXiv:1402.4119} \BibitemShut {NoStop}%
\bibitem [{\citenamefont {{Boyarsky}}\ \emph
  {et~al.}(2014{\natexlab{b}})\citenamefont {{Boyarsky}}, \citenamefont
  {{Franse}}, \citenamefont {{Iakubovskyi}},\ and\ \citenamefont
  {{Ruchayskiy}}}]{Boyarsky:2014qy}%
  \BibitemOpen
  \bibfield  {author} {\bibinfo {author} {\bibfnamefont {A.}~\bibnamefont
  {{Boyarsky}}}, \bibinfo {author} {\bibfnamefont {J.}~\bibnamefont
  {{Franse}}}, \bibinfo {author} {\bibfnamefont {D.}~\bibnamefont
  {{Iakubovskyi}}}, \ and\ \bibinfo {author} {\bibfnamefont {O.}~\bibnamefont
  {{Ruchayskiy}}},\ }\href@noop {} {\bibfield  {journal} {\bibinfo  {journal}
  {ArXiv e-prints}\ } (\bibinfo {year} {2014}{\natexlab{b}})},\ \Eprint
  {http://arxiv.org/abs/1408.2503} {arXiv:1408.2503} \BibitemShut {NoStop}%
\bibitem [{\citenamefont {{Abazajian}}(2014{\natexlab{b}})}]{Abazajian:2014ys}%
  \BibitemOpen
  \bibfield  {author} {\bibinfo {author} {\bibfnamefont {K.~N.}\ \bibnamefont
  {{Abazajian}}},\ }\href {\doibase 10.1103/Physics.7.128} {\bibfield
  {journal} {\bibinfo  {journal} {Physics Online Journal}\ }\textbf {\bibinfo
  {volume} {7}},\ \bibinfo {eid} {128} (\bibinfo {year}
  {2014}{\natexlab{b}})}\BibitemShut {NoStop}%
\bibitem [{\citenamefont {{Malyshev}}\ \emph {et~al.}(2014)\citenamefont
  {{Malyshev}}, \citenamefont {{Neronov}},\ and\ \citenamefont
  {{Eckert}}}]{Malyshev:2014fr}%
  \BibitemOpen
  \bibfield  {author} {\bibinfo {author} {\bibfnamefont {D.}~\bibnamefont
  {{Malyshev}}}, \bibinfo {author} {\bibfnamefont {A.}~\bibnamefont
  {{Neronov}}}, \ and\ \bibinfo {author} {\bibfnamefont {D.}~\bibnamefont
  {{Eckert}}},\ }\href {\doibase 10.1103/PhysRevD.90.103506} {\bibfield
  {journal} {\bibinfo  {journal} {\prd}\ }\textbf {\bibinfo {volume} {90}},\
  \bibinfo {eid} {103506} (\bibinfo {year} {2014})},\ \Eprint
  {http://arxiv.org/abs/1408.3531} {arXiv:1408.3531 [astro-ph.HE]} \BibitemShut
  {NoStop}%
\bibitem [{\citenamefont {{Lovell}}\ \emph {et~al.}(2014)\citenamefont
  {{Lovell}}, \citenamefont {{Bertone}}, \citenamefont {{Boyarsky}},
  \citenamefont {{Jenkins}},\ and\ \citenamefont
  {{Ruchayskiy}}}]{Lovell:2014zr}%
  \BibitemOpen
  \bibfield  {author} {\bibinfo {author} {\bibfnamefont {M.~R.}\ \bibnamefont
  {{Lovell}}}, \bibinfo {author} {\bibfnamefont {G.}~\bibnamefont {{Bertone}}},
  \bibinfo {author} {\bibfnamefont {A.}~\bibnamefont {{Boyarsky}}}, \bibinfo
  {author} {\bibfnamefont {A.}~\bibnamefont {{Jenkins}}}, \ and\ \bibinfo
  {author} {\bibfnamefont {O.}~\bibnamefont {{Ruchayskiy}}},\ }\href@noop {}
  {\bibfield  {journal} {\bibinfo  {journal} {ArXiv e-prints}\ } (\bibinfo
  {year} {2014})},\ \Eprint {http://arxiv.org/abs/1411.0311} {arXiv:1411.0311}
  \BibitemShut {NoStop}%
\bibitem [{\citenamefont {{Anderson}}\ \emph {et~al.}(2014)\citenamefont
  {{Anderson}}, \citenamefont {{Churazov}},\ and\ \citenamefont
  {{Bregman}}}]{Anderson:2014mz}%
  \BibitemOpen
  \bibfield  {author} {\bibinfo {author} {\bibfnamefont {M.~E.}\ \bibnamefont
  {{Anderson}}}, \bibinfo {author} {\bibfnamefont {E.}~\bibnamefont
  {{Churazov}}}, \ and\ \bibinfo {author} {\bibfnamefont {J.~N.}\ \bibnamefont
  {{Bregman}}},\ }\href@noop {} {\bibfield  {journal} {\bibinfo  {journal}
  {ArXiv e-prints}\ } (\bibinfo {year} {2014})},\ \Eprint
  {http://arxiv.org/abs/1408.4115} {arXiv:1408.4115 [astro-ph.HE]} \BibitemShut
  {NoStop}%
\bibitem [{\citenamefont {{Jeltema}}\ and\ \citenamefont
  {{Profumo}}(2014{\natexlab{a}})}]{Jeltema:2014ly}%
  \BibitemOpen
  \bibfield  {author} {\bibinfo {author} {\bibfnamefont {T.~E.}\ \bibnamefont
  {{Jeltema}}}\ and\ \bibinfo {author} {\bibfnamefont {S.}~\bibnamefont
  {{Profumo}}},\ }\href@noop {} {\bibfield  {journal} {\bibinfo  {journal}
  {ArXiv e-prints}\ } (\bibinfo {year} {2014}{\natexlab{a}})},\ \Eprint
  {http://arxiv.org/abs/1408.1699} {arXiv:1408.1699 [astro-ph.HE]} \BibitemShut
  {NoStop}%
\bibitem [{\citenamefont {{Riemer-Sorensen}}(2014)}]{Riemer-Sorensen:2014fp}%
  \BibitemOpen
  \bibfield  {author} {\bibinfo {author} {\bibfnamefont {S.}~\bibnamefont
  {{Riemer-Sorensen}}},\ }\href@noop {} {\bibfield  {journal} {\bibinfo
  {journal} {ArXiv e-prints}\ } (\bibinfo {year} {2014})},\ \Eprint
  {http://arxiv.org/abs/1405.7943} {arXiv:1405.7943} \BibitemShut {NoStop}%
\bibitem [{\citenamefont {{Urban}}\ \emph {et~al.}(2014)\citenamefont
  {{Urban}}, \citenamefont {{Werner}}, \citenamefont {{Allen}}, \citenamefont
  {{Simionescu}}, \citenamefont {{Kaastra}},\ and\ \citenamefont
  {{Strigari}}}]{Urban:2014db}%
  \BibitemOpen
  \bibfield  {author} {\bibinfo {author} {\bibfnamefont {O.}~\bibnamefont
  {{Urban}}}, \bibinfo {author} {\bibfnamefont {N.}~\bibnamefont {{Werner}}},
  \bibinfo {author} {\bibfnamefont {S.~W.}\ \bibnamefont {{Allen}}}, \bibinfo
  {author} {\bibfnamefont {A.}~\bibnamefont {{Simionescu}}}, \bibinfo {author}
  {\bibfnamefont {J.~S.}\ \bibnamefont {{Kaastra}}}, \ and\ \bibinfo {author}
  {\bibfnamefont {L.~E.}\ \bibnamefont {{Strigari}}},\ }\href@noop {}
  {\bibfield  {journal} {\bibinfo  {journal} {ArXiv e-prints}\ } (\bibinfo
  {year} {2014})},\ \Eprint {http://arxiv.org/abs/1411.0050} {arXiv:1411.0050}
  \BibitemShut {NoStop}%
\bibitem [{\citenamefont {{Bulbul}}\ \emph
  {et~al.}(2014{\natexlab{b}})\citenamefont {{Bulbul}}, \citenamefont
  {{Markevitch}}, \citenamefont {{Foster}}, \citenamefont {{Smith}},
  \citenamefont {{Loewenstein}},\ and\ \citenamefont
  {{Randall}}}]{Bulbul:2014xe}%
  \BibitemOpen
  \bibfield  {author} {\bibinfo {author} {\bibfnamefont {E.}~\bibnamefont
  {{Bulbul}}}, \bibinfo {author} {\bibfnamefont {M.}~\bibnamefont
  {{Markevitch}}}, \bibinfo {author} {\bibfnamefont {A.~R.}\ \bibnamefont
  {{Foster}}}, \bibinfo {author} {\bibfnamefont {R.~K.}\ \bibnamefont
  {{Smith}}}, \bibinfo {author} {\bibfnamefont {M.}~\bibnamefont
  {{Loewenstein}}}, \ and\ \bibinfo {author} {\bibfnamefont {S.~W.}\
  \bibnamefont {{Randall}}},\ }\href@noop {} {\bibfield  {journal} {\bibinfo
  {journal} {ArXiv e-prints}\ } (\bibinfo {year} {2014}{\natexlab{b}})},\
  \Eprint {http://arxiv.org/abs/1409.4143} {arXiv:1409.4143 [astro-ph.HE]}
  \BibitemShut {NoStop}%
\bibitem [{\citenamefont {{Boyarsky}}\ \emph
  {et~al.}(2014{\natexlab{c}})\citenamefont {{Boyarsky}}, \citenamefont
  {{Franse}}, \citenamefont {{Iakubovskyi}},\ and\ \citenamefont
  {{Ruchayskiy}}}]{Boyarsky:2014uo}%
  \BibitemOpen
  \bibfield  {author} {\bibinfo {author} {\bibfnamefont {A.}~\bibnamefont
  {{Boyarsky}}}, \bibinfo {author} {\bibfnamefont {J.}~\bibnamefont
  {{Franse}}}, \bibinfo {author} {\bibfnamefont {D.}~\bibnamefont
  {{Iakubovskyi}}}, \ and\ \bibinfo {author} {\bibfnamefont {O.}~\bibnamefont
  {{Ruchayskiy}}},\ }\href@noop {} {\bibfield  {journal} {\bibinfo  {journal}
  {ArXiv e-prints}\ } (\bibinfo {year} {2014}{\natexlab{c}})},\ \Eprint
  {http://arxiv.org/abs/1408.4388} {arXiv:1408.4388} \BibitemShut {NoStop}%
\bibitem [{\citenamefont {{Jeltema}}\ and\ \citenamefont
  {{Profumo}}(2014{\natexlab{b}})}]{Jeltema:2014qq}%
  \BibitemOpen
  \bibfield  {author} {\bibinfo {author} {\bibfnamefont {T.}~\bibnamefont
  {{Jeltema}}}\ and\ \bibinfo {author} {\bibfnamefont {S.}~\bibnamefont
  {{Profumo}}},\ }\href@noop {} {\bibfield  {journal} {\bibinfo  {journal}
  {ArXiv e-prints}\ } (\bibinfo {year} {2014}{\natexlab{b}})},\ \Eprint
  {http://arxiv.org/abs/1411.1759} {arXiv:1411.1759 [astro-ph.HE]} \BibitemShut
  {NoStop}%
\bibitem [{\citenamefont {{Iakubovskyi}}(2014)}]{Iakubovskyi:2014ek}%
  \BibitemOpen
  \bibfield  {author} {\bibinfo {author} {\bibfnamefont {D.~A.}\ \bibnamefont
  {{Iakubovskyi}}},\ }\href@noop {} {\bibfield  {journal} {\bibinfo  {journal}
  {Advances in Astronomy and Space Physics}\ }\textbf {\bibinfo {volume} {4}},\
  \bibinfo {pages} {9} (\bibinfo {year} {2014})},\ \Eprint
  {http://arxiv.org/abs/1410.2852} {arXiv:1410.2852 [astro-ph.HE]} \BibitemShut
  {NoStop}%
\bibitem [{\citenamefont {{Carlson}}\ \emph {et~al.}(2015)\citenamefont
  {{Carlson}}, \citenamefont {{Jeltema}},\ and\ \citenamefont
  {{Profumo}}}]{Carlson:2015gf}%
  \BibitemOpen
  \bibfield  {author} {\bibinfo {author} {\bibfnamefont {E.}~\bibnamefont
  {{Carlson}}}, \bibinfo {author} {\bibfnamefont {T.}~\bibnamefont
  {{Jeltema}}}, \ and\ \bibinfo {author} {\bibfnamefont {S.}~\bibnamefont
  {{Profumo}}},\ }\href {\doibase 10.1088/1475-7516/2015/02/009} {\bibfield
  {journal} {\bibinfo  {journal} {Journal of Cosmology and Astroparticle
  Physics}\ }\textbf {\bibinfo {volume} {2}},\ \bibinfo {eid} {009} (\bibinfo
  {year} {2015})},\ \Eprint {http://arxiv.org/abs/1411.1758} {arXiv:1411.1758
  [astro-ph.HE]} \BibitemShut {NoStop}%
\bibitem [{\citenamefont {{Tamura}}\ \emph {et~al.}(2015)\citenamefont
  {{Tamura}}, \citenamefont {{Iizuka}}, \citenamefont {{Maeda}}, \citenamefont
  {{Mitsuda}},\ and\ \citenamefont {{Yamasaki}}}]{Tamura:2015hl}%
  \BibitemOpen
  \bibfield  {author} {\bibinfo {author} {\bibfnamefont {T.}~\bibnamefont
  {{Tamura}}}, \bibinfo {author} {\bibfnamefont {R.}~\bibnamefont {{Iizuka}}},
  \bibinfo {author} {\bibfnamefont {Y.}~\bibnamefont {{Maeda}}}, \bibinfo
  {author} {\bibfnamefont {K.}~\bibnamefont {{Mitsuda}}}, \ and\ \bibinfo
  {author} {\bibfnamefont {N.~Y.}\ \bibnamefont {{Yamasaki}}},\ }\href
  {\doibase 10.1093/pasj/psu156} {\bibfield  {journal} {\bibinfo  {journal}
  {Publications of the Astronomical Society of Japan}\ }\textbf {\bibinfo
  {volume} {67}},\ \bibinfo {eid} {23} (\bibinfo {year} {2015})},\ \Eprint
  {http://arxiv.org/abs/1412.1869} {arXiv:1412.1869 [astro-ph.HE]} \BibitemShut
  {NoStop}%
\bibitem [{\citenamefont {{Boyarsky}}\ \emph
  {et~al.}(2007{\natexlab{b}})\citenamefont {{Boyarsky}}, \citenamefont {{den
  Herder}}, \citenamefont {{Neronov}},\ and\ \citenamefont
  {{Ruchayskiy}}}]{Boyarsky:2007lr}%
  \BibitemOpen
  \bibfield  {author} {\bibinfo {author} {\bibfnamefont {A.}~\bibnamefont
  {{Boyarsky}}}, \bibinfo {author} {\bibfnamefont {J.-W.}\ \bibnamefont {{den
  Herder}}}, \bibinfo {author} {\bibfnamefont {A.}~\bibnamefont {{Neronov}}}, \
  and\ \bibinfo {author} {\bibfnamefont {O.}~\bibnamefont {{Ruchayskiy}}},\
  }\href {\doibase 10.1016/j.astropartphys.2007.06.003} {\bibfield  {journal}
  {\bibinfo  {journal} {Astroparticle Physics}\ }\textbf {\bibinfo {volume}
  {28}},\ \bibinfo {pages} {303} (\bibinfo {year} {2007}{\natexlab{b}})},\
  \Eprint {http://arxiv.org/abs/astro-ph/0612219} {astro-ph/0612219}
  \BibitemShut {NoStop}%
\bibitem [{\citenamefont {{Figueroa-Feliciano}}\ \emph
  {et~al.}(2015)\citenamefont {{Figueroa-Feliciano}}, \citenamefont
  {{Anderson}}, \citenamefont {{Castro}}, \citenamefont {{Goldfinger}},
  \citenamefont {{Rutherford}}, \citenamefont {{Eckart}}, \citenamefont
  {{Kelley}}, \citenamefont {{Kilbourne}}, \citenamefont {{McCammon}},
  \citenamefont {{Morgan}}, \citenamefont {{Porter}},\ and\ \citenamefont
  {{Szymkowiak}}}]{Figueroa-Feliciano:2015fk}%
  \BibitemOpen
  \bibfield  {author} {\bibinfo {author} {\bibfnamefont {E.}~\bibnamefont
  {{Figueroa-Feliciano}}}, \bibinfo {author} {\bibfnamefont {A.~J.}\
  \bibnamefont {{Anderson}}}, \bibinfo {author} {\bibfnamefont
  {D.}~\bibnamefont {{Castro}}}, \bibinfo {author} {\bibfnamefont {D.~C.}\
  \bibnamefont {{Goldfinger}}}, \bibinfo {author} {\bibfnamefont
  {J.}~\bibnamefont {{Rutherford}}}, \bibinfo {author} {\bibfnamefont {M.~E.}\
  \bibnamefont {{Eckart}}}, \bibinfo {author} {\bibfnamefont {R.~L.}\
  \bibnamefont {{Kelley}}}, \bibinfo {author} {\bibfnamefont {C.~A.}\
  \bibnamefont {{Kilbourne}}}, \bibinfo {author} {\bibfnamefont
  {D.}~\bibnamefont {{McCammon}}}, \bibinfo {author} {\bibfnamefont
  {K.}~\bibnamefont {{Morgan}}}, \bibinfo {author} {\bibfnamefont {F.~S.}\
  \bibnamefont {{Porter}}}, \ and\ \bibinfo {author} {\bibfnamefont {A.~E.}\
  \bibnamefont {{Szymkowiak}}},\ }\href@noop {} {\bibfield  {journal} {\bibinfo
   {journal} {ArXiv e-prints}\ } (\bibinfo {year} {2015})},\ \Eprint
  {http://arxiv.org/abs/1506.05519} {arXiv:1506.05519} \BibitemShut {NoStop}%
\bibitem [{\citenamefont {{Speckhard}}\ \emph {et~al.}(2015)\citenamefont
  {{Speckhard}}, \citenamefont {{Ng}}, \citenamefont {{Beacom}},\ and\
  \citenamefont {{Laha}}}]{Speckhard:2015qy}%
  \BibitemOpen
  \bibfield  {author} {\bibinfo {author} {\bibfnamefont {E.~G.}\ \bibnamefont
  {{Speckhard}}}, \bibinfo {author} {\bibfnamefont {K.~C.~Y.}\ \bibnamefont
  {{Ng}}}, \bibinfo {author} {\bibfnamefont {J.~F.}\ \bibnamefont {{Beacom}}},
  \ and\ \bibinfo {author} {\bibfnamefont {R.}~\bibnamefont {{Laha}}},\
  }\href@noop {} {\bibfield  {journal} {\bibinfo  {journal} {ArXiv e-prints}\ }
  (\bibinfo {year} {2015})},\ \Eprint {http://arxiv.org/abs/1507.04744}
  {arXiv:1507.04744} \BibitemShut {NoStop}%
\bibitem [{\citenamefont {{Harrison}}\ \emph {et~al.}(2013)\citenamefont
  {{Harrison}}, \citenamefont {{Craig}}, \citenamefont {{Christensen}},
  \citenamefont {{Hailey}}, \citenamefont {{Zhang}}, \citenamefont {{Boggs}},
  \citenamefont {{Stern}}, \citenamefont {{Cook}}, \citenamefont {{Forster}},
  \citenamefont {{Giommi}}, \citenamefont {{Grefenstette}}, \citenamefont
  {{Kim}}, \citenamefont {{Kitaguchi}}, \citenamefont {{Koglin}}, \citenamefont
  {{Madsen}}, \citenamefont {{Mao}}, \citenamefont {{Miyasaka}}, \citenamefont
  {{Mori}}, \citenamefont {{Perri}}, \citenamefont {{Pivovaroff}},
  \citenamefont {{Puccetti}}, \citenamefont {{Rana}}, \citenamefont
  {{Westergaard}}, \citenamefont {{Willis}}, \citenamefont {{Zoglauer}},
  \citenamefont {{An}}, \citenamefont {{Bachetti}}, \citenamefont
  {{Barri{\`e}re}}, \citenamefont {{Bellm}}, \citenamefont {{Bhalerao}},
  \citenamefont {{Brejnholt}}, \citenamefont {{Fuerst}}, \citenamefont
  {{Liebe}}, \citenamefont {{Markwardt}}, \citenamefont {{Nynka}},
  \citenamefont {{Vogel}}, \citenamefont {{Walton}}, \citenamefont {{Wik}},
  \citenamefont {{Alexander}}, \citenamefont {{Cominsky}}, \citenamefont
  {{Hornschemeier}}, \citenamefont {{Hornstrup}}, \citenamefont {{Kaspi}},
  \citenamefont {{Madejski}}, \citenamefont {{Matt}}, \citenamefont
  {{Molendi}}, \citenamefont {{Smith}}, \citenamefont {{Tomsick}},
  \citenamefont {{Ajello}}, \citenamefont {{Ballantyne}}, \citenamefont
  {{Balokovic}}, \citenamefont {{Barret}}, \citenamefont {{Bauer}},
  \citenamefont {{Blandford}}, \citenamefont {{Brandt}}, \citenamefont
  {{Brenneman}}, \citenamefont {{Chiang}}, \citenamefont {{Chakrabarty}},
  \citenamefont {{Chenevez}}, \citenamefont {{Comastri}}, \citenamefont
  {{Dufour}}, \citenamefont {{Elvis}}, \citenamefont {{Fabian}}, \citenamefont
  {{Farrah}}, \citenamefont {{Fryer}}, \citenamefont {{Gotthelf}},
  \citenamefont {{Grindlay}}, \citenamefont {{Helfand}}, \citenamefont
  {{Krivonos}}, \citenamefont {{Meier}}, \citenamefont {{Miller}},
  \citenamefont {{Natalucci}}, \citenamefont {{Ogle}}, \citenamefont {{Ofek}},
  \citenamefont {{Ptak}}, \citenamefont {{Reynolds}}, \citenamefont {{Rigby}},
  \citenamefont {{Tagliaferri}}, \citenamefont {{Thorsett}}, \citenamefont
  {{Treister}},\ and\ \citenamefont {{Urry}}}]{Harrison:2013fv}%
  \BibitemOpen
  \bibfield  {author} {\bibinfo {author} {\bibfnamefont {F.~A.}\ \bibnamefont
  {{Harrison}}}, \bibinfo {author} {\bibfnamefont {W.~W.}\ \bibnamefont
  {{Craig}}}, \bibinfo {author} {\bibfnamefont {F.~E.}\ \bibnamefont
  {{Christensen}}}, \bibinfo {author} {\bibfnamefont {C.~J.}\ \bibnamefont
  {{Hailey}}}, \bibinfo {author} {\bibfnamefont {W.~W.}\ \bibnamefont
  {{Zhang}}}, \bibinfo {author} {\bibfnamefont {S.~E.}\ \bibnamefont
  {{Boggs}}}, \bibinfo {author} {\bibfnamefont {D.}~\bibnamefont {{Stern}}},
  \bibinfo {author} {\bibfnamefont {W.~R.}\ \bibnamefont {{Cook}}}, \bibinfo
  {author} {\bibfnamefont {K.}~\bibnamefont {{Forster}}}, \bibinfo {author}
  {\bibfnamefont {P.}~\bibnamefont {{Giommi}}}, \bibinfo {author}
  {\bibfnamefont {B.~W.}\ \bibnamefont {{Grefenstette}}}, \bibinfo {author}
  {\bibfnamefont {Y.}~\bibnamefont {{Kim}}}, \bibinfo {author} {\bibfnamefont
  {T.}~\bibnamefont {{Kitaguchi}}}, \bibinfo {author} {\bibfnamefont {J.~E.}\
  \bibnamefont {{Koglin}}}, \bibinfo {author} {\bibfnamefont {K.~K.}\
  \bibnamefont {{Madsen}}}, \bibinfo {author} {\bibfnamefont {P.~H.}\
  \bibnamefont {{Mao}}}, \bibinfo {author} {\bibfnamefont {H.}~\bibnamefont
  {{Miyasaka}}}, \bibinfo {author} {\bibfnamefont {K.}~\bibnamefont {{Mori}}},
  \bibinfo {author} {\bibfnamefont {M.}~\bibnamefont {{Perri}}}, \bibinfo
  {author} {\bibfnamefont {M.~J.}\ \bibnamefont {{Pivovaroff}}}, \bibinfo
  {author} {\bibfnamefont {S.}~\bibnamefont {{Puccetti}}}, \bibinfo {author}
  {\bibfnamefont {V.~R.}\ \bibnamefont {{Rana}}}, \bibinfo {author}
  {\bibfnamefont {N.~J.}\ \bibnamefont {{Westergaard}}}, \bibinfo {author}
  {\bibfnamefont {J.}~\bibnamefont {{Willis}}}, \bibinfo {author}
  {\bibfnamefont {A.}~\bibnamefont {{Zoglauer}}}, \bibinfo {author}
  {\bibfnamefont {H.}~\bibnamefont {{An}}}, \bibinfo {author} {\bibfnamefont
  {M.}~\bibnamefont {{Bachetti}}}, \bibinfo {author} {\bibfnamefont {N.~M.}\
  \bibnamefont {{Barri{\`e}re}}}, \bibinfo {author} {\bibfnamefont {E.~C.}\
  \bibnamefont {{Bellm}}}, \bibinfo {author} {\bibfnamefont {V.}~\bibnamefont
  {{Bhalerao}}}, \bibinfo {author} {\bibfnamefont {N.~F.}\ \bibnamefont
  {{Brejnholt}}}, \bibinfo {author} {\bibfnamefont {F.}~\bibnamefont
  {{Fuerst}}}, \bibinfo {author} {\bibfnamefont {C.~C.}\ \bibnamefont
  {{Liebe}}}, \bibinfo {author} {\bibfnamefont {C.~B.}\ \bibnamefont
  {{Markwardt}}}, \bibinfo {author} {\bibfnamefont {M.}~\bibnamefont
  {{Nynka}}}, \bibinfo {author} {\bibfnamefont {J.~K.}\ \bibnamefont
  {{Vogel}}}, \bibinfo {author} {\bibfnamefont {D.~J.}\ \bibnamefont
  {{Walton}}}, \bibinfo {author} {\bibfnamefont {D.~R.}\ \bibnamefont {{Wik}}},
  \bibinfo {author} {\bibfnamefont {D.~M.}\ \bibnamefont {{Alexander}}},
  \bibinfo {author} {\bibfnamefont {L.~R.}\ \bibnamefont {{Cominsky}}},
  \bibinfo {author} {\bibfnamefont {A.~E.}\ \bibnamefont {{Hornschemeier}}},
  \bibinfo {author} {\bibfnamefont {A.}~\bibnamefont {{Hornstrup}}}, \bibinfo
  {author} {\bibfnamefont {V.~M.}\ \bibnamefont {{Kaspi}}}, \bibinfo {author}
  {\bibfnamefont {G.~M.}\ \bibnamefont {{Madejski}}}, \bibinfo {author}
  {\bibfnamefont {G.}~\bibnamefont {{Matt}}}, \bibinfo {author} {\bibfnamefont
  {S.}~\bibnamefont {{Molendi}}}, \bibinfo {author} {\bibfnamefont {D.~M.}\
  \bibnamefont {{Smith}}}, \bibinfo {author} {\bibfnamefont {J.~A.}\
  \bibnamefont {{Tomsick}}}, \bibinfo {author} {\bibfnamefont {M.}~\bibnamefont
  {{Ajello}}}, \bibinfo {author} {\bibfnamefont {D.~R.}\ \bibnamefont
  {{Ballantyne}}}, \bibinfo {author} {\bibfnamefont {M.}~\bibnamefont
  {{Balokovic}}}, \bibinfo {author} {\bibfnamefont {D.}~\bibnamefont
  {{Barret}}}, \bibinfo {author} {\bibfnamefont {F.~E.}\ \bibnamefont
  {{Bauer}}}, \bibinfo {author} {\bibfnamefont {R.~D.}\ \bibnamefont
  {{Blandford}}}, \bibinfo {author} {\bibfnamefont {W.~N.}\ \bibnamefont
  {{Brandt}}}, \bibinfo {author} {\bibfnamefont {L.~W.}\ \bibnamefont
  {{Brenneman}}}, \bibinfo {author} {\bibfnamefont {J.}~\bibnamefont
  {{Chiang}}}, \bibinfo {author} {\bibfnamefont {D.}~\bibnamefont
  {{Chakrabarty}}}, \bibinfo {author} {\bibfnamefont {J.}~\bibnamefont
  {{Chenevez}}}, \bibinfo {author} {\bibfnamefont {A.}~\bibnamefont
  {{Comastri}}}, \bibinfo {author} {\bibfnamefont {F.}~\bibnamefont
  {{Dufour}}}, \bibinfo {author} {\bibfnamefont {M.}~\bibnamefont {{Elvis}}},
  \bibinfo {author} {\bibfnamefont {A.~C.}\ \bibnamefont {{Fabian}}}, \bibinfo
  {author} {\bibfnamefont {D.}~\bibnamefont {{Farrah}}}, \bibinfo {author}
  {\bibfnamefont {C.~L.}\ \bibnamefont {{Fryer}}}, \bibinfo {author}
  {\bibfnamefont {E.~V.}\ \bibnamefont {{Gotthelf}}}, \bibinfo {author}
  {\bibfnamefont {J.~E.}\ \bibnamefont {{Grindlay}}}, \bibinfo {author}
  {\bibfnamefont {D.~J.}\ \bibnamefont {{Helfand}}}, \bibinfo {author}
  {\bibfnamefont {R.}~\bibnamefont {{Krivonos}}}, \bibinfo {author}
  {\bibfnamefont {D.~L.}\ \bibnamefont {{Meier}}}, \bibinfo {author}
  {\bibfnamefont {J.~M.}\ \bibnamefont {{Miller}}}, \bibinfo {author}
  {\bibfnamefont {L.}~\bibnamefont {{Natalucci}}}, \bibinfo {author}
  {\bibfnamefont {P.}~\bibnamefont {{Ogle}}}, \bibinfo {author} {\bibfnamefont
  {E.~O.}\ \bibnamefont {{Ofek}}}, \bibinfo {author} {\bibfnamefont
  {A.}~\bibnamefont {{Ptak}}}, \bibinfo {author} {\bibfnamefont {S.~P.}\
  \bibnamefont {{Reynolds}}}, \bibinfo {author} {\bibfnamefont {J.~R.}\
  \bibnamefont {{Rigby}}}, \bibinfo {author} {\bibfnamefont {G.}~\bibnamefont
  {{Tagliaferri}}}, \bibinfo {author} {\bibfnamefont {S.~E.}\ \bibnamefont
  {{Thorsett}}}, \bibinfo {author} {\bibfnamefont {E.}~\bibnamefont
  {{Treister}}}, \ and\ \bibinfo {author} {\bibfnamefont {C.~M.}\ \bibnamefont
  {{Urry}}},\ }\href {\doibase 10.1088/0004-637X/770/2/103} {\bibfield
  {journal} {\bibinfo  {journal} {\apj}\ }\textbf {\bibinfo {volume} {770}},\
  \bibinfo {eid} {103} (\bibinfo {year} {2013})},\ \Eprint
  {http://arxiv.org/abs/1301.7307} {arXiv:1301.7307 [astro-ph.IM]} \BibitemShut
  {NoStop}%
\bibitem [{\citenamefont {{Abazajian}}(2006{\natexlab{b}})}]{Abazajian:2006yg}%
  \BibitemOpen
  \bibfield  {author} {\bibinfo {author} {\bibfnamefont {K.}~\bibnamefont
  {{Abazajian}}},\ }\href {\doibase 10.1103/PhysRevD.73.063513} {\bibfield
  {journal} {\bibinfo  {journal} {\prd}\ }\textbf {\bibinfo {volume} {73}},\
  \bibinfo {eid} {063513} (\bibinfo {year} {2006}{\natexlab{b}})},\ \Eprint
  {http://arxiv.org/abs/astro-ph/0512631} {astro-ph/0512631} \BibitemShut
  {NoStop}%
\bibitem [{\citenamefont {{Abazajian}}\ and\ \citenamefont
  {{Koushiappas}}(2006)}]{Abazajian:2006yz}%
  \BibitemOpen
  \bibfield  {author} {\bibinfo {author} {\bibfnamefont {K.}~\bibnamefont
  {{Abazajian}}}\ and\ \bibinfo {author} {\bibfnamefont {S.~M.}\ \bibnamefont
  {{Koushiappas}}},\ }\href {\doibase 10.1103/PhysRevD.74.023527} {\bibfield
  {journal} {\bibinfo  {journal} {\prd}\ }\textbf {\bibinfo {volume} {74}},\
  \bibinfo {eid} {023527} (\bibinfo {year} {2006})},\ \Eprint
  {http://arxiv.org/abs/astro-ph/0605271} {astro-ph/0605271} \BibitemShut
  {NoStop}%
\bibitem [{\citenamefont {{Boyarsky}}\ \emph
  {et~al.}(2009{\natexlab{a}})\citenamefont {{Boyarsky}}, \citenamefont
  {{Lesgourgues}}, \citenamefont {{Ruchayskiy}},\ and\ \citenamefont
  {{Viel}}}]{Boyarsky:2009fk}%
  \BibitemOpen
  \bibfield  {author} {\bibinfo {author} {\bibfnamefont {A.}~\bibnamefont
  {{Boyarsky}}}, \bibinfo {author} {\bibfnamefont {J.}~\bibnamefont
  {{Lesgourgues}}}, \bibinfo {author} {\bibfnamefont {O.}~\bibnamefont
  {{Ruchayskiy}}}, \ and\ \bibinfo {author} {\bibfnamefont {M.}~\bibnamefont
  {{Viel}}},\ }\href {\doibase 10.1088/1475-7516/2009/05/012} {\bibfield
  {journal} {\bibinfo  {journal} {Journal of Cosmology and Astroparticle
  Physics}\ }\textbf {\bibinfo {volume} {5}},\ \bibinfo {eid} {012} (\bibinfo
  {year} {2009}{\natexlab{a}})},\ \Eprint {http://arxiv.org/abs/0812.0010}
  {arXiv:0812.0010} \BibitemShut {NoStop}%
\bibitem [{\citenamefont {{Boyarsky}}\ \emph
  {et~al.}(2009{\natexlab{b}})\citenamefont {{Boyarsky}}, \citenamefont
  {{Lesgourgues}}, \citenamefont {{Ruchayskiy}},\ and\ \citenamefont
  {{Viel}}}]{Boyarsky:2009jk}%
  \BibitemOpen
  \bibfield  {author} {\bibinfo {author} {\bibfnamefont {A.}~\bibnamefont
  {{Boyarsky}}}, \bibinfo {author} {\bibfnamefont {J.}~\bibnamefont
  {{Lesgourgues}}}, \bibinfo {author} {\bibfnamefont {O.}~\bibnamefont
  {{Ruchayskiy}}}, \ and\ \bibinfo {author} {\bibfnamefont {M.}~\bibnamefont
  {{Viel}}},\ }\href {\doibase 10.1103/PhysRevLett.102.201304} {\bibfield
  {journal} {\bibinfo  {journal} {Physical Review Letters}\ }\textbf {\bibinfo
  {volume} {102}},\ \bibinfo {eid} {201304} (\bibinfo {year}
  {2009}{\natexlab{b}})},\ \Eprint {http://arxiv.org/abs/0812.3256}
  {arXiv:0812.3256 [hep-ph]} \BibitemShut {NoStop}%
\bibitem [{\citenamefont {{Viel}}\ \emph {et~al.}(2005)\citenamefont {{Viel}},
  \citenamefont {{Lesgourgues}}, \citenamefont {{Haehnelt}}, \citenamefont
  {{Matarrese}},\ and\ \citenamefont {{Riotto}}}]{Viel:2005fv}%
  \BibitemOpen
  \bibfield  {author} {\bibinfo {author} {\bibfnamefont {M.}~\bibnamefont
  {{Viel}}}, \bibinfo {author} {\bibfnamefont {J.}~\bibnamefont
  {{Lesgourgues}}}, \bibinfo {author} {\bibfnamefont {M.~G.}\ \bibnamefont
  {{Haehnelt}}}, \bibinfo {author} {\bibfnamefont {S.}~\bibnamefont
  {{Matarrese}}}, \ and\ \bibinfo {author} {\bibfnamefont {A.}~\bibnamefont
  {{Riotto}}},\ }\href {\doibase 10.1103/PhysRevD.71.063534} {\bibfield
  {journal} {\bibinfo  {journal} {\prd}\ }\textbf {\bibinfo {volume} {71}},\
  \bibinfo {eid} {063534} (\bibinfo {year} {2005})},\ \Eprint
  {http://arxiv.org/abs/astro-ph/0501562} {astro-ph/0501562} \BibitemShut
  {NoStop}%
\bibitem [{\citenamefont {{Viel}}\ \emph {et~al.}(2013)\citenamefont {{Viel}},
  \citenamefont {{Becker}}, \citenamefont {{Bolton}},\ and\ \citenamefont
  {{Haehnelt}}}]{Viel:2013eu}%
  \BibitemOpen
  \bibfield  {author} {\bibinfo {author} {\bibfnamefont {M.}~\bibnamefont
  {{Viel}}}, \bibinfo {author} {\bibfnamefont {G.~D.}\ \bibnamefont
  {{Becker}}}, \bibinfo {author} {\bibfnamefont {J.~S.}\ \bibnamefont
  {{Bolton}}}, \ and\ \bibinfo {author} {\bibfnamefont {M.~G.}\ \bibnamefont
  {{Haehnelt}}},\ }\href {\doibase 10.1103/PhysRevD.88.043502} {\bibfield
  {journal} {\bibinfo  {journal} {\prd}\ }\textbf {\bibinfo {volume} {88}},\
  \bibinfo {eid} {043502} (\bibinfo {year} {2013})},\ \Eprint
  {http://arxiv.org/abs/1306.2314} {arXiv:1306.2314 [astro-ph.CO]} \BibitemShut
  {NoStop}%
\bibitem [{\citenamefont {{Bode}}\ \emph {et~al.}(2001)\citenamefont {{Bode}},
  \citenamefont {{Ostriker}},\ and\ \citenamefont {{Turok}}}]{Bode:2001zl}%
  \BibitemOpen
  \bibfield  {author} {\bibinfo {author} {\bibfnamefont {P.}~\bibnamefont
  {{Bode}}}, \bibinfo {author} {\bibfnamefont {J.~P.}\ \bibnamefont
  {{Ostriker}}}, \ and\ \bibinfo {author} {\bibfnamefont {N.}~\bibnamefont
  {{Turok}}},\ }\href {\doibase 10.1086/321541} {\bibfield  {journal} {\bibinfo
   {journal} {\apj}\ }\textbf {\bibinfo {volume} {556}},\ \bibinfo {pages} {93}
  (\bibinfo {year} {2001})},\ \Eprint {http://arxiv.org/abs/astro-ph/0010389}
  {astro-ph/0010389} \BibitemShut {NoStop}%
\bibitem [{\citenamefont {{Schneider}}\ \emph {et~al.}(2012)\citenamefont
  {{Schneider}}, \citenamefont {{Smith}}, \citenamefont {{Macci{\`o}}},\ and\
  \citenamefont {{Moore}}}]{Schneider:2012ul}%
  \BibitemOpen
  \bibfield  {author} {\bibinfo {author} {\bibfnamefont {A.}~\bibnamefont
  {{Schneider}}}, \bibinfo {author} {\bibfnamefont {R.~E.}\ \bibnamefont
  {{Smith}}}, \bibinfo {author} {\bibfnamefont {A.~V.}\ \bibnamefont
  {{Macci{\`o}}}}, \ and\ \bibinfo {author} {\bibfnamefont {B.}~\bibnamefont
  {{Moore}}},\ }\href {\doibase 10.1111/j.1365-2966.2012.21252.x} {\bibfield
  {journal} {\bibinfo  {journal} {Mon. Not. Royal Astron. Soc.}\ }\textbf
  {\bibinfo {volume} {424}},\ \bibinfo {pages} {684} (\bibinfo {year}
  {2012})},\ \Eprint {http://arxiv.org/abs/1112.0330} {arXiv:1112.0330
  [astro-ph.CO]} \BibitemShut {NoStop}%
\bibitem [{\citenamefont {{Strigari}}\ \emph {et~al.}(2006)\citenamefont
  {{Strigari}}, \citenamefont {{Bullock}}, \citenamefont {{Kaplinghat}},
  \citenamefont {{Kravtsov}}, \citenamefont {{Gnedin}}, \citenamefont
  {{Abazajian}},\ and\ \citenamefont {{Klypin}}}]{Strigari:2006rm}%
  \BibitemOpen
  \bibfield  {author} {\bibinfo {author} {\bibfnamefont {L.~E.}\ \bibnamefont
  {{Strigari}}}, \bibinfo {author} {\bibfnamefont {J.~S.}\ \bibnamefont
  {{Bullock}}}, \bibinfo {author} {\bibfnamefont {M.}~\bibnamefont
  {{Kaplinghat}}}, \bibinfo {author} {\bibfnamefont {A.~V.}\ \bibnamefont
  {{Kravtsov}}}, \bibinfo {author} {\bibfnamefont {O.~Y.}\ \bibnamefont
  {{Gnedin}}}, \bibinfo {author} {\bibfnamefont {K.}~\bibnamefont
  {{Abazajian}}}, \ and\ \bibinfo {author} {\bibfnamefont {A.~A.}\ \bibnamefont
  {{Klypin}}},\ }\href {\doibase 10.1086/506381} {\bibfield  {journal}
  {\bibinfo  {journal} {\apj}\ }\textbf {\bibinfo {volume} {652}},\ \bibinfo
  {pages} {306} (\bibinfo {year} {2006})},\ \Eprint
  {http://arxiv.org/abs/astro-ph/0603775} {astro-ph/0603775} \BibitemShut
  {NoStop}%
\bibitem [{\citenamefont {{Boylan-Kolchin}}\ \emph {et~al.}(2012)\citenamefont
  {{Boylan-Kolchin}}, \citenamefont {{Bullock}},\ and\ \citenamefont
  {{Kaplinghat}}}]{Boylan-Kolchin:2012nr}%
  \BibitemOpen
  \bibfield  {author} {\bibinfo {author} {\bibfnamefont {M.}~\bibnamefont
  {{Boylan-Kolchin}}}, \bibinfo {author} {\bibfnamefont {J.~S.}\ \bibnamefont
  {{Bullock}}}, \ and\ \bibinfo {author} {\bibfnamefont {M.}~\bibnamefont
  {{Kaplinghat}}},\ }\href {\doibase 10.1111/j.1365-2966.2012.20695.x}
  {\bibfield  {journal} {\bibinfo  {journal} {Mon. Not. Royal Astron. Soc.}\
  }\textbf {\bibinfo {volume} {422}},\ \bibinfo {pages} {1203} (\bibinfo {year}
  {2012})},\ \Eprint {http://arxiv.org/abs/1111.2048} {arXiv:1111.2048
  [astro-ph.CO]} \BibitemShut {NoStop}%
\bibitem [{\citenamefont {{Evoli}}\ \emph {et~al.}(2014)\citenamefont
  {{Evoli}}, \citenamefont {{Mesinger}},\ and\ \citenamefont
  {{Ferrara}}}]{Evoli:2014lr}%
  \BibitemOpen
  \bibfield  {author} {\bibinfo {author} {\bibfnamefont {C.}~\bibnamefont
  {{Evoli}}}, \bibinfo {author} {\bibfnamefont {A.}~\bibnamefont {{Mesinger}}},
  \ and\ \bibinfo {author} {\bibfnamefont {A.}~\bibnamefont {{Ferrara}}},\
  }\href {\doibase 10.1088/1475-7516/2014/11/024} {\bibfield  {journal}
  {\bibinfo  {journal} {Journal of Cosmology and Astroparticle Physics}\
  }\textbf {\bibinfo {volume} {11}},\ \bibinfo {eid} {024} (\bibinfo {year}
  {2014})},\ \Eprint {http://arxiv.org/abs/1408.1109} {arXiv:1408.1109
  [astro-ph.HE]} \BibitemShut {NoStop}%
\bibitem [{\citenamefont {{Sekiguchi}}\ and\ \citenamefont
  {{Tashiro}}(2014)}]{Sekiguchi:2014qf}%
  \BibitemOpen
  \bibfield  {author} {\bibinfo {author} {\bibfnamefont {T.}~\bibnamefont
  {{Sekiguchi}}}\ and\ \bibinfo {author} {\bibfnamefont {H.}~\bibnamefont
  {{Tashiro}}},\ }\href {\doibase 10.1088/1475-7516/2014/08/007} {\bibfield
  {journal} {\bibinfo  {journal} {Journal of Cosmology and Astroparticle
  Physics}\ }\textbf {\bibinfo {volume} {8}},\ \bibinfo {eid} {007} (\bibinfo
  {year} {2014})},\ \Eprint {http://arxiv.org/abs/1401.5563} {arXiv:1401.5563}
  \BibitemShut {NoStop}%
\bibitem [{\citenamefont {{Sitwell}}\ \emph {et~al.}(2014)\citenamefont
  {{Sitwell}}, \citenamefont {{Mesinger}}, \citenamefont {{Ma}},\ and\
  \citenamefont {{Sigurdson}}}]{Sitwell:2014vn}%
  \BibitemOpen
  \bibfield  {author} {\bibinfo {author} {\bibfnamefont {M.}~\bibnamefont
  {{Sitwell}}}, \bibinfo {author} {\bibfnamefont {A.}~\bibnamefont
  {{Mesinger}}}, \bibinfo {author} {\bibfnamefont {Y.-Z.}\ \bibnamefont
  {{Ma}}}, \ and\ \bibinfo {author} {\bibfnamefont {K.}~\bibnamefont
  {{Sigurdson}}},\ }\href {\doibase 10.1093/mnras/stt2392} {\bibfield
  {journal} {\bibinfo  {journal} {Mon. Not. Royal Astron. Soc.}\ }\textbf
  {\bibinfo {volume} {438}},\ \bibinfo {pages} {2664} (\bibinfo {year}
  {2014})},\ \Eprint {http://arxiv.org/abs/1310.0029} {arXiv:1310.0029}
  \BibitemShut {NoStop}%
\bibitem [{\citenamefont {{Shimabukuro}}\ \emph {et~al.}(2014)\citenamefont
  {{Shimabukuro}}, \citenamefont {{Ichiki}}, \citenamefont {{Inoue}},\ and\
  \citenamefont {{Yokoyama}}}]{Shimabukuro:2014rm}%
  \BibitemOpen
  \bibfield  {author} {\bibinfo {author} {\bibfnamefont {H.}~\bibnamefont
  {{Shimabukuro}}}, \bibinfo {author} {\bibfnamefont {K.}~\bibnamefont
  {{Ichiki}}}, \bibinfo {author} {\bibfnamefont {S.}~\bibnamefont {{Inoue}}}, \
  and\ \bibinfo {author} {\bibfnamefont {S.}~\bibnamefont {{Yokoyama}}},\
  }\href {\doibase 10.1103/PhysRevD.90.083003} {\bibfield  {journal} {\bibinfo
  {journal} {\prd}\ }\textbf {\bibinfo {volume} {90}},\ \bibinfo {eid} {083003}
  (\bibinfo {year} {2014})},\ \Eprint {http://arxiv.org/abs/1403.1605}
  {arXiv:1403.1605} \BibitemShut {NoStop}%
\bibitem [{\citenamefont {{de Vega}}\ \emph {et~al.}(2013)\citenamefont {{de
  Vega}}, \citenamefont {{Moreno}}, \citenamefont {{Moya de Guerra}},
  \citenamefont {{Ram{\'o}n Medrano}},\ and\ \citenamefont
  {{S{\'a}nchez}}}]{de-Vega:2013tg}%
  \BibitemOpen
  \bibfield  {author} {\bibinfo {author} {\bibfnamefont {H.~J.}\ \bibnamefont
  {{de Vega}}}, \bibinfo {author} {\bibfnamefont {O.}~\bibnamefont {{Moreno}}},
  \bibinfo {author} {\bibfnamefont {E.}~\bibnamefont {{Moya de Guerra}}},
  \bibinfo {author} {\bibfnamefont {M.}~\bibnamefont {{Ram{\'o}n Medrano}}}, \
  and\ \bibinfo {author} {\bibfnamefont {N.~G.}\ \bibnamefont
  {{S{\'a}nchez}}},\ }\href {\doibase 10.1016/j.nuclphysb.2012.08.019}
  {\bibfield  {journal} {\bibinfo  {journal} {Nuclear Physics B}\ }\textbf
  {\bibinfo {volume} {866}},\ \bibinfo {pages} {177} (\bibinfo {year}
  {2013})},\ \Eprint {http://arxiv.org/abs/1109.3452} {arXiv:1109.3452
  [hep-ph]} \BibitemShut {NoStop}%
\bibitem [{\citenamefont {{Barry}}\ \emph {et~al.}(2014)\citenamefont
  {{Barry}}, \citenamefont {{Heeck}},\ and\ \citenamefont
  {{Rodejohann}}}]{Barry:2014ai}%
  \BibitemOpen
  \bibfield  {author} {\bibinfo {author} {\bibfnamefont {J.}~\bibnamefont
  {{Barry}}}, \bibinfo {author} {\bibfnamefont {J.}~\bibnamefont {{Heeck}}}, \
  and\ \bibinfo {author} {\bibfnamefont {W.}~\bibnamefont {{Rodejohann}}},\
  }\href {\doibase 10.1007/JHEP07(2014)081} {\bibfield  {journal} {\bibinfo
  {journal} {Journal of High Energy Physics}\ }\textbf {\bibinfo {volume}
  {7}},\ \bibinfo {eid} {81} (\bibinfo {year} {2014})},\ \Eprint
  {http://arxiv.org/abs/1404.5955} {arXiv:1404.5955 [hep-ph]} \BibitemShut
  {NoStop}%
\bibitem [{\citenamefont {{Mertens}}\ \emph
  {et~al.}(2015{\natexlab{a}})\citenamefont {{Mertens}}, \citenamefont
  {{Lasserre}}, \citenamefont {{Groh}}, \citenamefont {{Drexlin}},
  \citenamefont {{Gl{\"u}ck}}, \citenamefont {{Huber}}, \citenamefont {{Poon}},
  \citenamefont {{Steidl}}, \citenamefont {{Steinbrink}},\ and\ \citenamefont
  {{Weinheimer}}}]{Mertens:2015xe}%
  \BibitemOpen
  \bibfield  {author} {\bibinfo {author} {\bibfnamefont {S.}~\bibnamefont
  {{Mertens}}}, \bibinfo {author} {\bibfnamefont {T.}~\bibnamefont
  {{Lasserre}}}, \bibinfo {author} {\bibfnamefont {S.}~\bibnamefont {{Groh}}},
  \bibinfo {author} {\bibfnamefont {G.}~\bibnamefont {{Drexlin}}}, \bibinfo
  {author} {\bibfnamefont {F.}~\bibnamefont {{Gl{\"u}ck}}}, \bibinfo {author}
  {\bibfnamefont {A.}~\bibnamefont {{Huber}}}, \bibinfo {author} {\bibfnamefont
  {A.~W.~P.}\ \bibnamefont {{Poon}}}, \bibinfo {author} {\bibfnamefont
  {M.}~\bibnamefont {{Steidl}}}, \bibinfo {author} {\bibfnamefont
  {N.}~\bibnamefont {{Steinbrink}}}, \ and\ \bibinfo {author} {\bibfnamefont
  {C.}~\bibnamefont {{Weinheimer}}},\ }\href {\doibase
  10.1088/1475-7516/2015/02/020} {\bibfield  {journal} {\bibinfo  {journal}
  {Journal of Cosmology and Astroparticle Physics}\ }\textbf {\bibinfo {volume}
  {2}},\ \bibinfo {eid} {020} (\bibinfo {year} {2015}{\natexlab{a}})},\ \Eprint
  {http://arxiv.org/abs/1409.0920} {arXiv:1409.0920 [physics.ins-det]}
  \BibitemShut {NoStop}%
\bibitem [{\citenamefont {{Mertens}}\ \emph
  {et~al.}(2015{\natexlab{b}})\citenamefont {{Mertens}}, \citenamefont
  {{Dolde}}, \citenamefont {{Korzeczek}}, \citenamefont {{Glueck}},
  \citenamefont {{Groh}}, \citenamefont {{Martin}}, \citenamefont {{Poon}},\
  and\ \citenamefont {{Steidl}}}]{Mertens:2015uo}%
  \BibitemOpen
  \bibfield  {author} {\bibinfo {author} {\bibfnamefont {S.}~\bibnamefont
  {{Mertens}}}, \bibinfo {author} {\bibfnamefont {K.}~\bibnamefont {{Dolde}}},
  \bibinfo {author} {\bibfnamefont {M.}~\bibnamefont {{Korzeczek}}}, \bibinfo
  {author} {\bibfnamefont {F.}~\bibnamefont {{Glueck}}}, \bibinfo {author}
  {\bibfnamefont {S.}~\bibnamefont {{Groh}}}, \bibinfo {author} {\bibfnamefont
  {R.~D.}\ \bibnamefont {{Martin}}}, \bibinfo {author} {\bibfnamefont
  {A.~W.~P.}\ \bibnamefont {{Poon}}}, \ and\ \bibinfo {author} {\bibfnamefont
  {M.}~\bibnamefont {{Steidl}}},\ }\href {\doibase 10.1103/PhysRevD.91.042005}
  {\bibfield  {journal} {\bibinfo  {journal} {\prd}\ }\textbf {\bibinfo
  {volume} {91}},\ \bibinfo {eid} {042005} (\bibinfo {year}
  {2015}{\natexlab{b}})},\ \Eprint {http://arxiv.org/abs/1410.7684}
  {arXiv:1410.7684 [hep-ph]} \BibitemShut {NoStop}%
\bibitem [{\citenamefont {{Alekhin}}\ \emph {et~al.}(2015)\citenamefont
  {{Alekhin}}, \citenamefont {{Altmannshofer}}, \citenamefont {{Asaka}},
  \citenamefont {{Batell}}, \citenamefont {{Bezrukov}}, \citenamefont
  {{Bondarenko}}, \citenamefont {{Boyarsky}}, \citenamefont {{Craig}},
  \citenamefont {{Choi}}, \citenamefont {{Corral}}, \citenamefont {{Curtin}},
  \citenamefont {{Davidson}}, \citenamefont {{de Gouv{\^e}a}}, \citenamefont
  {{Dell'Oro}}, \citenamefont {{deNiverville}}, \citenamefont {{Bhupal Dev}},
  \citenamefont {{Dreiner}}, \citenamefont {{Drewes}}, \citenamefont
  {{Eijima}}, \citenamefont {{Essig}}, \citenamefont {{Fradette}},
  \citenamefont {{Garbrecht}}, \citenamefont {{Gavela}}, \citenamefont
  {{Giudice}}, \citenamefont {{Gorbunov}}, \citenamefont {{Gori}},
  \citenamefont {{Grojean}}, \citenamefont {{Goodsell}}, \citenamefont
  {{Guffanti}}, \citenamefont {{Hambye}}, \citenamefont {{Hansen}},
  \citenamefont {{Helo}}, \citenamefont {{Hernandez}}, \citenamefont
  {{Ibarra}}, \citenamefont {{Ivashko}}, \citenamefont {{Izaguirre}},
  \citenamefont {{Jaeckel}}, \citenamefont {{Jeong}}, \citenamefont
  {{Kahlhoefer}}, \citenamefont {{Kahn}}, \citenamefont {{Katz}}, \citenamefont
  {{Kim}}, \citenamefont {{Kovalenko}}, \citenamefont {{Krnjaic}},
  \citenamefont {{Lyubovitskij}}, \citenamefont {{Marcocci}}, \citenamefont
  {{Mccullough}}, \citenamefont {{McKeen}}, \citenamefont {{Mitselmakher}},
  \citenamefont {{Moch}}, \citenamefont {{Mohapatra}}, \citenamefont
  {{Morrissey}}, \citenamefont {{Ovchynnikov}}, \citenamefont {{Paschos}},
  \citenamefont {{Pilaftsis}}, \citenamefont {{Pospelov}}, \citenamefont {{Hall
  Reno}}, \citenamefont {{Ringwald}}, \citenamefont {{Ritz}}, \citenamefont
  {{Roszkowski}}, \citenamefont {{Rubakov}}, \citenamefont {{Ruchayskiy}},
  \citenamefont {{Shelton}}, \citenamefont {{Schienbein}}, \citenamefont
  {{Schmeier}}, \citenamefont {{Schmidt-Hoberg}}, \citenamefont {{Schwaller}},
  \citenamefont {{Senjanovic}}, \citenamefont {{Seto}}, \citenamefont
  {{Shaposhnikov}}, \citenamefont {{Shuve}}, \citenamefont {{Shrock}},
  \citenamefont {{Shchutska}}, \citenamefont {{Spannowsky}}, \citenamefont
  {{Spray}}, \citenamefont {{Staub}}, \citenamefont {{Stolarski}},
  \citenamefont {{Strassler}}, \citenamefont {{Tello}}, \citenamefont
  {{Tramontano}}, \citenamefont {{Tripathi}}, \citenamefont {{Tulin}},
  \citenamefont {{Vissani}}, \citenamefont {{Winkler}},\ and\ \citenamefont
  {{Zurek}}}]{Alekhin:2015dq}%
  \BibitemOpen
  \bibfield  {author} {\bibinfo {author} {\bibfnamefont {S.}~\bibnamefont
  {{Alekhin}}}, \bibinfo {author} {\bibfnamefont {W.}~\bibnamefont
  {{Altmannshofer}}}, \bibinfo {author} {\bibfnamefont {T.}~\bibnamefont
  {{Asaka}}}, \bibinfo {author} {\bibfnamefont {B.}~\bibnamefont {{Batell}}},
  \bibinfo {author} {\bibfnamefont {F.}~\bibnamefont {{Bezrukov}}}, \bibinfo
  {author} {\bibfnamefont {K.}~\bibnamefont {{Bondarenko}}}, \bibinfo {author}
  {\bibfnamefont {A.}~\bibnamefont {{Boyarsky}}}, \bibinfo {author}
  {\bibfnamefont {N.}~\bibnamefont {{Craig}}}, \bibinfo {author} {\bibfnamefont
  {K.-Y.}\ \bibnamefont {{Choi}}}, \bibinfo {author} {\bibfnamefont
  {C.}~\bibnamefont {{Corral}}}, \bibinfo {author} {\bibfnamefont
  {D.}~\bibnamefont {{Curtin}}}, \bibinfo {author} {\bibfnamefont
  {S.}~\bibnamefont {{Davidson}}}, \bibinfo {author} {\bibfnamefont
  {A.}~\bibnamefont {{de Gouv{\^e}a}}}, \bibinfo {author} {\bibfnamefont
  {S.}~\bibnamefont {{Dell'Oro}}}, \bibinfo {author} {\bibfnamefont
  {P.}~\bibnamefont {{deNiverville}}}, \bibinfo {author} {\bibfnamefont
  {P.~S.}\ \bibnamefont {{Bhupal Dev}}}, \bibinfo {author} {\bibfnamefont
  {H.}~\bibnamefont {{Dreiner}}}, \bibinfo {author} {\bibfnamefont
  {M.}~\bibnamefont {{Drewes}}}, \bibinfo {author} {\bibfnamefont
  {S.}~\bibnamefont {{Eijima}}}, \bibinfo {author} {\bibfnamefont
  {R.}~\bibnamefont {{Essig}}}, \bibinfo {author} {\bibfnamefont
  {A.}~\bibnamefont {{Fradette}}}, \bibinfo {author} {\bibfnamefont
  {B.}~\bibnamefont {{Garbrecht}}}, \bibinfo {author} {\bibfnamefont
  {B.}~\bibnamefont {{Gavela}}}, \bibinfo {author} {\bibfnamefont {G.~F.}\
  \bibnamefont {{Giudice}}}, \bibinfo {author} {\bibfnamefont {D.}~\bibnamefont
  {{Gorbunov}}}, \bibinfo {author} {\bibfnamefont {S.}~\bibnamefont {{Gori}}},
  \bibinfo {author} {\bibfnamefont {C.}~\bibnamefont {{Grojean}}}, \bibinfo
  {author} {\bibfnamefont {M.~D.}\ \bibnamefont {{Goodsell}}}, \bibinfo
  {author} {\bibfnamefont {A.}~\bibnamefont {{Guffanti}}}, \bibinfo {author}
  {\bibfnamefont {T.}~\bibnamefont {{Hambye}}}, \bibinfo {author}
  {\bibfnamefont {S.~H.}\ \bibnamefont {{Hansen}}}, \bibinfo {author}
  {\bibfnamefont {J.~C.}\ \bibnamefont {{Helo}}}, \bibinfo {author}
  {\bibfnamefont {P.}~\bibnamefont {{Hernandez}}}, \bibinfo {author}
  {\bibfnamefont {A.}~\bibnamefont {{Ibarra}}}, \bibinfo {author}
  {\bibfnamefont {A.}~\bibnamefont {{Ivashko}}}, \bibinfo {author}
  {\bibfnamefont {E.}~\bibnamefont {{Izaguirre}}}, \bibinfo {author}
  {\bibfnamefont {J.}~\bibnamefont {{Jaeckel}}}, \bibinfo {author}
  {\bibfnamefont {Y.~S.}\ \bibnamefont {{Jeong}}}, \bibinfo {author}
  {\bibfnamefont {F.}~\bibnamefont {{Kahlhoefer}}}, \bibinfo {author}
  {\bibfnamefont {Y.}~\bibnamefont {{Kahn}}}, \bibinfo {author} {\bibfnamefont
  {A.}~\bibnamefont {{Katz}}}, \bibinfo {author} {\bibfnamefont {C.~S.}\
  \bibnamefont {{Kim}}}, \bibinfo {author} {\bibfnamefont {S.}~\bibnamefont
  {{Kovalenko}}}, \bibinfo {author} {\bibfnamefont {G.}~\bibnamefont
  {{Krnjaic}}}, \bibinfo {author} {\bibfnamefont {V.~E.}\ \bibnamefont
  {{Lyubovitskij}}}, \bibinfo {author} {\bibfnamefont {S.}~\bibnamefont
  {{Marcocci}}}, \bibinfo {author} {\bibfnamefont {M.}~\bibnamefont
  {{Mccullough}}}, \bibinfo {author} {\bibfnamefont {D.}~\bibnamefont
  {{McKeen}}}, \bibinfo {author} {\bibfnamefont {G.}~\bibnamefont
  {{Mitselmakher}}}, \bibinfo {author} {\bibfnamefont {S.-O.}\ \bibnamefont
  {{Moch}}}, \bibinfo {author} {\bibfnamefont {R.~N.}\ \bibnamefont
  {{Mohapatra}}}, \bibinfo {author} {\bibfnamefont {D.~E.}\ \bibnamefont
  {{Morrissey}}}, \bibinfo {author} {\bibfnamefont {M.}~\bibnamefont
  {{Ovchynnikov}}}, \bibinfo {author} {\bibfnamefont {E.}~\bibnamefont
  {{Paschos}}}, \bibinfo {author} {\bibfnamefont {A.}~\bibnamefont
  {{Pilaftsis}}}, \bibinfo {author} {\bibfnamefont {M.}~\bibnamefont
  {{Pospelov}}}, \bibinfo {author} {\bibfnamefont {M.}~\bibnamefont {{Hall
  Reno}}}, \bibinfo {author} {\bibfnamefont {A.}~\bibnamefont {{Ringwald}}},
  \bibinfo {author} {\bibfnamefont {A.}~\bibnamefont {{Ritz}}}, \bibinfo
  {author} {\bibfnamefont {L.}~\bibnamefont {{Roszkowski}}}, \bibinfo {author}
  {\bibfnamefont {V.}~\bibnamefont {{Rubakov}}}, \bibinfo {author}
  {\bibfnamefont {O.}~\bibnamefont {{Ruchayskiy}}}, \bibinfo {author}
  {\bibfnamefont {J.}~\bibnamefont {{Shelton}}}, \bibinfo {author}
  {\bibfnamefont {I.}~\bibnamefont {{Schienbein}}}, \bibinfo {author}
  {\bibfnamefont {D.}~\bibnamefont {{Schmeier}}}, \bibinfo {author}
  {\bibfnamefont {K.}~\bibnamefont {{Schmidt-Hoberg}}}, \bibinfo {author}
  {\bibfnamefont {P.}~\bibnamefont {{Schwaller}}}, \bibinfo {author}
  {\bibfnamefont {G.}~\bibnamefont {{Senjanovic}}}, \bibinfo {author}
  {\bibfnamefont {O.}~\bibnamefont {{Seto}}}, \bibinfo {author} {\bibfnamefont
  {M.}~\bibnamefont {{Shaposhnikov}}}, \bibinfo {author} {\bibfnamefont
  {B.}~\bibnamefont {{Shuve}}}, \bibinfo {author} {\bibfnamefont
  {R.}~\bibnamefont {{Shrock}}}, \bibinfo {author} {\bibfnamefont
  {L.}~\bibnamefont {{Shchutska}}}, \bibinfo {author} {\bibfnamefont
  {M.}~\bibnamefont {{Spannowsky}}}, \bibinfo {author} {\bibfnamefont
  {A.}~\bibnamefont {{Spray}}}, \bibinfo {author} {\bibfnamefont
  {F.}~\bibnamefont {{Staub}}}, \bibinfo {author} {\bibfnamefont
  {D.}~\bibnamefont {{Stolarski}}}, \bibinfo {author} {\bibfnamefont
  {M.}~\bibnamefont {{Strassler}}}, \bibinfo {author} {\bibfnamefont
  {V.}~\bibnamefont {{Tello}}}, \bibinfo {author} {\bibfnamefont
  {F.}~\bibnamefont {{Tramontano}}}, \bibinfo {author} {\bibfnamefont
  {A.}~\bibnamefont {{Tripathi}}}, \bibinfo {author} {\bibfnamefont
  {S.}~\bibnamefont {{Tulin}}}, \bibinfo {author} {\bibfnamefont
  {F.}~\bibnamefont {{Vissani}}}, \bibinfo {author} {\bibfnamefont {M.~W.}\
  \bibnamefont {{Winkler}}}, \ and\ \bibinfo {author} {\bibfnamefont {K.~M.}\
  \bibnamefont {{Zurek}}},\ }\href@noop {} {\bibfield  {journal} {\bibinfo
  {journal} {ArXiv e-prints}\ } (\bibinfo {year} {2015})},\ \Eprint
  {http://arxiv.org/abs/1504.04855} {arXiv:1504.04855 [hep-ph]} \BibitemShut
  {NoStop}%
\bibitem [{\citenamefont {{Akhmedov}}\ \emph {et~al.}(2013)\citenamefont
  {{Akhmedov}}, \citenamefont {{Kartavtsev}}, \citenamefont {{Lindner}},
  \citenamefont {{Michaels}},\ and\ \citenamefont
  {{Smirnov}}}]{Akhmedov:2013xy}%
  \BibitemOpen
  \bibfield  {author} {\bibinfo {author} {\bibfnamefont {E.}~\bibnamefont
  {{Akhmedov}}}, \bibinfo {author} {\bibfnamefont {A.}~\bibnamefont
  {{Kartavtsev}}}, \bibinfo {author} {\bibfnamefont {M.}~\bibnamefont
  {{Lindner}}}, \bibinfo {author} {\bibfnamefont {L.}~\bibnamefont
  {{Michaels}}}, \ and\ \bibinfo {author} {\bibfnamefont {J.}~\bibnamefont
  {{Smirnov}}},\ }\href {\doibase 10.1007/JHEP05(2013)081} {\bibfield
  {journal} {\bibinfo  {journal} {Journal of High Energy Physics}\ }\textbf
  {\bibinfo {volume} {5}},\ \bibinfo {eid} {81} (\bibinfo {year} {2013})},\
  \Eprint {http://arxiv.org/abs/1302.1872} {arXiv:1302.1872 [hep-ph]}
  \BibitemShut {NoStop}%
\bibitem [{\citenamefont {{Abada}}\ \emph {et~al.}(2015)\citenamefont
  {{Abada}}, \citenamefont {{De Romeri}}, \citenamefont {{Monteil}},
  \citenamefont {{Orloff}},\ and\ \citenamefont {{Teixeira}}}]{Abada:2015lq}%
  \BibitemOpen
  \bibfield  {author} {\bibinfo {author} {\bibfnamefont {A.}~\bibnamefont
  {{Abada}}}, \bibinfo {author} {\bibfnamefont {V.}~\bibnamefont {{De
  Romeri}}}, \bibinfo {author} {\bibfnamefont {S.}~\bibnamefont {{Monteil}}},
  \bibinfo {author} {\bibfnamefont {J.}~\bibnamefont {{Orloff}}}, \ and\
  \bibinfo {author} {\bibfnamefont {A.~M.}\ \bibnamefont {{Teixeira}}},\ }\href
  {\doibase 10.1007/JHEP04(2015)051} {\bibfield  {journal} {\bibinfo  {journal}
  {Journal of High Energy Physics}\ }\textbf {\bibinfo {volume} {4}},\ \bibinfo
  {pages} {51} (\bibinfo {year} {2015})},\ \Eprint
  {http://arxiv.org/abs/1412.6322} {arXiv:1412.6322 [hep-ph]} \BibitemShut
  {NoStop}%
\bibitem [{\citenamefont {{Lopez-Pavon}}\ \emph {et~al.}(2015)\citenamefont
  {{Lopez-Pavon}}, \citenamefont {{Molinaro}},\ and\ \citenamefont
  {{Petcov}}}]{Lopez-Pavon:2015uq}%
  \BibitemOpen
  \bibfield  {author} {\bibinfo {author} {\bibfnamefont {J.}~\bibnamefont
  {{Lopez-Pavon}}}, \bibinfo {author} {\bibfnamefont {E.}~\bibnamefont
  {{Molinaro}}}, \ and\ \bibinfo {author} {\bibfnamefont {S.~T.}\ \bibnamefont
  {{Petcov}}},\ }\href@noop {} {\bibfield  {journal} {\bibinfo  {journal}
  {ArXiv e-prints}\ } (\bibinfo {year} {2015})},\ \Eprint
  {http://arxiv.org/abs/1506.05296} {arXiv:1506.05296 [hep-ph]} \BibitemShut
  {NoStop}%
\bibitem [{\citenamefont {{Kolb}}\ \emph {et~al.}(1996)\citenamefont {{Kolb}},
  \citenamefont {{Mohapatra}},\ and\ \citenamefont {{Teplitz}}}]{Kolb:1996sp}%
  \BibitemOpen
  \bibfield  {author} {\bibinfo {author} {\bibfnamefont {E.~W.}\ \bibnamefont
  {{Kolb}}}, \bibinfo {author} {\bibfnamefont {R.~N.}\ \bibnamefont
  {{Mohapatra}}}, \ and\ \bibinfo {author} {\bibfnamefont {V.~L.}\ \bibnamefont
  {{Teplitz}}},\ }\href {\doibase 10.1103/PhysRevLett.77.3066} {\bibfield
  {journal} {\bibinfo  {journal} {Physical Review Letters}\ }\textbf {\bibinfo
  {volume} {77}},\ \bibinfo {pages} {3066} (\bibinfo {year} {1996})},\ \Eprint
  {http://arxiv.org/abs/hep-ph/9605350} {hep-ph/9605350} \BibitemShut {NoStop}%
\bibitem [{\citenamefont {{Hidaka}}\ and\ \citenamefont
  {{Fuller}}(2006)}]{Hidaka:2006yq}%
  \BibitemOpen
  \bibfield  {author} {\bibinfo {author} {\bibfnamefont {J.}~\bibnamefont
  {{Hidaka}}}\ and\ \bibinfo {author} {\bibfnamefont {G.~M.}\ \bibnamefont
  {{Fuller}}},\ }\href {\doibase 10.1103/PhysRevD.74.125015} {\bibfield
  {journal} {\bibinfo  {journal} {\prd}\ }\textbf {\bibinfo {volume} {74}},\
  \bibinfo {pages} {125015} (\bibinfo {year} {2006})},\ \Eprint
  {http://arxiv.org/abs/arXiv:astro-ph/0609425} {arXiv:astro-ph/0609425}
  \BibitemShut {NoStop}%
\bibitem [{\citenamefont {{Fryer}}\ and\ \citenamefont
  {{Kusenko}}(2006)}]{Fryer:2006rt}%
  \BibitemOpen
  \bibfield  {author} {\bibinfo {author} {\bibfnamefont {C.~L.}\ \bibnamefont
  {{Fryer}}}\ and\ \bibinfo {author} {\bibfnamefont {A.}~\bibnamefont
  {{Kusenko}}},\ }\href {\doibase 10.1086/500933} {\bibfield  {journal}
  {\bibinfo  {journal} {Astrophs. J. Suppl.}\ }\textbf {\bibinfo {volume}
  {163}},\ \bibinfo {pages} {335} (\bibinfo {year} {2006})},\ \Eprint
  {http://arxiv.org/abs/arXiv:astro-ph/0512033} {arXiv:astro-ph/0512033}
  \BibitemShut {NoStop}%
\bibitem [{\citenamefont {{Hidaka}}\ and\ \citenamefont
  {{Fuller}}(2007)}]{Hidaka:2007kx}%
  \BibitemOpen
  \bibfield  {author} {\bibinfo {author} {\bibfnamefont {J.}~\bibnamefont
  {{Hidaka}}}\ and\ \bibinfo {author} {\bibfnamefont {G.~M.}\ \bibnamefont
  {{Fuller}}},\ }\href {\doibase 10.1103/PhysRevD.76.083516} {\bibfield
  {journal} {\bibinfo  {journal} {\prd}\ }\textbf {\bibinfo {volume} {76}},\
  \bibinfo {pages} {083516} (\bibinfo {year} {2007})},\ \Eprint
  {http://arxiv.org/abs/0706.3886} {arXiv:0706.3886} \BibitemShut {NoStop}%
\bibitem [{\citenamefont {{Choubey}}\ \emph {et~al.}(2007)\citenamefont
  {{Choubey}}, \citenamefont {{Harries}},\ and\ \citenamefont
  {{Ross}}}]{Choubey:2007bs}%
  \BibitemOpen
  \bibfield  {author} {\bibinfo {author} {\bibfnamefont {S.}~\bibnamefont
  {{Choubey}}}, \bibinfo {author} {\bibfnamefont {N.~P.}\ \bibnamefont
  {{Harries}}}, \ and\ \bibinfo {author} {\bibfnamefont {G.~G.}\ \bibnamefont
  {{Ross}}},\ }\href {\doibase 10.1103/PhysRevD.76.073013} {\bibfield
  {journal} {\bibinfo  {journal} {\prd}\ }\textbf {\bibinfo {volume} {76}},\
  \bibinfo {eid} {073013} (\bibinfo {year} {2007})},\ \Eprint
  {http://arxiv.org/abs/hep-ph/0703092} {hep-ph/0703092} \BibitemShut {NoStop}%
\bibitem [{\citenamefont {{Fuller}}\ \emph {et~al.}(2009)\citenamefont
  {{Fuller}}, \citenamefont {{Kusenko}},\ and\ \citenamefont
  {{Petraki}}}]{Fuller:2009uq}%
  \BibitemOpen
  \bibfield  {author} {\bibinfo {author} {\bibfnamefont {G.~M.}\ \bibnamefont
  {{Fuller}}}, \bibinfo {author} {\bibfnamefont {A.}~\bibnamefont {{Kusenko}}},
  \ and\ \bibinfo {author} {\bibfnamefont {K.}~\bibnamefont {{Petraki}}},\
  }\href {\doibase 10.1016/j.physletb.2008.11.016} {\bibfield  {journal}
  {\bibinfo  {journal} {Physics Letters B}\ }\textbf {\bibinfo {volume}
  {670}},\ \bibinfo {pages} {281} (\bibinfo {year} {2009})},\ \Eprint
  {http://arxiv.org/abs/0806.4273} {arXiv:0806.4273} \BibitemShut {NoStop}%
\bibitem [{\citenamefont {{Raffelt}}\ and\ \citenamefont
  {{Zhou}}(2011)}]{Raffelt:2011ij}%
  \BibitemOpen
  \bibfield  {author} {\bibinfo {author} {\bibfnamefont {G.~G.}\ \bibnamefont
  {{Raffelt}}}\ and\ \bibinfo {author} {\bibfnamefont {S.}~\bibnamefont
  {{Zhou}}},\ }\href {\doibase 10.1103/PhysRevD.83.093014} {\bibfield
  {journal} {\bibinfo  {journal} {\prd}\ }\textbf {\bibinfo {volume} {83}},\
  \bibinfo {eid} {093014} (\bibinfo {year} {2011})},\ \Eprint
  {http://arxiv.org/abs/1102.5124} {arXiv:1102.5124 [hep-ph]} \BibitemShut
  {NoStop}%
\bibitem [{\citenamefont {{Warren}}\ \emph {et~al.}(2014)\citenamefont
  {{Warren}}, \citenamefont {{Meixner}}, \citenamefont {{Mathews}},
  \citenamefont {{Hidaka}},\ and\ \citenamefont {{Kajino}}}]{Warren:2014dp}%
  \BibitemOpen
  \bibfield  {author} {\bibinfo {author} {\bibfnamefont {M.~L.}\ \bibnamefont
  {{Warren}}}, \bibinfo {author} {\bibfnamefont {M.}~\bibnamefont {{Meixner}}},
  \bibinfo {author} {\bibfnamefont {G.}~\bibnamefont {{Mathews}}}, \bibinfo
  {author} {\bibfnamefont {J.}~\bibnamefont {{Hidaka}}}, \ and\ \bibinfo
  {author} {\bibfnamefont {T.}~\bibnamefont {{Kajino}}},\ }\href {\doibase
  10.1103/PhysRevD.90.103007} {\bibfield  {journal} {\bibinfo  {journal}
  {\prd}\ }\textbf {\bibinfo {volume} {90}},\ \bibinfo {eid} {103007} (\bibinfo
  {year} {2014})},\ \Eprint {http://arxiv.org/abs/1405.6101} {arXiv:1405.6101
  [astro-ph.HE]} \BibitemShut {NoStop}%
\bibitem [{\citenamefont {{Zhou}}(2015)}]{Zhou:2015th}%
  \BibitemOpen
  \bibfield  {author} {\bibinfo {author} {\bibfnamefont {S.}~\bibnamefont
  {{Zhou}}},\ }\href {\doibase 10.1142/S0217751X15300331} {\bibfield  {journal}
  {\bibinfo  {journal} {International Journal of Modern Physics A}\ }\textbf
  {\bibinfo {volume} {30}},\ \bibinfo {eid} {1530033} (\bibinfo {year}
  {2015})},\ \Eprint {http://arxiv.org/abs/1504.02729} {arXiv:1504.02729
  [hep-ph]} \BibitemShut {NoStop}%
\bibitem [{\citenamefont {{Kusenko}}\ and\ \citenamefont
  {{Segr{\`e}}}(1999)}]{Kusenko:1999hc}%
  \BibitemOpen
  \bibfield  {author} {\bibinfo {author} {\bibfnamefont {A.}~\bibnamefont
  {{Kusenko}}}\ and\ \bibinfo {author} {\bibfnamefont {G.}~\bibnamefont
  {{Segr{\`e}}}},\ }\href {\doibase 10.1103/PhysRevD.59.061302} {\bibfield
  {journal} {\bibinfo  {journal} {\prd}\ }\textbf {\bibinfo {volume} {59}},\
  \bibinfo {eid} {061302} (\bibinfo {year} {1999})},\ \Eprint
  {http://arxiv.org/abs/astro-ph/9811144} {astro-ph/9811144} \BibitemShut
  {NoStop}%
\bibitem [{\citenamefont {{Fuller}}\ \emph {et~al.}(2003)\citenamefont
  {{Fuller}}, \citenamefont {{Kusenko}}, \citenamefont {{Mocioiu}},\ and\
  \citenamefont {{Pascoli}}}]{Fuller:2003vn}%
  \BibitemOpen
  \bibfield  {author} {\bibinfo {author} {\bibfnamefont {G.~M.}\ \bibnamefont
  {{Fuller}}}, \bibinfo {author} {\bibfnamefont {A.}~\bibnamefont {{Kusenko}}},
  \bibinfo {author} {\bibfnamefont {I.}~\bibnamefont {{Mocioiu}}}, \ and\
  \bibinfo {author} {\bibfnamefont {S.}~\bibnamefont {{Pascoli}}},\ }\href
  {\doibase 10.1103/PhysRevD.68.103002} {\bibfield  {journal} {\bibinfo
  {journal} {\prd}\ }\textbf {\bibinfo {volume} {68}},\ \bibinfo {pages}
  {103002} (\bibinfo {year} {2003})},\ \Eprint
  {http://arxiv.org/abs/arXiv:astro-ph/0307267} {arXiv:astro-ph/0307267}
  \BibitemShut {NoStop}%
\bibitem [{\citenamefont {{Kusenko}}\ \emph {et~al.}(2008)\citenamefont
  {{Kusenko}}, \citenamefont {{Mandal}},\ and\ \citenamefont
  {{Mukherjee}}}]{Kusenko:2008lr}%
  \BibitemOpen
  \bibfield  {author} {\bibinfo {author} {\bibfnamefont {A.}~\bibnamefont
  {{Kusenko}}}, \bibinfo {author} {\bibfnamefont {B.~P.}\ \bibnamefont
  {{Mandal}}}, \ and\ \bibinfo {author} {\bibfnamefont {A.}~\bibnamefont
  {{Mukherjee}}},\ }\href {\doibase 10.1103/PhysRevD.77.123009} {\bibfield
  {journal} {\bibinfo  {journal} {\prd}\ }\textbf {\bibinfo {volume} {77}},\
  \bibinfo {eid} {123009} (\bibinfo {year} {2008})},\ \Eprint
  {http://arxiv.org/abs/0801.4734} {arXiv:0801.4734} \BibitemShut {NoStop}%
\bibitem [{\citenamefont {{Kishimoto}}(2011)}]{Kishimoto:2011fk}%
  \BibitemOpen
  \bibfield  {author} {\bibinfo {author} {\bibfnamefont {C.~T.}\ \bibnamefont
  {{Kishimoto}}},\ }\href@noop {} {\bibfield  {journal} {\bibinfo  {journal}
  {ArXiv e-prints}\ } (\bibinfo {year} {2011})},\ \Eprint
  {http://arxiv.org/abs/1101.1304} {arXiv:1101.1304 [astro-ph.HE]} \BibitemShut
  {NoStop}%
\end{thebibliography}%

\end{document}